\documentclass[12pt]{article}

\usepackage{setspace} 
\usepackage[vmargin = 1.25in, hmargin =1.25in]{geometry}

\usepackage[dvipsnames,svgnames]{xcolor}
\usepackage{amsmath, amssymb, amsfonts, graphicx, tikz,pdflscape, mathtools, amsthm, upgreek, bm,pgfplots,csvsimple,multirow, multicol, booktabs,bbm}
\usepackage{verbatim}
\usepackage[longnamesfirst]{natbib}
\setcitestyle{citesep={,}}
\usepackage{accents}
\newcommand{\ubar}[1]{\underaccent{\bar}{#1}}
\usepackage{needspace}

\usepackage{comment}
\usepackage[inline]{enumitem}
\usepackage{tocvsec2}
\usepackage{threeparttable}
\usetikzlibrary{calc, cd, arrows}
\usetikzlibrary{decorations.pathreplacing}
\usetikzlibrary{positioning}
\usetikzlibrary{calc} 
\usetikzlibrary{arrows}
\usetikzlibrary{patterns}
\usetikzlibrary{decorations.markings}
\usetikzlibrary{shapes.misc}
\usetikzlibrary{matrix,shapes,arrows,fit,tikzmark}
\usetikzlibrary{cd}
\usepgfplotslibrary{fillbetween}
\usetikzlibrary{intersections}
\usepackage[skip = 10pt]{caption}
\usepackage{subcaption}
\allowdisplaybreaks[1]
\allowdisplaybreaks[1]

\definecolor{Blue}{RGB}{86,180,233}
\definecolor{Orange}{RGB}{230,159,0}
\definecolor{Green}{RGB}{0,158,115}
\definecolor{GmailBlue}{RGB}{42, 93, 176} 
\usepackage[
	pagebackref,
	colorlinks=true,
	citecolor= GmailBlue,
	linkcolor=GmailBlue,
	urlcolor = GmailBlue
]{hyperref}

\newcommand{\bibtexorder}[1]{}

\usepackage{pgfplots}
\usepgfplotslibrary{groupplots,colorbrewer}
\pgfplotsset{compat=newest}
\pgfplotsset{cycle list/Set1}
\usepackage{tikz}
\usetikzlibrary{matrix,calc,shapes,arrows.meta,positioning}
\tikzset{
    vertex/.style = {shape=circle,draw, minimum size = 1.8em, inner sep = 0pt},
    edge/.style = {->,> = latex}
}

\usepackage[capitalize,noabbrev]{cleveref}

\makeatletter
\AddToHook{cmd/appendix/before}{\def\cref@section@alias{appendix}\def\cref@subsection@alias{appendix}}
\makeatother

\newtheoremstyle{break}
{}
{}
{\itshape}
{}
{\bfseries}
{}
{\newline}
{}

\theoremstyle{break}
\newtheorem{thm}{Theorem}
\newtheorem*{theorem*}{Theorem}
\newtheorem*{cor*}{Corollary}

\newtheorem{prop}{Proposition}
\newtheorem{lem}{Lemma}

\crefname{prop}{Proposition}{Propositions}
\crefname{thm}{Theorem}{Theorems}
\crefname{lem}{Lemma}{Lemmas}
\crefname{blem}{Lemma}{Lemmas}

\theoremstyle{definition}
\newtheorem{defn}{Definition}

\newtheorem*{rem*}{Remark}
\newtheorem*{claim*}{Claim}

\newtheorem{as}{Assumption}

\def\b{\beta}
\def\g{\gamma}
\def\d{\delta}
\def\e{\varepsilon}

\def\th{\theta}

\def\k{\kappa}

\def\s{\sigma}

\def\G{\Gamma}
\def\D{\Delta}
\def\Th{\Theta}

\def\R{\mathbf{R}}

\def\QQ{\mathcal{Q}}

\def\P{\mathbf{P}}

\def\pd{\partial}

\DeclareMathOperator{\E}{\mathbb{E}}

\DeclareMathOperator*{\argmax}{argmax}
\DeclareMathOperator*{\maz}{maximize}

\newcommand{\abs}[1]{\lvert #1 \rvert} 

\newcommand{\Paren}[1]{\left( #1 \right)}

\newcommand{\Brac}[1]{\left[ #1 \right]}

\newcommand{\Set}[1]{\left\{ #1 \right\}}

\newcommand{\de}{\mathop{}\!\mathrm{d}}

\usepackage{datetime}
\newdateformat{specialdate}{\THEDAY~\monthname[\THEMONTH] \THEYEAR}

\title{Competitive Sequential Screening\thanks{We thank Honjar Xing for excellent research assistance. For helpful comments, we thank Mark Armstrong, Alessandro Bonatti, Laura Doval, Glenn Ellison, Drew Fudenberg, Krittanai Laohakunakorn, Stephen Morris, Ellen Muir, Jean Tirole, Juuso V\"alim\"aki, Alex Wolitzky, and Jidong Zhou. For useful feedback, we thank seminar audiences at Cornell, MIT, Oslo, Berkeley, Stanford, Penn State, WUSTL, Notre Dame, UGA, Georgetown, Oxford, Western Ontario, and Surrey.}}

\author{Ian Ball\thanks{Department of Economics, MIT, \texttt{ianball@mit.edu}.}
\and Deniz Kattwinkel\thanks{Department of Economics, UCL, \texttt{d.kattwinkel@ucl.ac.uk}.}
\and Jan Knoepfle\thanks{School of Economics and Finance, Queen Mary University of London, \texttt{j.knoepfle@qmul.ac.uk}.}}

\date{\specialdate\today}

\begin{document}

\begin{titlepage} 

\maketitle \thispagestyle{empty} 

\begin{abstract}
We study competition between firms that contract with consumers before the consumers fully learn their product preferences. In a Hotelling duopoly, firms screen consumers by offering menus of option contracts. We characterize the unique equilibrium. Consumers select contracts from both firms. Each consumer is endogenously locked into the firm from which he chooses an option with a lower strike price. Lock-in yields inefficient consumption. Yet earlier contracting stiffens competition because less informed consumers are more homogeneous. Sufficiently early contracting raises consumer surplus relative to spot pricing---reversing the ranking under monopoly. Exclusive contracting further increases consumer surplus by intensifying competition.



\end{abstract}



\end{titlepage}


\section{Introduction}


When consumers learn their valuations for a product over time, a monopolist can increase revenue by contracting in advance, before the consumers' uncertainty is resolved \citep{BaronBesanko1984,courty2000sequential}. In these settings, it is optimal for a monopolist to offer a menu of option contracts so that consumers self-select according to their beliefs about their valuation for the product. In practice, \emph{competing} firms often offer such
advance contracts. For example, households buy annual subscriptions to streaming platforms such as Netflix and Disney\textsuperscript{+} without knowing what future releases they will watch on each platform.\footnote{According to some estimates, the worldwide subscription market generates $\$500$ billion in annual revenue \citep{DMR2025}. Certain subscriptions can be explained by behavioral biases, e.g., where consumers ``pay not to go to the gym'' \citep{DellaVignaMalmendier2004,DellaVignaMalmendier2006} or fail to cancel unused subscriptions due to inattention \citep{EinavEtal2025}. Abstracting from these effects in particular examples, we analyze the general structure of competition between firms that sequentially screen consumers who learn over time.} Skiers select season passes from Epic and Ikon before learning their future availability or the weather conditions at different mountains.\footnote{This is distinct from standard quantity discounts because consumers' information arrives sequentially. Indeed, each pass is for sale only until early December, before the ski season begins.} In the business-to-business context, firms often pay upfront for reserved capacity from multiple input suppliers before they learn their exact production requirements.

We introduce a model of competition between firms that offer menus of option contracts to sequentially screen consumers. While sequential screening by a monopolist tends to harm consumers, we find that sequential screening by competing firms can benefit consumers. In our duopoly model, consumers are initially uncertain of which product they will prefer, so they hold contracts at both firms even though they can use only one product at a time. Holding a more premium contract at one firm partially locks the consumer into that firm, 
distorting future consumption. On the other hand, we find that earlier contracting intensifies competition relative to classical spot pricing because consumers are less informed and hence less heterogeneous. Under sufficiently early contracting, the benefit from this competition outweighs the loss from allocative distortion, so consumer surplus increases. 

In our model, two firms, $A$ and $B$, offer horizontally differentiated products, $A$ and $B$, to a population of consumers with unit demand. We formulate the model in terms of a single consumer drawn from this population. The consumer privately observes an imperfect signal of his valuations for the products. The firms simultaneously post menus of option contracts (i.e., subscriptions), where each contract specifies a subscription fee and a strike price. A spot price is the special case of a contract in which the subscription fee is zero. In the first period, the consumer chooses one contract from each firm, based on his belief about his future preferences. In the second period, the consumer learns his exact valuations for the products and then chooses which product to purchase. Formally, the consumer's preferences are determined by his \emph{position} on the Hotelling line, which is the sum of his initial signal (i.e., type) and an independent taste shock that is realized after contracting.

Sequential screening introduces new strategic effects. Even if the consumer ultimately uses only one product, he is initially uncertain of which product he will prefer. To retain the option to use either product, the consumer enters contracts with both firms. This multi-homing in contracts creates strategic complexity for the firms. When a firm designs its menu of contracts, it must take into account that its menu influences the consumer's choice of contracts at \emph{both} firms. 

In \cref{res:equilibrium}, we identify the essentially unique equilibrium of this game. To ensure that the product market is covered, we assume that the consumer's average valuation for the two products is sufficiently high. Every consumer type chooses one contract from each firm. Consumers who expect to prefer product $B$ select more ``premium'' contracts at firm $B$ (with lower strike prices and higher subscription fees) and less ``premium'' contracts at firm $A$ (with higher strike prices and lower subscription fees). Even if the firms are symmetric, the resulting allocation is inefficient because the consumer is endogenously \emph{locked into} the firm where he holds a more premium subscription. The consumer sometimes purchases that firm's product, because of its lower strike price, even when he likes the other product more. 

 By comparing our equilibrium with the outcome if only one firm is present, we derive testable implications for how an incumbent monopolist's pricing changes when a new, horizontally differentiated firm enters. Subscription fees \emph{decrease}, but strike prices \emph{increase}. 
 This prediction is consistent with the evolution of the streaming market. In 2022, Netflix first introduced a basic subscription plan with advertisements (effectively a higher usage price) after losing subscribers to new entrants such as Disney\textsuperscript{+}.\footnote{See \url{https://tinyurl.com/2ypanxz6}.} 
 Strike prices increase because, under competition, downward distortion reduces the consumer's information rents more effectively. The reason is that consumption shifts to the other firm's product, rather than to non-consumption, and consumers who are more eager for one firm are less eager for the other.

To characterize the equilibrium, we explicitly account for how each firm's menu affects the consumer's choices from the other firm's menu. For concreteness, consider firm $B$. To compute firm $B$'s best response to firm $A$'s menu, we consider an arbitrary deviation by firm $B$ to a general stochastic mechanism, together with an obedient recommendation of which contract each type should select from firm $A$'s menu. We obtain a dynamic virtual value expression that depends on these recommendations. We then maximize this expression pointwise to determine firm $B$'s best response. Our proof therefore establishes a robustness property of the equilibrium: firms cannot profit even by deviating to general mechanisms involving lotteries.

In order to isolate the roles of contract-timing, multi-homing, and competition, we next analyze three benchmark settings: spot pricing, exclusive contracting, and multi-product monopoly. 

\emph{Spot pricing} is the classical benchmark in which firms post prices at the time of exchange, after the consumer learns his valuations for the products. We characterize the unique equilibrium (\cref{res:Hotelling}). If the firms are symmetric, then they charge the same price, so the resulting allocation is efficient, provided that the market is covered. 

Under \emph{exclusive contracting}, each firm offers a menu of contracts, as in the main model, but the consumer is prohibited from entering contracts with more than one firm. We characterize the unique Pareto-dominant equilibrium for firms (\cref{res:exclusive}). In this equilibrium, the firms split the subscription market.  The consumer chooses his exclusive supplier based on his expected preferences, and he is completely locked into his chosen firm.  In equilibrium, each firm charges each consumer the same strike price that it would charge that consumer if it were a monopolist. If the consumer experiences an extreme taste shock in favor of the other firm, then he does not purchase either product. When the firms are symmetric, the resulting allocation is less efficient than in the main model with non-exclusive contracting.

In the \emph{multi-product monopoly} setting, a joint monopolist controls both products $A$ and $B$. This monopolist can offer joint option contracts, which specify strike prices for both products. If the firms are symmetric and the average product valuation is sufficiently high, then we show (\cref{res:multi-product_monopoly}) that it is optimal for the monopolist to offer a single inclusive contract that entitles the consumer to purchase either product at no additional cost. The resulting allocation is efficient, but consumer surplus is lower than in all other settings. 

Finally, we compare consumer and producer surplus across the three competitive settings: non-exclusive contracting, exclusive contracting, and spot pricing. The consumer surplus ranking generally depends on the balance between two countervailing effects: efficiency and competition. Relative to spot pricing, subscriptions reduce efficiency by locking in consumers---partially with non-exclusive contracts, or completely with exclusive contracts. On the other hand, competition for consumers is stiffer when consumers are less informed, and hence less heterogeneous. Competition is further intensified under exclusivity because each firm's offer must be better than its competitor's, rather than simply good enough to induce the consumer to multi-home.


If contracting takes place sufficiently early, so  consumers' signals are sufficiently uninformative, then the competition effect dominates: consumer surplus strictly increases (and producer surplus strictly decreases) from spot pricing to non-exclusive contracting to exclusive contracting (\cref{res:surplus}). 
This result reverses the classical finding under monopoly. Relative to spot pricing, early contracting reduces consumer surplus under monopoly but increases consumer surplus under competition.

The rest of the paper is organized as follows. \cref{sec:related_literature} discusses related literature. \cref{sec:model} introduces the main model. As a warm-up, \cref{sec:single-firm} considers the case in which only one firm is present. \cref{sec:equilibrium_analysis} characterizes the equilibrium of the main model. \cref{sec:other_regimes} analyzes the alternative settings: spot pricing, exclusive contracting, and multi-product monopoly. \cref{sec:welfare_comparison} compares consumer and producer surplus across these settings. \cref{sec:conclusion} is the conclusion. Proofs are in \cref{sec:proofs}.


\subsection{Related literature} \label{sec:related_literature}

 Our paper provides the first analysis of competition between firms that can sequentially screen consumers by offering unrestricted menus of option contracts. We obtain new insights about efficiency and consumer welfare because of the combination of multi-homing (in first-period contract choices) and single-homing (in second-period product purchases). This combination is not present in models of competitive static price discrimination or monopolistic sequential screening.
 


The literature on competitive intertemporal pricing generally considers behavior-based price discrimination, where firms condition prices on consumers' purchase histories; see \cite{FudenbergVillas-Boas2007} for a survey. In our model, firms screen consumers according to the consumers' beliefs (about their future preferences), not their past purchases; such screening is possible in our setting because consumers are \emph{partially} informed about their future preferences. Most relevant to our work, \cite{FudenbergTirole2000} study behavior-based pricing in a symmetric Hotelling duopoly.\footnote{Earlier, \cite{CaminalMatutes1990} study a similar problem with more restrictive preferences. \cite{Villas-Boas1999} studies a similar model with overlapping generations and an infinite horizon, but that model does not allow firms to offer intertemporal contracts.} Each firm offers a differentiated product. In the first period, the consumer chooses which product to purchase. Firms observe this choice, so in the second period each firm can set one price for ``loyalists'' and another for ``newcomers'' poached from the competitor. In their long-term contracting specification, which is closest to our model, firms can commit to contracts specifying prices in both periods, but firms \emph{cannot} commit to not poach the other firm's consumers in the second period. Their model allows consumer preferences to be either constant or independent across periods. In each case, there is no scope to screen consumers with option contracts.\footnote{If preferences are constant, the consumer has no residual uncertainty at the time of contracting. If preferences are independent, all consumer types have the same belief about future preferences.}  

A few papers study competition between two firms that offer advance-purchase discounts. Formally, each firm offers its product at one price in advance and another price on the spot. The main difference from our model is that firms cannot offer option contracts, and consumers cannot select contracts at both firms. In \cite{gale1993price}, every consumer believes, prior to contracting, that he is equally likely to prefer either product. Because there is no initial horizontal heterogeneity across consumers, the equilibrium advance-purchase prices equal marginal cost. In \cite{CachonFeldman2017}, consumers differ in their likelihood of preferring each firm, so advance-purchase prices are strictly positive in equilibrium. Complementary to our results, they find that the equilibrium with advance purchases generally yields lower profits than the spot-pricing equilibrium.

A larger literature studies \emph{static} models of competitive nonlinear pricing; for surveys, see \cite{Stole2007} and \cite{Armstrong2016}. In these papers, the consumer has no uncertainty about his preferences at the time of contracting, so the consumer does not value future \emph{flexibility} to make different consumption choices.\footnote{\cite{ArmstrongZhou2022} study how the consumer's information about his valuations influences the resulting pricing equilibrium. In their model, however, consumers do not learn over time; consumers make purchase decisions based only upon their initial signals.} 
Models with one-stop shopping include \citet{ArmstrongVickers2001}, \cite{RochetStole2002}, \cite{YangYe2008}, \cite{Bonatti2011}, \cite{Gomes2022}, and \cite{TamayoTan2025}. Each firm essentially faces a monopolistic nonlinear pricing problem, subject to a type-dependent outside option (determined by the other firm's schedule).\footnote{See also \cite{guerrieri2010adverse, garrett2019competitive} where consumers search to uncover the menu of quality-price pairs offered by competing sellers.} In \cite{armstrong2010competitive}, two competing sellers offer two-part tariffs for \emph{two goods} with tariffs depending on whether the consumer purchases one or both goods from that seller. By offering a fixed discount whenever the consumer purchases both goods from the same firm (independent of the precise quantities purchased), the sellers partially separate the (two-dimensional) horizontal types. 

The mechanism design literature studies sequential screening by a \emph{monopolist}; see \cite{bergemann2019dynamic} and \cite{pavan2025dynamic} for surveys. \cite{BaronBesanko1984} first study how a regulator can optimally sequentially screen 
a natural monopolist. \cite{courty2000sequential} provide a more general analysis of optimal sequential screening by the seller of a single product.\footnote{\cite{Miravete1996} studies a similar model with a more restrictive class of contracts.} Their characterization serves as a useful benchmark for our model; see our \cref{res:monopoly_benchmark}.  Subsequent work extends these two-period, single-agent models in various directions, while retaining the assumption of a single monopolist.\footnote{See   \cite{deb2015dynamic}, \cite{grubb2015cellular}, and \cite{krahmer2015optimal}.}
In particular, \cite{esHo2007optimal} consider multiple consumers competing for a single good. \cite{pavan2014dynamic} consider a more general multi-agent model with Markovian types and an infinite horizon.

Finally, there is a more methodological literature on competing mechanisms, surveyed by \cite{Martimort2006}. Revelation and delegation principles for common agency settings appear in \cite{MartimortStole2002} and \cite{Peters2001}. \cite{pavan2010truthful} propose a revelation game between principals in which the agent reports to each principal his private type and his choice at the other principal. In our solution approach, we allow each firm to design a mechanism that specifies, for each consumer type, an allocation at that firm and a recommended choice from the other firm's menu, subject to obedience constraints.
\cite{AttarEtal2025} analyze a model in which competing principals can make private disclosures to the agents.  Motivated by applications, we allow firms to post menus of contracts; we confirm that our equilibrium remains an equilibrium if firms can offer stochastic mechanisms. 


\section{Model} \label{sec:model}

We introduce the model, define our equilibrium notion, and discuss modeling choices. 

\subsection{Setting}

\paragraph{Environment} There are two firms, denoted $A$ and $B$. Each firm $i$ produces a differentiated product, called product $i$. There is a single consumer, with quasilinear utility. The consumer's \emph{valuations} for the products are given by
\[
    v_A (\th) = v_0 - \th, 
    \qquad
    v_B( \th) = v_0 + \th, 
\]
where $v_0$ is a known constant and the parameter $\th$ is private information that the consumer learns sequentially, as described below. These valuations are induced by the classical Hotelling line specification with linear transportation costs. Under this interpretation, firm $A$ is located at the left endpoint and firm $B$ is located at the right endpoint.\footnote{The Hotelling interpretation requires that every consumer lies between the two firms, so  the support of $\th$ must be bounded. We will assume for convenience that the support of $\th$ is unbounded. Our equilibrium construction works if the support of $\th$ is bounded but sufficiently large.}
The parameter $\th$ is the consumer's \emph{position} on the line. If the consumer receives product $i$ only, then his consumption utility is $\max \{ 0, v_i (\th) \}$; the maximum reflects \emph{free disposal}. The consumer has \emph{unit demand}: he can only use one of the products. If he receives both products, then his consumption utility is $\max \{ 0, v_A (\th), v_B( \th) \}$.


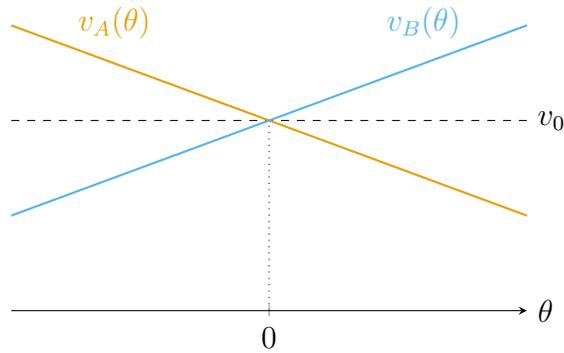
\begin{figure}
\centering
\begin{tikzpicture}
    \begin{axis}[
        xmin = -1, xmax = 1,
        ymin = 1.5, ymax = 6,
        axis y line = none,
        axis x line = middle,
        xlabel = {$\th$},
        xlabel style={at=(current axis.right of origin), anchor=west}, 
        ymajorticks = false,
        xtick = {0},
        xticklabels = {$0$},
        clip = false,
    ]
        \addplot[dashed] coordinates { (-1,3.5) (1,3.5)} node [pos = 1, anchor = west] {$v_0$};
        \addplot[thick, Orange] coordinates { (-1,4.5) (1,2.5)} node [pos = 0.2, yshift = 5pt, anchor = south] {$v_A(\th)$};  
        \addplot[thick, Blue] coordinates { (-1,2.5) (1,4.5)} node [pos = 0.8 , yshift = 5pt, anchor = south] {$v_B(\th)$};

        \addplot[dotted] coordinates {(0,1.5)  (0,3.5)};
    \end{axis}
\end{tikzpicture}
\caption{Consumer's valuations along the Hotelling line}
\label{fig:valuations}
\end{figure}

\cref{fig:valuations} plots the functions $v_A$ and $v_B$. Note that $v_A(\th) + v_B(\th) = 2 v_0$ and $v_B (\th) - v_A( \th) = 2 \th$. Thus,  $v_0$ is the average valuation across the two products. The consumer's position $\th$ determines his relative preference for product $B$ over product $A$.  
If $\th > 0$, the consumer values $B$ more; if $\th < 0$, the consumer values $A$ more.

\paragraph{Information} In the first period, the consumer privately learns his \emph{type} $\g \in \G \coloneq [\ubar{\g}, \bar{\g}]$, where $\ubar{\g} < \bar{\g}$. The consumer's type $\g$ is drawn from a distribution $G$, which has continuous, strictly positive density $g \colon [\ubar{\g}, \bar{\g}] \to (0, \infty)$. In the second period, the consumer learns his position
\[
    \th = \g + \e,
\]
where $\e$ is a taste shock that is drawn independently of $\g$ from an integrable, full-support distribution $F$, which has continuous, strictly positive density  $f \colon \R \to (0, \infty)$.\footnote{The independent components $\g$ and $\e$ both capture horizontal taste. By contrast, \citet[p.~616]{Ellison05} observes that in models with both vertical and horizontal heterogeneity, the conclusions can be sensitive to the common assumption that vertical and horizontal tastes are independent.}  Denote the set of consumer positions by $\Th = \R$. Without loss, we assume $\E [\e] = 0$. Thus, $\E [ \th | \g ] = \g$, so the consumer's type $\g$ equals his expected position. We impose the following standing assumption on the densities $g$ and $f$.

\begin{as}[Regularity] \label{as:regularity}~
\begin{enumerate}
    \item The  density $g$ is log-concave. 
    \item The density $f$ is log-concave and symmetric.
\end{enumerate}
\end{as}

Some of our results still hold if \cref{as:regularity} is weakened; we will comment on this throughout the paper. Log-concavity ensures that each firm's best response is determined by local optimality conditions. Many standard distributions have log-concave densities, including  uniform, normal, exponential, logistic, gamma, and chi-squared distributions \citep[Table 1, p.~455]{BagnoliBergstrom2005}. \cref{as:regularity} imposes symmetry on the distribution of the taste shock, but not on the distribution of the consumer's type. Thus, the two firms can be asymmetric. 

\paragraph{Contracting} A \emph{contract} (with firm $i$) is a pair $(p_i, s_i)$, which specifies that the consumer pays the subscription fee $s_i$ to firm $i$ in the first period for the right (but not the obligation) to purchase product $i$ at price $p_i$ in the second period. In the language of option contracts, $p_i$ is the \emph{strike price}. This class of contracts includes two important special cases. If $p_i = 0$, then the consumer receives the product in the second period at no additional cost; this can be interpreted as an advance purchase or an inclusive plan. If $s_i = 0$, then the consumer pays only if he receives the product; this is a spot price. Our model can be re-interpreted to capture repeated consumption; see \cref{sec:discussion}. 

Each firm offers a menu of contracts for the consumer to choose from. This menu is represented by a subscription schedule. Formally, a \emph{subscription schedule} for firm $i$ is a lower semicontinuous function 
\begin{equation} \label{eq:s_i}
    s_i \colon [0, \infty) \to [0, \infty], 
\end{equation}
assigning to each strike price $p_i \in [0, \infty)$ a subscription fee $s_i (p_i) \in [0,  \infty]$. In particular, $s_i (p_i) = \infty$ if firm $i$ does not offer a contract with strike price $p_i$. Lower semi\-continuity ensures that the consumer has a best response to any pair of subscription schedules.

We restrict strike prices to be nonnegative. Since the consumer can freely dispose of each product, the outcome of any contract with a negative strike price can be replicated by a contract with zero strike price.\footnote{Formally, suppose the consumer selects a contract $(p_i, s_i)$ with $p_i < 0$. It is strictly optimal for the consumer to always purchase product $i$, so his total payment to firm $i$ is $s_i + p_i$. Therefore, the contract $(p_i, s_i)$ can be replaced by the contract $(0, s_i + p_i)$, without changing final payoffs. Technically, we could have $s_i + p_i \leq 0$, but we separately argue that allowing contracts with negative subscription fees does not change the equilibrium characterization.} We also require subscription fees to be nonnegative; allowing negative subscription fees would not change our equilibrium characterization.

The consumer can choose not to enter a contract with firm $i$. We represent this choice as the \emph{null contract} with $p_i = \infty$ and $s_i = 0$. To accommodate this null contract, we adopt the following convention: for any subscription contract $s_i$, the domain is extended to $[0, \infty]$ by setting $s_i (\infty) = 0$. Note that this extension preserves lower semicontinuity.\footnote{Here and below, we endow $[0, \infty]$ with the order topology.}
 
\paragraph{Timing} The game proceeds as follows over two periods. 
\begin{enumerate}
    \item[1.a] \label{it:schedule} Simultaneously, each firm $i$ posts a subscription schedule $s_i$.
    \item[1.b] The consumer observes his type $\g$ and the subscription-schedule pair $(s_A, s_B)$.  
    \item[1.c] \label{it:contract} The consumer selects, for each $i$, a strike price $p_i$ from schedule $s_i$. 
    \item[2.a] \label{it:new} The consumer observes his position $\th$. 
    \item[2.b] The consumer chooses, for each $i$, whether to purchase product $i$. 
\end{enumerate}

\paragraph{Payoffs} Final payoffs are determined by the selected contracts, $(p_A,s_A(p_A))$ and $(p_B, s_B( p_B))$, the purchased quantity vector $q = (q_A, q_B) \in \{0,1\}^2$, and the consumer's position $\th$. There is no discounting. Firm $i$'s payoff is its revenue 
\[
    \pi_i = s_i (p_i) + p_i q_i,
\]
which includes upfront revenue from the subscription fee and subsequent revenue if its product is purchased. The consumer's utility equals consumption utility net of total payments (from subscription fees and strike prices):
\[
u = - s_A (p_A) - s_B (p_B) - p_A q_A - p_B q_B + \max \{ 0, v_A(\th) q_A, v_B(\th) q_B \}.
\]

\subsection{Equilibrium} 

To define our solution concept, we begin by defining strategies. For each firm $i$, a strategy is a subscription schedule $s_i$, as defined in \eqref{eq:s_i}. Facing a pair of subscription schedules, the consumer chooses a strategy $(p,q)$,\footnote{Technically, $(p,q)$ is a strategy in a particular continuation game. Our use of the term strategy is consistent with the convention in (single-principal) mechanism design, where a player's strategy is defined as a map from types to messages \emph{within} the mechanism chosen by the principal.} which consists of a contract-selection strategy 
\[
p = (p_A, p_B) \colon \G \to [0, \infty]^2,
\]
and a product-purchase strategy 
\[
q = (q_A, q_B) \colon \G \times \Th \to \{0,1\}^2.
\]
The contract-selection strategy specifies which strike price the consumer selects from each firm's subscription schedule after learning his type $\g$. The consumer can choose not to enter a contract with firm $i$ by selecting the  null-contract with strike price $p_i = \infty$. The product-purchase strategy specifies whether the consumer purchases each firm's product after learning his position $\th$.


To define equilibrium, we first define a best response for the consumer. Below, $\E_{\th |\g}$ denotes an expectation over the random position $\th$ given the type realization $\g$. 

\begin{defn}[Consumer best response] A consumer strategy $(p, q)$ is a \emph{best response} to a subscription-schedule pair $(s_A, s_B)$ if the following hold.
\begin{enumerate}
    \item \label{it:BR_EP} For  each $(\g, \th) \in \G \times \Th$, 
    \[
        q(\g, \th) \in \argmax_{q' \in \{0,1\}^2} ~ \bigl( - p_A(\g) q'_A - p_B(\g) q'_B +  \max \{ 0, v_A(\th) q'_A, v_B (\th ) q'_B \} \bigr).
    \]
    \item \label{it:BR_EI} For each $\g \in \G$, 
    \[
        p(\g) \in \argmax_{ p' \in [0,\infty]^2} ~ \Paren{ - s_A (p_A') - s_B (p_B') + \E_{\th | \g} \Brac{  \max \{ 0, v_A(\th) - p'_A, v_B( \th) - p'_B \} }}.
    \]
\end{enumerate}
\end{defn}

The conditions proceed backward in time. 
Condition~\ref{it:BR_EP} requires that the consumer purchase products optimally in the second period, given the contracts selected in the first period. The subscription fees $s_A( p_A(\g))$ and $s_B( p_B(\g))$ do not appear in the objective because these payments are sunk by the second period. Since the consumer has unit demand, purchasing both products is never strictly optimal. Condition~\ref{it:BR_EI} requires that the consumer select contracts optimally in the first period, anticipating that he will subsequently purchase products optimally in the second period. The expression $\max\{ 0, v_A( \th) - p_A', v_B( \th) - p_B' \}$ is the consumer's second-period utility from purchasing the products optimally, given that the product prices are $p_A'$ and $p_B'$ and that his position is $\th$. 

Now we define our solution concept. To \emph{explicitly} impose sequential rationality on the consumer, we would need to specify the consumer's response to each pair of subscription schedules. In order to simplify the equilibrium object, we specify the consumer's response only to the on-path subscription-schedule pair, and we impose sequential rationality at off-path schedules \emph{implicitly} via condition \ref{it:firm} below. We write $\E_{\g, \th}$ to denote expectations over the random vector $(\g, \th)$. 

\begin{defn}[Equilibrium]  \label{def:equilibrium} A strategy profile $(s_A^\ast, s_B^\ast, p^\ast,q^\ast)$ is an \emph{equilibrium} if the following hold. 
\begin{enumerate}
    \item \label{it:consumer} The consumer's strategy $(p^\ast,q^\ast)$ is a best response to the schedule pair $(s_A^\ast, s_B^\ast)$.
    \item \label{it:firm} For each firm $i \in \{A, B\}$ and each alternative subscription schedule $s_i' \neq s_i^\ast$, there exists a consumer best response $(p', q')$ to $(s_i', s_{-i}^\ast)$ such that
    \begin{equation*}
      \E_{\g, \th} [ s_i' (p_i' (\g)) + p_i' (\g) q_i' (\g, \th) ] \leq  \E_{\g, \th} [ s_i^\ast (p_i^\ast (\g)) + p_i^\ast (\g) q_i^\ast(\g, \th) ].
    \end{equation*}
\end{enumerate}
\end{defn}

Condition~\ref{it:consumer} is the best response condition for the consumer. Condition~\ref{it:firm} requires that neither firm has a profitable unilateral deviation. Suppose that firm $i$ unilaterally deviates from the subscription schedule $s_i^\ast$ to an alternative subscription schedule $s_i'$. The consumer, now facing the schedule pair $(s_i', s_{-i}^\ast)$ rather than $(s_i^\ast, s_{-i}^\ast)$, will generally find it optimal to choose different strike prices at \emph{both} firms. To evaluate the firms' resulting payoffs, we must select a consumer best response to $(s_i', s_{-i}^\ast)$. We do not specify this selection as part of the equilibrium, but condition~\ref{it:firm} requires that there \emph{exists} a consumer best response that makes firm $i$'s deviation to $s_i'$ unprofitable. It turns out that for any subscription-schedule pair, the consumer's optimal choice of contracts is unique for almost every type (\cref{res:BR_structure}, \cref{sec:proof_equilibrium_uniqueness}). Thus, each firm's expected revenue is independent of which best response the consumer plays. Therefore, in condition~\ref{it:firm}, we could equivalently replace ``there exists a best response $(p',q')$ to $(s_i', s_{-i}^\ast)$'' with ``for every best response $(p',q')$ to $(s_i', s_{-i}^\ast)$.''\footnote{\label{ft:NE_firms}In fact, \cref{res:revenue} implies that each subscription-schedule pair $s = (s_A, s_B)$ can be associated with a unique payoff vector $\hat{\pi}(s) = (\hat{\pi}_A( s), \hat{\pi}_B(s))$ for the firms. Thus, condition~\ref{it:firm} could also be replaced with the equivalent condition that $(s_A^\ast, s_B^\ast)$ is a Nash equilibrium of the strategic-form game (between the firms) induced by the payoff functions in $\hat{\pi}$.}


\subsection{Discussion} \label{sec:discussion}

We discuss interpretations of our model and highlight important modeling choices. 

\paragraph{Repeated consumption} In our model, consumption occurs only once. With the following interpretation, however, our model can capture repeated consumption, which is common in some of our motivating applications. Formally, the following model with repeated consumption is payoff-equivalent to our model. Partition the second period into finitely many subperiods, indexed by $t \in T$. In each subperiod $t$, the consumer experiences an independent taste shock $\e_t$, drawn from the distribution $F$.  The consumer's valuation for product consumption in subperiod $t$ is determined by 
his position $\th_t = \g + \e_t$, where the type $\g$ is persistent. Contracts and subscription schedules are defined as in the main model. In the first period, the consumer learns his type $\g$ and then selects a contract $(p_i, s_i)$ from each firm $i$. In each subperiod $t$, the consumer observes his position $\th_t$ and then chooses, for each $i$, whether to purchase product $i$ at the strike price $p_i$. Define second-period payoffs as the average over the subperiods. The expected payoffs in this model are identical to those in the main model. 

Under this interpretation, the unit-demand assumption means that the consumer cannot use both products within the same subperiod. Under a finer partition into subperiods, the unit-demand assumption is more innocuous, but the independence of taste shocks is more restrictive. In many applications, there is a suitable subperiod length for which both assumptions are reasonable.\footnote{For streaming subscriptions, a reasonable subperiod may be the time it takes a consumer to complete a season of a show. For ski passes, a subperiod could be one day (or weekend) of skiing.}

\paragraph{The class of contracts} The contracts in our model encompass a range of common pricing formats. A contract $(p_i, s_i)$ is literally an option contract with strike price $p_i$, but it can also represent a subscription contract with per-use price $p_i$. In the context of streaming subscriptions, this price $p_i$ takes the form of the length of ads that the user must watch with each show. A contract $(p_i, s_i)$ can also represent a partially refundable reservation with sticker price $p_i + s_i$ that can be canceled for a refund of $p_i$ (or equivalently, a cancellation fee of $s_i$).  In particular, a contract with $p_i = 0$ corresponds to an advance purchase or an inclusive subscription. A contract with $s_i = 0$ corresponds to a fully refundable reservation or spot pricing.

Our definition of a contract does not allow for stochastic contracts such as random strike prices or product lotteries. We show below, however, that the equilibrium from \cref{res:equilibrium} remains an equilibrium of the alternative game in which each firm can offer a  stochastic mechanism. 

\paragraph{Product demand system} We consider the canonical Hotelling duopoly, where consumers have unit product demand and consumer heterogeneity is horizontal. This combination allows us to illustrate an important new effect: consumers shift from single-homing, under spot pricing,  to multi-homing under subscriptions. Consumers are willing to pay for both subscriptions because they are uncertain of \emph{which} product they will prefer. Horizontal heterogeneity drives the equilibrium subscription pattern among consumers---more premium subscriptions at one firm are paired with less premium subscriptions at the other. 

\paragraph{Production costs} In our model, firms maximize revenue, so production costs are implicitly assumed to be zero. Production costs can be accommodated through the following standard normalization. If product $i$ costs $c_i$ to produce, then we interpret  $v_i(\th)$ as the consumer's valuation for product $i$ net of the cost $c_i$, and we interpret the price $p_i$ as the markup above $c_i$.\footnote{Formally, in order to maintain our assumption that prices (not markups) are nonnegative, we must enlarge each firm $i$'s strategy set to include contracts with markups in $[-c_i, \infty)$. This standard normalization preserves payoffs as long as the agent does not dispose of either product. But with nonnegative strike prices (even with negative mark-ups), disposal cannot occur with positive probability in any equilibrium.} 





\section{Warm-up: Single firm} \label{sec:single-firm}

In this section, we consider the benchmark in which only one of the two firms, say firm $i$, is present. The consumer's valuation for product $i$ is $v_i(\th)$, as in the main model. This benchmark allows us to highlight the effect of competition in the main model; in \cref{sec:multi_product_monopoly}, we consider the case in which a  monopolist controls both products.

With firm $i$ as the only principal, a sequential version of the revelation principle holds. Firm $i$ chooses an incentive-compatible direct mechanism, which consists of an allocation rule $q_i \colon \G \times \Th \to [0,1]$ and a transfer rule $t_i \colon \G \times \Th \to \R$. The allocation specifies the probability that the consumer receives firm $i$'s product. The transfer is the total payment from the consumer to firm $i$. This is a monopolistic sequential screening problem, as analyzed in \cite{courty2000sequential}. To facilitate comparison with the main model, we will describe an indirect implementation of firm $i$'s optimal mechanism, as follows. Firm $i$ offers a subscription schedule $s_i$, as defined in \eqref{eq:s_i}. Given this schedule, a strategy for the consumer consists of a contract-selection strategy $p_i \colon \G \to [0,\infty]$ and a product-purchase strategy $q_i \colon \G \times \Th \to \{0,1\}$. 

To describe the optimal mechanism, we introduce notation for the consumer's demand from the monopolist firm $i$. Suppose the strike price in the second period is $p_i$. Then it is optimal for the consumer with position $\th$ to purchase product $i$ if and only if $v_i (\th) \geq p_i$.  Define the interim demand function $Q_i^M \colon [0, \infty) \times \G \to [0,1]$ by
\begin{equation} \label{eq:interim_monop_demand}
    Q_i^M(p_i| \g) = \P_{\th| \g} ( v_i ( \th) \geq p_i ),
\end{equation}
where the probability is evaluated with respect to the random variable $\th$, conditional on the type realization $\g$. Note that $Q_A^M$ is strictly decreasing in $\g$ and $Q_B^M$ is strictly increasing in $\g$. 

The next result is essentially a special case of the solution in \cite{courty2000sequential}.\footnote{In particular, our additive specification satisfies their first-order stochastic dominance condition; see \citet[Lemma 3.3, p.~708]{courty2000sequential}.  There is one technical difference. 
In their model, valuations are bounded; in our model, valuations are unbounded.} A mechanism is \emph{uniquely optimal} if it is optimal and it agrees almost surely with all other optimal mechanisms. To minimize notation, we state the result with firm $B$ as the monopolist; a symmetric result holds for firm $A$. As $\g$ increases, the conditional distribution of $v_B(\th)$ increases in the sense of first-order stochastic dominance. Thus, it is meaningful to speak of ``high'' and ``low'' types.

\begin{prop}[Single-firm solution] \label{res:monopoly_benchmark}
Suppose that firm $B$ is a monopolist. Firm $B$'s uniquely optimal mechanism is implemented by the following profile, $(s_B^M, p_B^M, q_B^M)$.  Let\/ $\bar{p}_B^M = 1 / g(\ubar{\g})$. 
\begin{itemize}
        \item In the first period, firm $B$'s subscription schedule $s_B^M \colon [0, \infty) \to [0, \infty]$ and the consumer's contract-selection strategy $p_B^M \colon \G \to [0, \infty]$ are given by
    \begin{equation*}
    \begin{aligned}
             s_B^M (p_B) &= \E_{\th | \ubar{\g}} \Brac{ ( v_B (\th) - \bar{p}_B^M)_+} + \int_{p_B \wedge \bar{p}_B^M}^{\bar{p}_B^M} \hat{Q}_B^M (p_B') \de p_B',  \\
   p_B^M ( \g) &= \frac{1 - G(\g)}{g(\g)},
    \end{aligned}
    \end{equation*}
where the function $\hat{Q}_B^M \colon [0, \bar{p}_B^M] \to [0,1]$ is defined by $\hat{Q}_B^M (p_B^M (\g)) = Q_B^M ( p_B^M(\g) | \g)$ for all types $\g$.\footnote{Since $p_B^M$ is monotone  \citep[][Theorems 1 \& 3, p.~446--448]{BagnoliBergstrom2005} and $p_B^M(\G)= [0, \bar{p}_B^M]$, the function $\hat{Q}_B^M$ is well-defined at all but at most countably many prices. Hence, $s_B^M$ is well-defined at \emph{every} price.}
    \item In the second period, the consumer's product-purchase strategy $q_B^M$ is given by
    \[
    q_B^M(\g, \th) 
    =
    \begin{cases}
        1 &\text{if}~ v_B(\th) \geq p_B^M (\g), \\
        0 &\text{otherwise}.
    \end{cases}
    \]
\end{itemize}
\end{prop}

First, we describe the consumer's choices, given firm $B$'s subscription schedule $s_B^M$. In the first period, type $\g$ selects the strike price $p_B^M(\g)$ and pays the associated subscription fee $s_B^M (p_B^M(\g))$.  In the second period, type $\g$ learns his position $\th$ and then purchases the product if and only if his consumption value $v_B(\th)$ exceeds the strike price $p_B^M(\g)$. The function $p_B^M$ is weakly decreasing because the density $g$ is log-concave (by \cref{as:regularity}).\footnote{See, e.g., \citet[][Theorems 1 \& 3, p.~446--448]{BagnoliBergstrom2005}.} Thus, higher types select contracts with lower strike prices and higher subscription fees. Intuitively, higher types are willing to pay more for a reduction in the strike price because they are more likely to purchase the product in the second period (at any fixed price). Relative to the efficient allocation, there is downward distortion for all types $\g < \bar{\g}$. The product is allocated efficiently to the highest type, $\bar{\g}$, who selects the strike price $p_B^M(\bar{\g}) = 0$. The lowest type, $\ubar{\g}$, selects the strike price $p_B^M (\ubar{\g}) = \bar{p}_B^M$. Strike prices above $\bar{p}_B^M$ are never chosen.

We provide intuition for the optimality of these strike prices using a perturbation argument. Consider a fixed type $\g$. Under the optimal mechanism, type $\g$ purchases at price $p_B^M(\g)$ if and only if the taste shock $\e$ satisfies $v_0 + \g + \e \geq p_B^M(\g)$, which holds with probability $Q_B^M( p_B^M(\g) | \g)$. Suppose that the principal slightly reduces the strike price $p_B^M(\g)$, while keeping the interim utility of type $\g$ constant. To preserve the interim utility of type $\g$, the revenue lost on the inframarginal taste-shock realizations is recouped by increasing the subscription fee for type $\g$. Therefore, the net change in revenue from type $\g$ comes from the strike price paid by the marginal taste-shock realizations. The principal's first-order condition requires that the revenue gain from type $\g$ equals the total revenue loss from all higher types, who now earn higher information rents because they benefit more than type $\g$ from the reduction in the strike price $p_B^M(\g)$. Writing $Q_{B B}^M$ and $Q_{B\g}^M$ for the partial derivatives of $Q_B^M$ with respect to the price of product $B$ and the consumer's type, respectively, the first-order condition is given by
\begin{equation} \label{eq:single_FOC}
   - g(\g)  p_B^M ( \g) Q_{B B}^M ( p_B^M (\g) | \g) = Q_{B\g}^M ( p_B^M(\g) | \g) (1 - G( \g)).
\end{equation}
It is easily verified that
\[
-Q_{BB}^M ( p_B^M (\g) | \g) =  f ( p_B^M(\g) - v_0 - \g) = Q_{B\g}^M ( p_B^M(\g) | \g),
\]
so we get $p_B^M(\g) = (1 - G(\g))/ g(\g)$. The formal proof covers stochastic mechanisms and takes into account global deviations. 

To induce each type $\g$ to select the strike price $p_B^M(\g)$, the subscription schedule is determined, up to a constant, by the envelope formula. Intuitively, the first-order condition for type $\g$ requires that if the strike price changes slightly from $p_B^M(\g)$ to $p_B^M(\g) + \d$, then the associated subscription fee must change by $- \d  Q_B^M ( p_B^M (\g) | \g)$ to offset the change in type $\g$'s expected net utility in the second period. The constant in the subscription schedule ensures that the participation constraint holds with equality for the lowest type, $\ubar{\g}$. In particular, the subscription fee $s_B^M (\bar{p}_B^M)$ equals $\E_{\th | \ubar{\g}} [( v_B (\th) - \bar{p}_B^M)_+]$, which is how much
type $\ubar{\g}$ values the option to purchase product $B$ at price $\bar{p}_B^M$. 


\cref{res:monopoly_benchmark} is stated with firm $B$ as the monopolist. If firm $A$ is the monopolist, then a symmetric analog of \cref{res:monopoly_benchmark} holds, with $\bar{\g}$ in place of $\ubar{\g}$, and $G(\g)$ in place of $1- G(\g)$. Hereafter, we use the following notation for the \emph{monopoly strike prices}:
\begin{equation} \label{eq:monopoly_equations}
    p_A^M(\g) = \frac{G(\g)}{g(\g)} 
    \quad
    \text{and}
    \quad
    p_B^M(\g) = \frac{1 - G(\g)}{g(\g)}. 
\end{equation}

\section{Equilibrium analysis} \label{sec:equilibrium_analysis}

We now return to the main model with two firms. Our main theorem characterizes the essentially unique equilibrium.  

\subsection{Equilibrium characterization}

To begin, we introduce notation for the consumer's product demand. In the second period, suppose that the strike prices are $p_A$ and $p_B$. Then it is strictly optimal for the consumer with position $\th$ to purchase product $i$ if and only if 
\begin{equation} \label{eq:purchase_condition}
    v_i(\th) - p_i > (v_{-i}(\th) - p_{-i})_+.
\end{equation}
Condition \eqref{eq:purchase_condition} says that purchasing product $i$ is strictly preferred to (a) purchasing product $-i$ and (b) purchasing neither product. Purchasing both products is never strictly optimal. The subscription fees do not enter condition~\eqref{eq:purchase_condition} because they are sunk by the second period. For each firm $i$, define the interim demand function $Q_i \colon [0, \infty)^2 \times \G \to [0,1]$ by
\[
    Q_i ( p_i, p_{-i} | \g) = \P_{\th| \g} \bigl( v_i ( \th) - p_i > (v_{-i} (\th) - p_{-i})_+ \bigr),
\]
where the probability is evaluated with respect to the random position $\th$, conditional on the realized type $\g$.\footnote{Ties occur with probability zero, so the demand function is unchanged if the strict inequality is replaced with a weak inequality.} Thus, $Q_i( p_i, p_{-i} | \g)$ is the interim probability that type $\g$ purchases product $i$, given that he selects the strike price $p_i$ at firm $i$ and the strike price $p_{-i}$ at firm $-i$. Note that $Q_i$ is strictly decreasing in $p_i$ and weakly increasing in $p_{-i}$ (strictly so if $(p_i + p_{-i})/2 \leq v_0$ so that the market is covered). Moreover, $Q_A$ is strictly decreasing in $\g$, and $Q_B$ is strictly increasing in $\g$.

To state the main equilibrium characterization, we define a suitable notion of uniqueness for equilibria. Two equilibria are \emph{equivalent} if all but finitely many types select the same pair of contracts in the two equilibria.\footnote{That is, 
the equilibria $(s_A, s_B, p, q)$ and $(s_A', s_B', p', q')$ are equivalent if there exists a finite subset $\G_0$ of $\G$ such that for all types $\g \in \G \setminus \G_0$ and for each firm $i$, we have $(p_i(\g), s_i ( p_i(\g))) = (p_i'(\g), s_i' (p_i'(\g)))$; by second-period optimality, it then follows that the functions $q_i ( \g, \cdot)$ and $q_i' (\g, \cdot)$ agree, except possibly where the consumer is indifferent.} This notion of equivalence does not require the subscription schedules to agree at unselected strike prices. An equilibrium is \emph{essentially unique} if it is equivalent to all other equilibria. For the next theorem, recall the monopoly strike prices defined in \eqref{eq:monopoly_equations}. As is standard, we characterize equilibrium in the case that
$v_0$ is large enough that the product market is covered.\footnote{For example, \cite{armstrong2010competitive}, \cite{ArmstrongZhou2022}, and \cite{RochetStole2002} all focus on covered markets. See \cite{YangYe2008} for a static nonlinear pricing problem with partial coverage.}


\begin{thm}[Equilibrium characterization] \label{res:equilibrium}
Assume $v_0 \geq  \max_{\g' \in \G}  (1/g(\g'))$. The following strategy profile, $(s_A^\ast, s_B^\ast, p^\ast, q^\ast)$, is an equilibrium.  Let $\bar{p}_A = 2/ g( \bar{\g})$ and $\bar{p}_B  = 2 /g( \ubar{\g})$. 
\begin{itemize}
    \item In the first period, the firms' subscription schedules, $s_A^\ast \colon [0, \infty) \to [0, \infty]$ and $s_B^\ast \colon  [0, \infty) \to [0, \infty]$, and the consumer's contract-selection strategy $p^\ast \colon \G \to [0, \infty]^2$ are given by 
    \begin{equation*}
    \begin{aligned}
        s_A^\ast (p_A) &= \E_{\th |\bar \g} \Brac{ \Paren{ v_{A} (\th) - \bar{p}_A -  v_{B}(\th)_+}_+} + \int_{p_A \wedge \bar{p}_A}^{\bar{p}_A} Q_A^\ast (p_A') \de p_A',  \\
         s_B^\ast (p_B) &= \E_{\th | \ubar \g} \Brac{ \Paren{ v_B (\th) - \bar{p}_B -  v_{A}(\th)_+}_+} + \int_{p_B \wedge \bar{p}_B}^{\bar{p}_B} Q_B^\ast (p_B') \de p_B',  \\
        p^\ast(\g) &= ( 2p_A^M(\g), 2p_B^M(\g)), 
       \end{aligned}
    \end{equation*}
    where for each firm $i$, the function $Q_i^\ast \colon [0, \bar{p}_i] \to [0,1]$ is defined by $Q_i^\ast ( p_i^\ast (\g)) = Q_i (  p_i^\ast(\g), p_{-i}^\ast(\g) | \g)$ for all types $\g$.\footnote{Since $p_i^\ast$ is monotone \citep[][Theorems 1 \& 3, p.~446--448]{BagnoliBergstrom2005} and $p_i^\ast(\G)= [0, \bar{p}_i]$, the function $Q_i^\ast$ is well-defined at all but at most countably many prices. Hence, $s_i^\ast$ is well-defined at \emph{every} price.} 
    \item In the second period, the consumer's product-purchase strategy $q^\ast$ is given by
\[
    q^\ast (\g, \th) =
    \begin{cases}
        (1,0) &\text{if}~ v_A(\th) - p_A^\ast (\g) > v_B(\th) - p_B^\ast (\g), \\
        (0,1) &\text{if}~ v_B(\th) - p_B^\ast (\g) \geq v_A(\th) - p_A^\ast(\g).
    \end{cases}
\]
\end{itemize}
Moreover, if $v_0 \geq (7/2) \max_{\g' \in \G} (1/ g(\g'))$, then $(s_A^\ast, s_B^\ast, p^\ast, q^\ast)$ is the essentially unique equilibrium. 
\end{thm}

First, we describe the consumer's choices, given the firms' subscription schedules $s_A^\ast$ and $s_B^\ast$. In the first period, type $\g$ selects the strike price $p_A^\ast(\g) = 2 p_A^M(\g)$ from firm $A$ and the strike price $p_B^\ast (\g) = 2p_B^M(\g)$ from firm $B$. The reason for the factor of $2$ will be discussed after we describe the full equilibrium. Type $\g$ pays the associated subscription fees $s_A^\ast (p_A^\ast (\g))$ to firm $A$ and $s_B^\ast (p_B^\ast(\g))$ to firm $B$.  Both subscription fees are strictly positive. In the second period, type $\g$ learns his position $\th$ and then purchases whichever product yields higher net utility (i.e., consumption value net of the strike price). The assumption that $v_0 \geq \max_{\g' \in \G} (1 /g(\g'))$ ensures that the average net utility of the two products is nonnegative since (i) the average valuation $(v_A(\th) + v_B(\th))/2$ equals $v_0$ for every position $\th$, and (ii) the average price $(p_A^\ast(\g) + p_B^\ast (\g))/2$ equals $1/g(\g)$ for every type $\g$. Thus, the net utility of at least one product is nonnegative, and the consumer always purchases a product. 

The function $p_A^\ast$ is weakly increasing and the function $p_B^\ast$ is weakly decreasing because the density $g$ is log-concave (by \cref{as:regularity}).\footnote{See, e.g., \citet[][Theorems 1 \& 3, p.~446--448]{BagnoliBergstrom2005}.}  Say that one contract is more \emph{premium} than another if it has a lower strike price and a higher subscription fee. More rightward types expect to have stronger preferences for firm $B$ over firm $A$,\footnote{Formally, the distribution of the utility difference $v_B(\th) - v_A( \th)$ increases, in the sense of first-order stochastic dominance, as the consumer's type $\g$ moves to the right.} so they select more premium subscriptions from  firm $B$ and less premium subscriptions from firm $A$. Intuitively, more rightward types are willing to pay more for a reduction in the strike price of product $B$ because they are more likely to purchase product $B$ in the second period, given any fixed prices. The rightmost type, $\bar{\g}$, selects the strike price $p_B^\ast(\bar{\g}) = 0$ from firm $B$ and the strike price $p_A^\ast( \bar{\g}) = \bar{p}_A$ from firm $A$. Strike prices above $\bar{p}_A$ are never chosen at firm $A$; strike prices above $\bar{p}_B$ are never chosen at firm $B$.


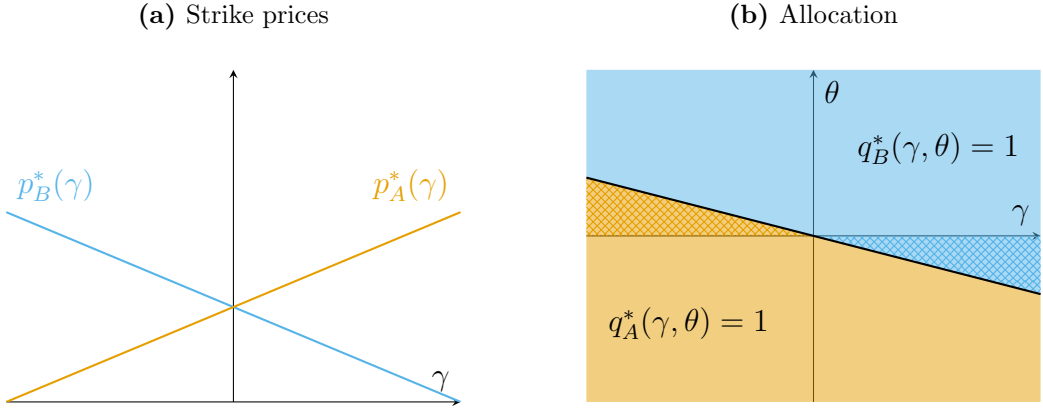
\begin{figure}
\begin{subfigure}{.5\textwidth}
\caption{Strike prices}
\begin{center}
\begin{tikzpicture}
  \begin{axis}[width = \textwidth, 
  height = 6cm,
    xmin=-1, xmax=1, ymin=0, ymax=7,
    axis lines=middle,
    xlabel={$\gamma$},
    ylabel={},
    tick style={black},
    tick align=inside,
    xtick=\empty,
    ytick=\empty,
    legend cell align=left,
  ]


  \addplot[
    Blue, thick, smooth,
    domain=-1:1,
    samples=2
  ] { 2*(1 - x) } node [pos = 0, anchor = south west] {$p_B^\ast(\g)$};
  
  \addplot[
    Orange, thick, smooth,
    domain=-1:1,
    samples=2
  ] { 2*(1 + x) }  node [pos = 1, anchor = south east] {$p_A^\ast(\g)$};



  \pgfmathsetmacro{\xB}{-0.7}
  \pgfmathsetmacro{\yB}{2*(1-\xB)}   
  \pgfmathsetmacro{\xA}{0.7}
  \pgfmathsetmacro{\yA}{2*(1+\xA)}   

  \end{axis}
\end{tikzpicture}
\end{center}
\label{fig:subscriptions}
\end{subfigure}
\begin{subfigure}{.5\textwidth}
\caption{Allocation}
\begin{center}
\begin{tikzpicture}
\pgfmathsetmacro{\Ymax}{5.7} 

\begin{axis}[
  width = \textwidth, height = 6cm,
  axis lines=middle,
  xmin=-1, xmax=1,
  ymin=-\Ymax, ymax=\Ymax,
  restrict y to domain=-\Ymax:\Ymax,
  xlabel={$\gamma$}, ylabel={$\theta$},
  samples=200, no marks,
  clip=false,
  xtick=\empty, ytick=\empty,
]

\path[name path=lower] (axis cs:-1,-\Ymax) -- (axis cs:1,-\Ymax);
\path[name path=upper] (axis cs:-1, \Ymax) -- (axis cs:1, \Ymax);
\path[name path=zero]  (axis cs:-1,0)      -- (axis cs:1,0);

\addplot[black, thick, name path=keycurve, domain=-1:1]
  ({x},{-2*x});

\addplot[Orange, fill opacity=0.5, draw=none]
  fill between[of=keycurve and lower, soft clip={domain=-1:1}];

\addplot[Blue, fill opacity=0.5, draw=none]
  fill between[of=keycurve and upper, soft clip={domain=-1:1}];


\addplot[
  draw=none,
  pattern=crosshatch,
  pattern color=Orange
]
fill between[
  of=keycurve and zero,
  soft clip={domain=-1:0}
];

\addplot[
  draw=none,
  pattern=crosshatch,
  pattern color=Blue
]
fill between[
  of=keycurve and zero,
  soft clip={domain=0:1}
];

\node[black] at (axis cs:0.55,3)   {$q_B^\ast(\gamma,\theta) = 1$};
\node[black] at (axis cs:-0.55,-3) {$q_A^\ast(\gamma,\theta) = 1$};

\end{axis}
\end{tikzpicture}
\end{center}
\label{fig:subscription_allocation}
\end{subfigure}
\caption{Equilibrium with the uniform type distribution}
\label{fig:equilibrium_uniform}
\end{figure}

\cref{fig:equilibrium_uniform} illustrates the equilibrium in an example where the consumer's type $\g$ is uniformly distributed over $\G = [-1,1]$; this distribution will serve as a running example throughout the paper. Since $G(\g) = (1+ \g)/2$, the equilibrium strike prices take a simple form: $p_A^\ast(\g) = 2(1 + \g)$ and $p_B^\ast(\g) = 2(1 - \g)$. \cref{fig:subscriptions} plots these strike prices as a function of the consumer's type $\g$. The median type, here $\g = 0$, selects the same strike price at both firms. Types to the left of the median select a lower strike price at firm $A$ than at firm $B$. Types to the right of the median select a lower strike price at firm $B$ than at firm $A$. 

\cref{fig:subscription_allocation} plots the equilibrium allocation. Each pair $(\g, \th) \in \G \times \Th$ is shaded orange or blue according to whether the consumer with type $\g$ and position $\th$ purchases product $A$ or $B$.\footnote{The distribution $F$ of the taste shock does not affect the equilibrium allocation rule  $q^\ast$, but $F$ does affect the distribution of consumers over $\G \times \Th$.} We cross-hatch the pairs $(\g, \th)$ at which the consumer purchases his less-preferred product. The consumer purchases product $B$ if and only if $v_B(\th) - v_A(\th) \geq p_B^\ast(\g) - p_A^\ast(\g)$, or equivalently, $\th \geq (p_B^\ast (\g) - p_A^\ast(\g))/2$.  It is \emph{efficient} to consume product $B$ if and only if  $v_B(\th) - v_A(\th) \geq 0$, or equivalently, $\th \geq 0$. Thus, the price difference $p_B^\ast(\g) - p_A^\ast(\g)$ creates a wedge away from efficient consumption. Even though the product market is covered, only the median type, here $\g = 0$, consumes efficiently in the second period. Types to the left of the median consume product $A$ more often than is efficient. Types to the right of the median consume product $B$ more often than is efficient. To see why, consider some consumer type $\g$ to the right of the median. If that consumer subsequently learns that he values product $A$ slightly more than product $B$, he will still purchase product $B$ if the valuation difference is smaller than the price difference. This consumer is partially \emph{locked into} firm $B$ in the second period because, in the first period, he selects a lower strike price at firm $B$ than at firm $A$. The larger is the price difference, the more severe is the lock-in. Symmetrically, types to the left of the median are partially locked into firm $A$.


Comparing the equilibrium in \cref{res:equilibrium} with the single-firm solution from \cref{res:monopoly_benchmark} yields testable predictions about how the prices at an incumbent monopolist change when a new firm enters. First, we compare the  equilibrium subscription schedule $s_i^\ast$ with the monopoly subscription schedule $s_i^M$ from \cref{res:monopoly_benchmark}.

\begin{prop}[Subscription schedules] \label{res:subscription_schedules}
Assume $v_0 \geq (7/2) \max_{\g' \in \G} (1/ g(\g'))$. For each firm $i$ and each price $p_i \in [0, \bar{p}_i^M]$, we have $s_i^\ast (p_i) < s_i^M (p_i)$.
\end{prop}

For concreteness, take firm $B$ as the incumbent. When firm $A$ enters, firm $B$ reduces all of its subscription fees and also expands its range of contracts to include new low-tier subscriptions with higher strike prices. The strike price that type $\g$ selects at firm $B$ doubles from $p_B^M (\g)$ before firm $A$ enters to $p_B^\ast(\g) = 2 p_B^M(\g)$ after firm $A$ enters.\footnote{Before firm $A$ enters, type $\g$ either purchases product $B$ at the selected price $p_B^M(\g)$ or does not purchase at all. After firm $A$ enters, type $\g$ purchases either product $B$ at the selected price $2 p_B^M(\g)$ or product $A$ at the selected strike price $2 p_A^M(\g)$.} This directional prediction is consistent with the evolution of the streaming market. In 2022, Netflix first introduced a basic subscription plan with advertisements (effectively a usage price) after losing subscribers to new entrants such as Disney\textsuperscript{+}.\footnote{See \url{https://tinyurl.com/2ypanxz6}.}

The formal proof of optimality is sketched in \cref{sec:proof_sketch_equilibrium,sec:proof_sketch_essential_uniqueness}. To provide intuition for the doubling of the selected strike prices, we analyze a perturbation of $p_B^\ast(\g)$ by firm $B$. We obtain a first-order condition that parallels condition \eqref{eq:single_FOC} from the single-firm case. With both firms present, the relevant alternative for the consumer is purchasing the other product instead of not purchasing at all. Therefore, the sensitivity of product $B$ demand to product $B$'s price equals half the density of consumers at the margin, rather than the whole density as in the monopoly case. Consider a fixed type $\g$. In equilibrium, type $\g$ purchases product $B$ at price $p_B^\ast(\g)$ if and only if the taste shock $\e$ satisfies $\g + \e \geq (p_B^\ast(\g) - p_A^\ast(\g))/2$, which holds with probability $Q_B( p_B^\ast(\g), p_A^\ast(\g) | \g)$. Suppose that firm $B$ slightly reduces the strike price $p_B^\ast(\g)$ while keeping the interim utility of type $\g$ constant. To preserve the interim utility of type $\g$, the revenue lost on the inframarginal taste-shock realizations is recouped by increasing the subscription fee for type $\g$. Therefore, the net change in revenue from type $\g$ comes from the strike price paid by the marginal taste-shock realizations. Firm $B$'s first-order condition requires that the revenue gain from type $\g$ equals the total revenue loss from all more rightward types, who now earn higher information rents because they benefit more than type $\g$ from the reduction in the strike price $p_B^\ast(\g)$. Writing $Q_{B B}$ and $Q_{B\g}$ for the partial derivatives of $Q_B$ with respect to the price of product $B$ and the consumer's type, respectively, the first-order condition is given by
\[
   - g(\g)  p_B^\ast ( \g) Q_{B B} ( p_B^\ast (\g), p_A^\ast(\g) | \g) = Q_{B\g} ( p_B^\ast(\g), p_A^\ast(\g) | \g) (1 - G( \g)).
\]
It is easily verified that
\[
    -Q_{BB} ( p_B^\ast(\g), p_A^\ast(\g) | \g)= (1/2) f ( (p_B^\ast(\g) - p_A^\ast(\g))/2 - \g) = (1/2) Q_{B\g} ( p_B^\ast(\g), p_A^\ast(\g) | \g),
\]
so we get $p_B^\ast(\g) = 2(1 - G(\g))/ g(\g)$.

Finally, we comment on the equilibrium subscription schedules. In order for each type $\g$ to select the strike-price pair $(p_A^\ast (\g), p_B^\ast(\g))$, the subscription schedule at each firm is determined, up to a constant, by the envelope formula. Intuitively, the first-order condition for type $\g$ requires that if type $\g$ selects from firm $i$ a slightly perturbed strike price $p_i^\ast(\g) + \d$ rather than $p_i^\ast(\g)$, then the associated subscription fee at firm $i$ must change by $-\d Q_i ( p_i^\ast (\g), p_{-i}^\ast (\g) | \g)$ to offset type $\g$'s change in expected net utility in the second period. The constant in each firm $i$'s subscription schedule is chosen so that the type who is most pessimistic about product $i$ is indifferent between entering and not entering a contract with firm $i$. For concreteness, consider firm $B$. Type $\ubar{\g}$ selects the strike price $p_B^\ast (\ubar{\g}) = \bar{p}_B$ at firm $B$ and the strike price $p_A^\ast (\ubar{\g}) = 0$ at firm $A$. Given that type $\ubar{\g}$ holds the option to purchase product $A$ at strike price $0$, his valuation for the option to purchase product $B$ at price $\bar{p}_B$ is $\E_{\th | \ubar{\g}} [( v_B (\th) - \bar{p}_B - v_A(\th)_+)_+]$, which is exactly the subscription fee $s_B^\ast (\bar{p}_B)$.\footnote{Moreover, if type $\ubar{\g}$ contracts with firm $A$ only, the strike price of $0$ remains optimal since it is the most premium subscription offered by firm $A$.}

\subsection{Proof approach: Equilibrium verification}
\label{sec:proof_sketch_equilibrium}

While the form of the equilibrium in \cref{res:equilibrium} is similar to the optimal mechanism in \cref{res:monopoly_benchmark}, proving \cref{res:equilibrium} requires different techniques because we are solving for a fixed point, rather than a maximizer. To prove that $(s_A^\ast, s_B^\ast, p^\ast, q^\ast)$ is an equilibrium, we must check that the consumer and the firms are all playing best responses, as required by \cref{def:equilibrium}. 

To prove that $(p^\ast, q^\ast)$ is a consumer best response to $(s_A^\ast, s_B^\ast)$, the key step is checking that the contract-selection strategy $p^\ast$ is optimal. Given $p^\ast$, it is immediate that the product-purchase strategy $q^\ast$ is optimal. As argued above, the subscription-schedule  pair ($s_A^\ast, s_B^\ast)$ ensures that the consumer cannot benefit from selecting a perturbed strike price at either firm. We must show, further, that no type $\g$ can profit from deviating to any price vector $(p_A', p_B') \neq (p_A^\ast (\g), p_B^\ast(\g))$. The difficulty is that type $\g$ can mimic a different type at each firm.  In the proof, we analyze each type's optimal choice of strike price at one firm, given each strike price selected at the other firm. 


The heart of the proof is showing that in the strategy profile $(s_A^\ast, s_B^\ast, p^\ast, q^\ast)$, each firm is playing a best response. In fact, we prove a stronger result: neither firm $i$ can profit by unilaterally deviating to any stochastic mechanism $(q_i,t_i) \colon \G \times \Th \to [0,1]\times \R$. Without loss of generality, we consider firm $B$'s best response to firm $A$'s schedule $s_A^*$.
The key difficulty is that firm $B$'s schedule affects the consumer's contract choices at both firms.
Rather than directly characterizing consumer best responses to every alternative subscription schedule for firm $B$, we introduce an \emph{auxiliary design problem}: Firm $B$ chooses a mechanism and a type-contingent recommendation for the consumer's interaction with firm $A$, subject to truthtelling and obedience constraints. We show that if firm $B$'s deviation induces the allocation rule $q = (q_A, q_B)$, then the consumer's interim utility $U$ satisfies
\begin{equation} \label{eq:envelope_main}
    U(\g) = U(\ubar{\g}) + \int_{\ubar{\g}}^{\g} \E_{\th | \g'} [ q_B(\g', \th) - q_A( \g', \th)] \de \g'.
\end{equation}
The interim utility takes this form because, as $\g$ moves to the right, the distribution of the consumer's valuation for product $B$ shifts up, and the distribution of the consumer's valuation for product $A$ shifts down. Firm $B$'s expected revenue equals total expected surplus net of the consumer's expected utility and net of firm $A$'s expected revenue. After changing the order of integration, we can express firm $B$'s expected revenue, up to boundary terms, as the expectation over $\g$ of the expression
\begin{equation}
\label{eq:pointwise}
\begin{aligned}
   &\E_{\th| \g} \Brac{ \Paren{ v_A(\th) - p_A(\g) + \frac{1-G(\g)}{g(\g)}} q_A(\g, \th) + 
   \Paren{ v_B(\th) - \frac {1- G(\g)}{g(\g)}} q_B(\g, \th)} \\
   &\quad -s_A^\ast( p_A(\g)),
\end{aligned}
\end{equation}
where $p_A(\g)$ is the recommended strike price for type $\g$ at firm $A$. The terms involving $v_A(\th)$ and $v_B(\th)$ come from the total surplus; the terms involving $(1 - G(\g))/g(\g)$ come from the consumer's utility; and the terms
involving $p_A(\g)$ and $s_A^\ast(p_A(\g))$ come from firm $A$'s revenue.  

Next, we consider the relaxed problem of pointwise maximizing the expression in \eqref{eq:pointwise}, without imposing global incentive constraints.\footnote{This way, we do not need to characterize the class of all functions $p_A$ and $q$ that firm $B$ can induce through its choice of subscription schedule. Indeed, since firm $B$ uses a single instrument---its subscription schedule---to influence the consumption of both products, it cannot unilaterally induce every monotone allocation $(q_A,q_B)$.} This expression looks similar to the usual dynamic virtual surplus, but the coefficient on the allocation probability $q_A( \g, \th)$ is not exogenous---it depends on the recommended strike price at firm $A$. We prove that for each type $\g$, the expression in \eqref{eq:pointwise} is maximized by the equilibrium choices $p_A(\g) = p_A^\ast(\g)$ and $q(\g, \cdot) = q^\ast( \g, \cdot)$. Here is a sketch of the argument. Under our assumption that $v_0 \geq  \max_{\g' \in \G}  (1/g(\g'))$, it can be shown that the pointwise maximizer, over $p_A(\g)$ and $q(\g, \cdot)$, induces full coverage, i.e., $q_A( \g, \th) + q_B(\g, \th) = 1$. Thus, the expression in \eqref{eq:pointwise} simplifies to
\begin{equation} \label{eq:firm_B_max}
\begin{aligned}
   &\E_{\th|\g} \Brac{ \Paren{ v_A(\th) - p_A(\g)} q_A(\g, \th) + 
   \Paren{ v_B(\th) - 2\frac {1- G(\g)}{g(\g)}} q_B(\g, \th)} \\
   &\quad  -s_A^\ast( p_A(\g)) +\frac{1 - G(\g)}{g(\g)}.
\end{aligned}
\end{equation}
Suppose type $\g$ chooses the strike price $p_B^\ast (\g) = 2 (1 - G(\g))/g(\g)$ from firm $B$. Then subtracting the constant $s_B^\ast (p_B^\ast(\g)) + (1 - G(\g))/ g(\g)$ from the expression in \eqref{eq:firm_B_max} yields the consumer's utility from his remaining choices, i.e., the strike price $p_A(\g)$ at firm $A$ and the product-purchase rule $q (\g, \cdot)$. Since $(p^\ast, q^\ast)$ is a consumer best response to $(s_A^\ast, s_B^\ast)$, the expression in \eqref{eq:firm_B_max} is maximized by $p_A(\g) = p_A^\ast(\g)$ and $q(\g, \cdot) = q^\ast( \g, \cdot)$.


\subsection{Proof approach: Essential uniqueness} \label{sec:proof_sketch_essential_uniqueness}

For essential uniqueness, we prove that all equilibria are equivalent to $(s_A^\ast, s_B^\ast, p^\ast, q^\ast)$. The stronger assumption that $v_0 \geq (7/2) \max_{\g' \in \G} (1/ g(\g'))$ serves to rule out certain pathological candidate equilibria, as we describe below. The key step is showing that in all equilibria, the consumer's contract-selection strategy must essentially agree with $p^\ast$. It is then straightforward to pin down the rest of the equilibrium strategy profile. Here is a sketch of the proof.  We suppose for a contradiction that there exists an equilibrium $(\hat{s}_A, \hat{s}_B, \hat{p}, \hat{q})$ with $\hat{p} \neq p^\ast$. Without loss, suppose that $\hat{p}_B \neq p_B^\ast$. We show that firm $B$ has a profitable deviation. There are two cases.

The main case is that the function $\hat{p}_A$ never takes very large values. In this case, we show that firm $B$ can deviate to a subscription schedule that induces each type $\g$ to select the strike price $p_B^\ast (\g)$ from firm $B$, while preserving second-period coverage in the product market. Our argument must handle the case that $\hat{s}_A$ is discontinuous, so we build upon results from monotone comparative statics, which do not rely on continuity. Next, we follow our analysis of \eqref{eq:pointwise}--\eqref{eq:firm_B_max} above, with $\hat{s}_A$ in place of $s_A^\ast$, to conclude that this deviation by firm $B$ is strictly profitable. 

There is also an edge case in which $\hat{p}_A$ takes very large values. In this edge case, the deviation we constructed above may leave the product market uncovered, invalidating a key step in our analysis of \eqref{eq:pointwise}--\eqref{eq:firm_B_max}. Instead, we prove that some firm has a strictly profitable local deviation, but the argument is quite intricate. The main difficulty is ruling out equilibria in which many types select strike-price pairs where the interim demand curve is kinked (because the market is exactly covered).\footnote{This is a ``well-known technical drawback'' of the Hotelling model \citep[p.~589]{ArmstrongVickers2001}. \cite{MerelSexton2010} analyze Hotelling equilibria in which each firm prices at the kink in its demand curve.} This is where we use the stronger lower bound on $v_0$ and the symmetry of the taste shock distribution (from \cref{as:regularity}). 


\section{Other contracting settings} \label{sec:other_regimes}

In this section, we analyze three alternative contracting settings: spot pricing (SP), exclusive contracting (E), and multi-product monopoly (MM). These settings illustrate how the equilibrium of the main model with non-exclusive contracting (NE) is shaped, respectively, by the timing of contracts, multi-homing, and competition. 

\subsection{Spot pricing}  \label{sec:spot_pricing}

In the  main model, the firms contract with the consumer \emph{before} the consumer perfectly learns his valuations for the products. Here, we suppose instead that the firms interact with the consumer only \emph{after} the consumer learns his valuations. Formally, each firm posts a spot price, and then the consumer, knowing his position $\th$, chooses which product to purchase. We call this the \emph{spot-pricing game}. In this game, the relevant primitive is the distribution of $\th$, denoted by $H$, which has density $h(\th) = \int_{\ubar \g}^{\bar \g} f(\th-\g)g(\g)\de \g$. The density $h$ is log-concave because $f$ and $g$ are log-concave (by \cref{as:regularity}) and log-concavity is preserved by convolution.\footnote{See, e.g., \citet[Proposition 3.5, p.~60]{SaumardWellner2014}.} It follows that each firm's revenue is log-concave in its own price, so the equilibrium can be characterized by first-order conditions. As in the main model, we focus on the case in which $v_0$ is large enough that the market is covered. 


\begin{prop}[Spot pricing] \label{res:Hotelling} There is a unique position $\th^\ast \in \R$ satisfying
\begin{equation} \label{eq:spot_pricing_critical}
    \th^\ast = \frac{1 - 2 H(\th^\ast)}{h(\th^\ast)}.
\end{equation}
If $v_0 \geq 1/h(\th^\ast)$, then the  unique equilibrium of the spot-pricing game is given by
\[
    p_A^\ast = 2 \frac{ H(\th^\ast) }{h(\th^\ast)},
    \qquad
    p_B^\ast  = 2 \frac{1 - H(\th^\ast)}{h(\th^\ast)}.
\]
\end{prop}


\cref{res:Hotelling} recovers the standard Hotelling equilibrium. Consumers to the left of $\th^\ast$ purchase product $A$, and consumers to the right of $\th^\ast$ purchase product $B$. By condition \eqref{eq:spot_pricing_critical}, type $\th^\ast$ is indifferent between purchasing product $A$ and product $B$. The assumption that $v_0 \geq 1/ h(\th^\ast)$ ensures that the consumer always finds it optimal to purchase one of the products, so the market is covered. Let $m$ denote the median of the distribution $H$. If $m = 0$, then $\th^\ast = 0$, so $p_A^\ast = p_B^\ast = 1 / h(0)$. In this case, the allocation is efficient. If $m > 0$, then it follows from \eqref{eq:spot_pricing_critical} that $0 < \th^\ast < m$, hence $p_A^\ast < p_B^\ast$. The more popular firm $B$ attracts more consumers than firm $A$ when prices are equal, so firm $B$ gains more revenue from the inframarginal consumers by raising its price. In equilibrium, consumers in the interval $(0, \th^\ast)$ purchase product $A$, even though they value product $B$ more. Thus, the allocation is inefficient. Firm $B$'s market share is strictly above half but strictly below $1 - H(0)$, the share of consumers who prefer product $B$. A symmetric result holds for firm $A$ if $m < 0$.

\subsection{Exclusive contracting} \label{sec:exclusive_contracts}

In the main model, we assume that the consumer can simultaneously enter contracts with both firms. Here, we suppose instead that contracts are \emph{exclusive}: if the consumer enters a contract with one firm, then he is prohibited from entering a contract with the other firm.\footnote{For a classical discussion of exclusive dealing, see \cite{BernheimWhinston1998}.} Each firm $i$ posts a subscription schedule $s_i$ as before, but now the consumer can select a contract from at most one firm.\footnote{Formally, by our null-contract convention, the consumer is restricted to selecting at most one finite price. With this modification, the equilibrium is defined as in the main model.} In the following equilibrium characterization, we use the notation $Q_i^M$ from \eqref{eq:interim_monop_demand} to denote the interim monopoly product demand. Say that an equilibrium is \emph{Pareto-dominant} if it gives both firms weakly higher payoffs than every other equilibrium.\footnote{Pareto dominance for firms is a natural criterion for equilibrium selection because the firms move before the consumer, and the firms' payoffs are independent of which best response the consumer plays; see \cref{ft:NE_firms} and \cref{res:revenue} (\cref{sec:proof_equilibrium_uniqueness}), which can be extended to the exclusive setting.} 

\begin{prop}[Exclusive contracting] \label{res:exclusive}
Assume $\ubar{\g} < 0 < \bar{\g}$. There is a unique type $\g^\dagger \in (\ubar{\g}, \bar{\g})$ satisfying
\begin{equation} \label{eq:exclusive_critical_position}
\begin{aligned}
    &\E_{\th | \g^\dagger} \Brac{ ( v_A(\th) - p^M_A(\g^\dagger))_+}  - p_{A}^M (\g^\dagger) Q_{B}^M (  p_B^M (\g^\dagger) | \g^\dagger) \\
    &=
    \E_{\th | \g^\dagger} \Brac{( v_B(\th) - p^M_B(\g^\dagger))_+}  - p_{B}^M (\g^\dagger) Q_{A}^M (  p_A^M (\g^\dagger) | \g^\dagger) .
\end{aligned}
\end{equation}
Let $p^\dagger_A = p_A^M ( \g^\dagger)$ and $p^\dagger_B = p_B^M(\g^\dagger)$. Under exclusive contracting, if $v_0 \geq 1/g(\g^\dagger) + \abs{\g^{\dagger}}$, then the following is the essentially unique Pareto-dominant equilibrium.
\begin{itemize}
    \item In the first period, the firms' subscription schedules $s_A^{\mathrm{E}}$ and  $s_B^{\mathrm{E}}$ are given by
    \begin{equation*}
    \begin{aligned}
        s_A^{\mathrm{E}} (p_A) &= p^\dagger_A   Q_{B}^M( p^\dagger_B | \g^\dagger) + \int_{p_A \wedge p^\dagger_A}^{p^\dagger_A} \hat{Q}_A^M (p_A') \de p_A', 
        \\
        s_B^{\mathrm{E}} (p_B) &= p^\dagger_B Q_A^M( p^\dagger_A | \g^\dagger)  + \int_{p_B \wedge p^\dagger_B}^{p^\dagger_B} \hat{Q}_B^M (p_B') \de p_B', 
    \end{aligned}
    \end{equation*}
    where for each firm $i$, the function $\hat Q_i^M \colon [0, p^\dagger_i] \to [0,1]$ is defined by $\hat{Q}_i^M (p_i^M(\g)) = Q_i^M (p_i^M(\g) | \g)$ for all $\g$ such that $p_i^M(\g) \leq p^\dagger_i$. For $\g < \g^\dagger$, type $\g$ selects $p_A^M(\g)$ from firm $A$. For $\g \geq \g^\dagger$, type $\g$ selects $p_B^M(\g)$ from firm $B$. 
    \item In the second period, the consumer's product-purchase strategy $q^{\mathrm{E}}$ is given by
    \[
    q^{\mathrm{E}}(\g, \th) = 
    \begin{cases}
        (1,0) &\text{if}~  \g < \g^\dagger~\text{and}~v_A(\th) \geq p_A^M (\g), \\
        (0,1) &\text{if}~ \g \geq \g^\dagger~\text{and}~v_B(\th) \geq p_B^M (\g), \\ 
        (0,0) &\text{otherwise}.
    \end{cases}
    \]
\end{itemize}
\end{prop}

In the equilibrium in \cref{res:exclusive}, the firms split the subscription market around the critical type $\g^\dagger$. Equation~\eqref{eq:exclusive_critical_position} ensures that the critical type $\g^{\dagger}$ is indifferent between the two firms. Each type $\g$ to the left of $\g^\dagger$ selects the strike price $p_A^M(\g)$ from firm $A$ and pays the associated subscription fee $s_A^{\mathrm{E}} ( p_A^M(\g))$. Among the types choosing firm $A$, more leftward types select more premium subscriptions. Each type $\g$ to the right of $\g^\dagger$ selects the strike price $p_B^M(\g)$ from firm $B$ and pays the associated subscription fee $s_B^{\mathrm{E}} (p_B^M(\g))$. Among the types choosing firm $B$, more rightward types select more premium subscriptions. In both cases, the consumer purchases his exclusive supplier's product in the second period if and only if the consumption value exceeds the strike price. The constants in the subscription schedules $s_A^{\mathrm{E}}$ and $s_B^{\mathrm{E}}$ are derived from a first-order condition for each firm, as we discuss below. The assumption that $v_0 \geq 1 / g(\g^\dagger) + |\g^{\dagger}|$ ensures that the consumer always selects a contract in the first period. Nevertheless, for each consumer type, there is a positive probability of not purchasing a product in the second period. The assumption that $\ubar{\g} < 0 < \bar{\g}$ means that some types expect to prefer product $A$, and some types expect to prefer product $B$. This mild assumption rules out equilibria in which all consumers choose the same firm as their exclusive supplier. 

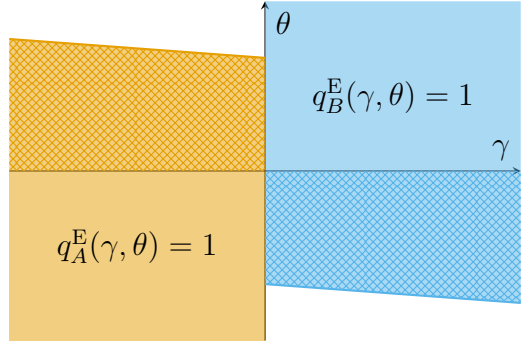
\begin{figure}
\begin{center}
\begin{tikzpicture}
\pgfmathsetmacro{\vzero}{7} 
\pgfmathsetmacro{\Ymax}{9}
\pgfmathsetmacro{\Ymin}{-9}

\begin{axis}[width = 0.55*\textwidth, height = 0.4*\textwidth,
  axis lines=middle,
  xmin=-1, xmax=1,
  ymin=\Ymin, ymax=\Ymax,
  restrict y to domain=\Ymin:\Ymax,
  xlabel={$\gamma$}, ylabel={$\theta$},
  samples=200, no marks,
  clip=false,
  xtick=\empty, ytick=\empty,
]

\path[name path=axisLeft]  (axis cs:-1,0) -- (axis cs:0,0);
\path[name path=axisRight] (axis cs:0,0)  -- (axis cs:1,0);

\path[name path=lowerLeft]  (axis cs:-1,\Ymin) -- (axis cs:0,\Ymin);
\path[name path=upperRight] (axis cs:0,\Ymax)  -- (axis cs:1,\Ymax);

\addplot[Orange, thick, name path=orangeB, domain=-1:0]
  ({x},{\vzero - x - 1});

\addplot[Orange, fill opacity=0.5, draw=none]
  fill between[of=axisLeft and lowerLeft];

\addplot[Orange, fill opacity=0.5, draw=none]
  fill between[of=orangeB and axisLeft];
\addplot[draw=none, pattern=crosshatch, pattern color=Orange]
  fill between[of=orangeB and axisLeft];

\addplot[Blue, thick, name path=blueB, domain=0:1]
  ({x},{1 - x - \vzero});

\addplot[Blue, fill opacity=0.5, draw=none]
  fill between[of=blueB and upperRight];

  
\addplot[draw=none, pattern=crosshatch, pattern color=Blue]
  fill between[of=axisRight and blueB];

\node at (axis cs:-0.5,-4) {$q_A^{\mathrm{E}}(\gamma,\theta) = 1$};
\node at (axis cs: 0.5, 4) {$q_B^{\mathrm{E}}(\gamma,\theta) = 1$};

\end{axis}
\end{tikzpicture}
\end{center}
\caption{Exclusive equilibrium allocation with the uniform type distribution}
\label{fig:Exclusive_alloc}
\end{figure}

If the distribution $G$ is symmetric, then it follows from \eqref{eq:exclusive_critical_position} that $\g^\dagger= 0$, hence $p^\dagger_A = p^\dagger_B = 1/(2g(0))$. \cref{fig:Exclusive_alloc} plots the equilibrium allocation in the running example where the consumer's type $\g$ is uniformly distributed over $\G = [-1,1]$ and $v_0 = 7$.\footnote{Unlike in the non-exclusive case, the equilibrium allocation now varies with $v_0$. As $v_0$ increases, the consumption region shifts outward.}  Each pair $(\g, \th) \in \G \times \Th$ is shaded orange or blue (or white) according to whether the consumer with type $\g$ and position $\th$ purchases product $A$ or $B$ (or neither). We cross-hatch the pairs $(\g, \th)$ where the consumer purchases the product from his exclusive supplier, even though he prefers the other product.

To compare the efficiency of the exclusive equilibrium in \cref{res:exclusive} with that of the non-exclusive equilibrium in \cref{res:equilibrium}, we introduce the following definition. Given allocation rules $q \colon \G \times \Th \to \{0,1\}^2$ and $q' \colon \G \times \Th \to \{0,1\}^2$, we say that $q$ is \emph{pointwise more efficient} than $q'$ if for almost every $(\g,\th)$, the allocation $q(\g, \th)$ yields weakly higher surplus than $q'(\g, \th)$, and this inequality holds strictly  with positive probability.\footnote{The surplus for $(\g, \th)$ under $q$ and $q'$ is given by $\max\{0, v_A(\th) q_A(\g, \th), v_B(\th) q_B ( \g ,\th) \}$ and $\max\{0, v_A(\th) q_A'(\g, \th), v_B(\th) q_B' ( \g ,\th) \}$.} If $q$ is pointwise more efficient than $q'$, then ex-ante total surplus is strictly higher under $q$ than under $q'$. For the efficiency comparison, we impose the assumptions from both \cref{res:equilibrium} and \cref{res:exclusive}, so that in each setting, the referenced equilibrium exists and is unique. We get a clear efficiency ranking when $G$ is symmetric. 

\begin{prop}[Efficiency] \label{res:efficiency} Assume $G$ is symmetric and $v_0 \geq (7/2) \max_{\g' \in \G} (1 / g(\g'))$. The allocation $q^\ast$ under the non-exclusive equilibrium (in \cref{res:equilibrium}) is pointwise more efficient than the allocation $q^{\mathrm{E}}$ under the Pareto-dominant exclusive equilibrium (in \cref{res:exclusive}).
\end{prop}

Suppose that $G$ is symmetric. In the non-exclusive equilibrium, types to the left of $0$ are partially locked into firm $A$, and types to the right of $0$ are partially locked into firm $B$. The lock-in is partial in the sense that the consumer may purchase from either firm, depending on his taste shock. In the Pareto-dominant exclusive equilibrium, the lock-in is complete, so the distortion is exacerbated. Once the consumer chooses one firm as his exclusive supplier, he cannot purchase the other firm's product.

Next, we sketch our derivation of the Pareto-dominant equilibrium in \cref{res:exclusive}. The consumer's choice of exclusive supplier depends on both firms' subscription schedules, but conditional upon choosing firm $i$, the consumer's optimal  strike price depends only on firm $i$'s schedule. Therefore, each firm faces a sequential screening problem subject to a type-dependent participation constraint (determined by the other firm's subscription schedule). For any pair of exclusive subscription schedules, the consumer's relative preference for firm $B$ is increasing as $\g$ moves to the right. Therefore, firm $B$'s best response to an arbitrary exclusive schedule $s_A$ offered by firm $A$ is determined by the following nested optimization problem. In the outer problem, firm $B$ chooses the leftmost type, $\ubar{\g}_B$, that it will serve. Given $\ubar{\g}_B$, the inner problem is a standard sequential screening problem over the restricted type space $[\ubar{\g}_B, \bar{\g}]$, subject to the constraint that type $\ubar{\g}_B$ 
chooses firm $B$. Therefore, it is optimal for firm $B$ to induce each type $\g$ in $[\ubar{\g}_B, \bar{\g}]$ to select the monopoly strike price $p_B^M(\g)$. The associated subscription schedule is determined by the envelope theorem, together with the binding participation constraint for type $\ubar{\g}_B$. Symmetrically, firm $A$ faces a nested optimization problem, where the outer problem consists of choosing the rightmost type, $\bar{\g}_A$, that it will serve.

Each firm's outer problem is quasiconcave, so its best response is determined by a first-order condition. 
The assumption that $v_0 \geq 1/g(\g^\dagger) + | \g^{\dagger}|$ ensures that in any equilibrium, every type selects some contract. We derive the equilibrium in \cref{res:exclusive} from the first-order conditions for each firm. Formally, if either firm marginally increases its market share by shifting its subscription schedule down, the additional revenue from the marginal consumer exactly offsets the revenue lost from the inframarginal consumers.

There is one subtlety, which explains why there can be multiple equilibria. If firm $i$ shifts its subscription schedule up, then some types will switch from firm $i$ to firm $-i$. These switching types may select strike prices at firm $-i$ that were never selected before firm $i$ deviated. Thus, firm $-i$'s subscription fees at \emph{unselected} strike prices affect the profitability of a deviation by firm $i$.\footnote{Indeed, this is an important reason why the revelation principle fails when there are multiple principals, as observed by \cite{MartimortStole2002}.} The equilibrium in \cref{res:exclusive} is the unique equilibrium with the property that the consumer's optimal strike price at each firm is continuous for \emph{all} types. There are other equilibria in which some firm offers ``bargain'' contracts with very high strike prices and very low subscription fees. These contracts are not selected, but would attract more consumers to switch if the other firm raised its subscription fees. Thus, these contracts deter the other firm from raising its subscription fees, thereby supporting an equilibrium with lower revenue for both firms.

\subsection{Multi-product monopolist} \label{sec:multi_product_monopoly}

In the main model, we assume that the two products are offered by competing firms. Here, we suppose instead that a monopolist controls both products. Now, a \emph{joint contract} is a triple $(p_A, p_B, s)$, which specifies that the consumer pays $s$ in the first period for the right (but not the obligation) to purchase either product $A$ at price $p_A$ or product $B$ at price $p_B$. Formally, the monopolist offers a joint subscription schedule, which is a lower semicontinuous function $s \colon [0, \infty]^2 \to [0, \infty]$ with $s(\infty, \infty) = 0$.\footnote{So, the null contract is now $(\infty, \infty, 0)$. In our result below, we will specify only the non-null contracts with finite subscription fees.} Such a subscription schedule can implement any direct incentive-compatible  deterministic mechanism. We characterize the monopolist's optimal deterministic mechanism under the assumption that $G$ is symmetric. 

\begin{prop}[Multi-product monopolist] \label{res:multi-product_monopoly} Assume that $G$ is symmetric. There exists a threshold $\bar{v}(F,G)$ such that if $v_0 \geq \bar{v}(F,G)$, then the uniquely optimal deterministic mechanism is implemented as follows. 
\begin{itemize}
    \item In the first period, the monopolist offers a single joint contract $(p_A^{\mathrm{MM}}, p_B^{\mathrm{MM}}, s^{\mathrm{MM}})$, where
\[
    p_A^{\mathrm{MM}} = p_B^{\mathrm{MM}} = 0 \quad \text{and} \quad 
    s^{\mathrm{MM}} = \E_{\th | 0} \Brac{ \max \{ v_A(\th), v_B( \th) \} }.
\]
Every type selects this contract. 
    \item In the second period, the consumer's product-purchase strategy $q^{\mathrm{MM}}$ is given by
\[
    q^{\mathrm{MM}} (\g, \th) =
    \begin{cases}
        (1,0) &\text{if}~ v_A(\th) > v_B(\th), \\
        (0,1) &\text{if}~ v_B(\th)\geq v_A(\th).
    \end{cases}
\]
\end{itemize}
\end{prop}

The monopolist offers a single contract that entitles the consumer to purchase either product at zero strike price. The subscription fee, $s^{\mathrm{MM}}$, is the largest fee that all types are willing to pay for this option. The consumer's ex-post value from this option, $\max\{ v_A(\th), v_B(\th )\}$, is a V-shaped function centered at $\th = 0$; see \cref{fig:valuations}. Since the taste-shock density $f$ is symmetric and single-peaked, it follows that type $\g = 0$ has the lowest willingness to pay for this option. Thus, $s^{\mathrm{MM}}$ equals type $0$'s valuation for the option. The consumer always selects this option and then consumes his preferred product in the second period. 

To prove \cref{res:multi-product_monopoly}, we cannot use the standard Myersonian approach. In the monopolist's multi-product sequential screening problem, global incentive constraints generally bind;  handling such constraints is challenging in sequential screening problems, even with a single product.\footnote{The difficulty is that first-period incentive compatibility implies that the allocation is integral monotone rather than pointwise monotone. \cite{BattagliniLamba2019} explore the limitations of the ``first-order approach'' in a multi-period setting with finitely many types. \cite{KrasikovLamba2026} consider a dynamic screening problem in which the buyer's valuation follows a Poisson renewal process; global incentive constraints bind, and they characterize the optimal deterministic mechanism. \cite{LiShi2025} characterize the optimal stochastic mechanism when the consumer has three types and continuous valuations.} Instead, we argue as follows. The mechanism in \cref{res:multi-product_monopoly} induces the efficient allocation, so it maximizes the total surplus among all mechanisms. We show that this mechanism minimizes the consumer's expected information rent among all deterministic mechanisms that induce full coverage in the product market. Intuitively, under this mechanism, the lowest information rents go to the types around $\g = 0$, where the density $g$ is highest. We further show that, as long as $v_0$ is large enough, full-coverage is optimal because exclusion reduces total surplus by more than it reduces the consumer's information rent. An expression for the threshold $\bar{v}(F, G)$ is given in the proof. 


Our proof approach immediately yields the following welfare comparison between multi-product monopoly and the equilibria of the other three settings. For this comparison, we impose the assumptions from \cref{res:equilibrium} and \cref{res:Hotelling,res:exclusive,res:multi-product_monopoly} so that each setting yields a unique prediction.\footnote{In particular, in the setting of exclusive subscriptions, the prediction is the unique Pareto-dominant equilibrium.} We label the four settings as non-exclusive contracting (NE), exclusive contracting (E), spot pricing (SP), and multi-product monopoly (MM).  For each setting $j \in \{ \mathrm{NE}, \mathrm{E}, \mathrm{SP}, \mathrm{MM} \}$, let $\mathrm{CS}^{j}$ and $\mathrm{PS}^{j}$ denote (expected) consumer surplus and producer surplus in setting $j$. 

\begin{prop}[Multi-product monopoly surplus comparison] \label{res:MM_welfare} Assume that $G$ is symmetric and $v_0 \geq \bar{v}(F,G) \vee (1/h(0)) \vee [ (7/2) \max_{\g' \in \G} (1/ g(\g'))]$. Then
\begin{equation*}
\begin{aligned}
    \mathrm{PS}^{\mathrm{MM}} &>   \max\{ \mathrm{PS}^{\mathrm{NE}}, \mathrm{PS}^{\mathrm{SP}}, \mathrm{PS}^{\mathrm{E}} \}, \\
   \mathrm{CS}^{\mathrm{MM}} &< \min\{ \mathrm{CS}^{\mathrm{NE}}, \mathrm{CS}^{\mathrm{SP}} , \mathrm{CS}^{\mathrm{E}} \}.
\end{aligned}
\end{equation*}
\end{prop}

We have restricted the multi-product monopolist to deterministic mechanisms. The deterministic benchmark is easier to interpret, and it yields a clear comparison with the subscriptions in the main model. However, the monopolist could generally increase revenue by offering product lotteries, similar to the \emph{static} multi-product Hotelling auction setting of \cite{LortscherMuir2025}.

\section{Contracting structure and welfare} \label{sec:welfare_comparison}

Having established that consumer surplus is lowest when the products are controlled by a joint monopolist, we turn to the more subtle comparison of consumer and producer surplus in the three competitive settings: non-exclusive contracting (NE), exclusive contracting (E), and spot pricing (SP). With symmetric firms, the spot-pricing equilibrium is efficient, so it follows from \cref{res:efficiency} that total surplus decreases from spot pricing to non-exclusive contracting to exclusive contracting. On the other hand, competition for consumers is stiffer when consumers are less informed, and hence less heterogeneous. Exclusivity further intensifies this competition because each firm's offer must be better than its competitor's, rather than simply good enough to induce the consumer to multi-home. The consumer surplus ranking generally depends on the balance between these countervailing efficiency and competition effects. In this section, we show that the competition effect dominates in the limiting case of ``early contracting.'' Then we compare the distribution of surplus over the consumer population. 





\subsection{Early contracting}

We obtain a sharp ranking of consumer and producer surplus when the firms can contract with the consumer sufficiently \emph{early}, before the consumer learns much about his preferences. This is the natural limit for distinguishing these settings; in the opposite limit where firms contract sufficiently \emph{late}, when the consumer has little residual uncertainty about his valuations, the three competitive settings converge. We formally model early contracting as follows. We retain the primitive distributions $G$ and $F$ of the random variables $\g$ and $\e$. For each positive $\s$,  we consider the \emph{$\s$-scaled environment} in which the consumer's position is given by $\th = \s \g + \e$. As before, the consumer learns his signal $\g$ (or equivalently, the scaling $\s \g$) before contracting. After contracting, the consumer learns his taste shock $\e$ (or equivalently, his position $\th$). In the spot-pricing game, firms still interact with the consumer after the consumer learns his position $\th$.

We say that a statement holds under \emph{sufficiently early contracting} if it holds in the $\s$-scaled environment for all sufficiently small $\s$. As $\s$ tends to $0$, the consumer's pre-contractual signal becomes less and less informative about his final preferences, and the distribution of the consumer's final position $\th$ converges to $F$. Our interpretation of this limit is that contracting takes place before the consumer has learned much about his preferences.\footnote{Under a strict learning interpretation, the position distribution would not change at all with the contract timing. We could obtain the same surplus rankings, while keeping fixed the distribution of $\th$, at the cost of some technical complications.} The surplus ranking is invariant to rescaling the valuation units, so in the $\s$-scaled environment, we can equivalently consider the rescaled position $\th / \s = \g  + \e/\s$ (with the rescaled constant $v_0/\s$). Thus, the early contracting limit (as $\s$ tends to $0$) admits the alternative interpretation that the distribution of the consumer's taste shock is sufficiently dispersed relative to the fixed signal distribution. 

We will impose the assumptions from \cref{res:equilibrium} and \cref{res:Hotelling,res:exclusive} so that each setting yields a unique prediction.\footnote{As before, in the setting of exclusive subscriptions, the prediction is the unique Pareto-dominant equilibrium.} Recall that $\mathrm{CS}^{j}$ and $\mathrm{PS}^{j}$ denote (expected) consumer surplus and producer surplus in setting $j$, for each $j \in \{ \mathrm{NE}, \mathrm{SP}, \mathrm{E} \}$.


\begin{prop}[Early contracting] \label{res:surplus}  Assume $\ubar{\g} < 0< \bar{\g}$ and $v_0 > 1/ f(0)$. Under sufficiently early contracting, the following hold:
\begin{equation*}
\begin{aligned}
    \mathrm{CS}^{\mathrm{E}} 
   &> \mathrm{CS}^{\mathrm{NE}}   
    > \mathrm{CS}^{\mathrm{SP}}, \\
    \mathrm{PS}^{\mathrm{E}} 
   &< \mathrm{PS}^{\mathrm{NE}}   
    < \mathrm{PS}^{\mathrm{SP}}. 
\end{aligned}
\end{equation*}
\end{prop}



First, consider the consumer surplus ranking between exclusive and non-exclusive contracting. This ranking is particularly striking given the interpretation that the consumer's taste shock becomes more \emph{dispersed} in the early contracting limit, thus exacerbating the total surplus loss from exclusivity. We show that the lower subscription fees under exclusivity more than offset this loss.  As $\s$ tends to $0$, the equilibrium strike prices in both settings tend to $0$. Intuitively, as consumers become homogeneous, the firms' downward distortion motive vanishes. In the exclusive equilibrium, the subscription fees also tend to $0$ by the logic of symmetric Bertrand competition. In the non-exclusive equilibrium, firm $i$'s subscription fee converges to $\E_{\th \sim F} [ (v_i(\th) - v_{-i}(\th)_+)_+]$. Thus, each firm extracts its expected contribution to the total surplus, lowering consumer surplus relative to the exclusive equilibrium.

\begin{figure}
\centering
\begin{tikzpicture}
    \begin{axis}[
        xmin = -4, xmax = 4,
        ymin = -2 , ymax = 8,
        axis y line = none,
        axis x line = middle,
        xlabel = {$\th$},
        xlabel style={at=(current axis.right of origin), anchor=west}, 
        ymajorticks = false,
        xtick = {0},
        xticklabels = {$0$},
        clip = false,
    ]

        \addplot[fill = Orange, fill opacity=0.5, draw=none] coordinates { (-3.5, 0) (-4, 0) (-4, 7.5) (0,3.5) (-3.5,0)};

        \addplot[fill = Blue, fill opacity=0.5, draw=none] coordinates { (3.5, 0) (4, 0) (4, 7.5) (0,3.5) (3.5,0)};

        \addplot[fill = Green, fill opacity=0.5, draw=none] coordinates { (-3.5, 0) (0,3.5) (3.5,0) (-3.5, 0)};
                
        \addplot[dashed] coordinates { (-4,3.5) (4,3.5)} node [pos = 1, anchor = west] {$v_0$};
        \addplot[thick] coordinates { (-4,7.5) (4,-0.5)} node [Orange, pos = 0.2, yshift = 10pt, anchor = south] {$v_A(\th)$};  
        \addplot[thick] coordinates { (-4,-0.5) (4,7.5)} node [Blue, pos = 0.8 , yshift = 10pt, anchor = south] {$v_B(\th)$};

        \addplot[dotted] coordinates {(0,0)  (0,3.5)};

    \end{axis}
\end{tikzpicture}
\caption{Limiting equilibrium under early contracting}
\label{fig:no_info}
\end{figure}
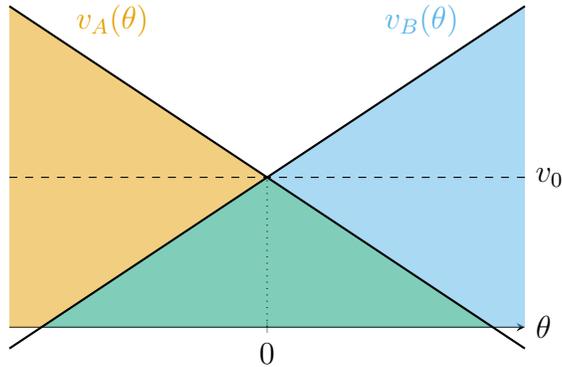

\cref{fig:no_info} illustrates the comparison graphically. The limiting total surplus under the non-exclusive equilibrium is represented by the total shaded region.\footnote{These regions extend over the entire horizontal axis;  the ``area'' of each shaded region should be interpreted as an integral with respect to the distribution $F$.} The orange (blue) region represents the revenue for firm $A$ ($B$). The remaining surplus (the green region) goes to the consumer. Under the limiting exclusive equilibrium, the consumer surplus is the exclusive total surplus, represented by the union of the green and blue regions (or, by symmetry, the green and orange regions).


Next, consider the consumer surplus ranking between  non-exclusive contracting and spot pricing. In the limiting non-exclusive equilibrium, the ex-ante outcome coincides with the equilibrium of the following ``personalized pricing'' game, analyzed in \cite{ThisseVives1988}. The consumer's position $\th$ becomes public, and then each firm posts a price. For each $\th$, valuations are common knowledge, so there is an associated complete-information (asymmetric) Bertrand game. If $v_i(\th) > v_{-i}(\th)_+$, then firm $i$ charges $v_i(\th) - v_{-i}(\th)_+$ and firm $-i$ charges $0$; the consumer purchases product $i$. The resulting allocation is efficient. The consumer's ex-ante expected payment to firm $i$ is $\E_{\th \sim F} [ (v_i(\th) - v_{-i}(\th)_+)_+]$, which is precisely firm $i$'s limiting subscription fee in our model (where the consumer's position is private). We conclude that the limiting consumer surplus comparison between non-exclusive contracting and spot pricing reduces to the classical  consumer surplus comparison between personalized and uniform pricing. \cite{ThisseVives1988} show that if consumers are uniformly distributed on the Hotelling line, then at every position, the consumer pays less under personalized pricing than under uniform pricing.\footnote{\cite{ZhouRhodes2024} compare uniform and personalized pricing in a more general oligopoly model. They find that consumer surplus is higher under personalized pricing whenever market coverage is sufficiently high. \cite{AliEtal2022} consider voluntary consumer disclosures to competing firms in a more general setting in which consumers are drawn from a symmetric, log-concave distribution on the Hotelling line.} We prove that the ex-ante \emph{expected} price paid by the consumer remains lower under personalized pricing as long as the distribution of the consumer's position is symmetric and log-concave.

An important implication of \cref{res:surplus} is that competition reverses the effect of early contracting on consumer surplus. In the early contracting limit, a monopolist can fully extract the surplus, leaving the consumer with no information rent. So early contracting, like personalized pricing, reduces consumer surplus under monopoly but increases consumer surplus under competition.

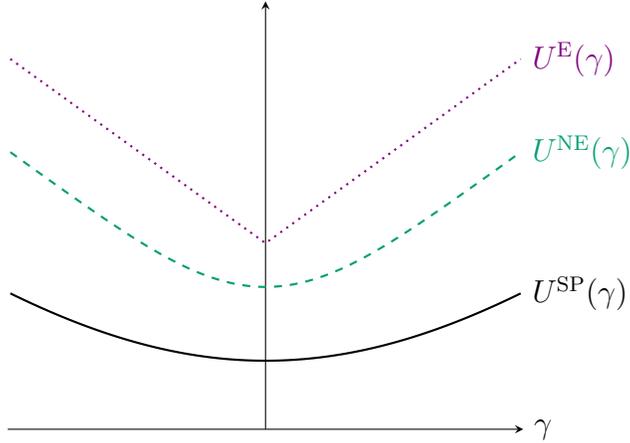
\begin{figure}
\begin{center}
\begin{tikzpicture}
  \begin{axis}[
    xmin=-1, xmax=1, ymin=4.5, ymax=6.8,
    axis lines=middle,
    xlabel={$\gamma$},
    ylabel={},
    xlabel style={at=(current axis.right of origin), anchor=west}, 
    tick style={black},
    tick align=inside,
    xtick=\empty,
    ytick=\empty,
    legend cell align=left,
    clip = false
  ]

\addplot[thick, Black] table[x index=0,y index=1,col sep=space] {Spot.tex}
node[pos=1, anchor=west, inner sep=4pt] {$U^{\mathrm{SP}}(\g)$};

\addplot[thick, dashed, Green] table[x index=0,y index=1,col sep=space] {Duopoly.tex}
node[pos= 1, anchor=west, inner sep=4pt] {$U^{\mathrm{NE}} (\g)$};

\addplot[thick, dotted, Purple] table[x index=0,y index=1,col sep=space] {Exclusive.tex}
node[pos=1, anchor=west, inner sep=4pt] {$U^{\mathrm{E}} (\g)$};

  \end{axis}
\end{tikzpicture}
\end{center}
\caption{Consumer utility functions in the three competitive settings, with a uniform type distribution and a standard normal taste shock}
\label{fig:consumer_utility}
\end{figure}

We confirm numerically that the consumer surplus ranking in \cref{res:surplus} often holds away from the limit. For each setting $j \in \{ \mathrm{NE}, \mathrm{SP}, \mathrm{E}\}$, let $U^j$ denote the consumer's interim utility function. \cref{fig:consumer_utility} plots these functions in our running example, where the type distribution is uniform over $\G = [-1,1]$, with $v_0 = 7$ and a standard normal taste shock. Here, the consumer surplus ranking even holds type-by-type: $U^{\mathrm{E}} (\g) > U^{\mathrm{NE}} (\g) > U^{\mathrm{SP}} (\g)$ for each type $\g$.

The producer surplus ranking is easily derived from the consumer surplus ranking. Under non-exclusive contracting and spot pricing, the allocation converges to efficiency in the early contracting limit. Under exclusive contracting, the limiting allocation is inefficient. Since producer surplus equals total surplus net of consumer surplus, it can be shown that the producer surplus ranking is exactly the reverse of the consumer surplus ranking. To be clear, a firm unilaterally benefits by moving from spot pricing to subscription pricing; this explains the popularity of subscriptions. But once this firm's competitor also adopts subscription pricing, producer surplus is lower in the resulting equilibrium than in the original spot-pricing equilibrium. In the race to the bottom of early contracting, consumers benefit at the expense of producers. 

\subsection{Distribution of consumer surplus}

 Next, we study the \emph{distribution} of surplus across the population of consumers. In \cref{fig:consumer_utility}, note that each utility function is single-dipped around type $\g = 0$, and the steepness of the functions increases from spot pricing to non-exclusive contracting to exclusive contracting. We now generalize these observations. 

We say that two utility functions $U$ and $V$ on the type space $\G$ are \emph{ordinally equivalent} if for all $\g, \g' \in \G$, we have 
\[
    U(\g) \geq U(\g') \iff V(\g) \geq V(\g').
\]
For ordinally equivalent functions, we introduce a natural analog of the dispersive order between distributions. Given ordinally equivalent functions $U$ and $V$ on $\G$, we say that $U$ is \emph{weakly more dispersed} than $V$, written $U \succeq_{\mathrm{disp}} V$, if for all $\g, \g' \in \G$, we have
\[
    |U(\g) - U(\g')| \geq |V(\g) - V(\g')|.
\]
Further, $U$ is \emph{strictly more dispersed} than $V$, written $U \succ_{\mathrm{disp}} V$, if $U \succeq_{\mathrm{disp}} V$ and $V \not\succeq_{\mathrm{disp}} U$. In words, one utility function is more dispersed than another if it amplifies the utility difference between any pair of types. 

\begin{prop}[Consumer surplus distribution] \label{res:welfare_distribution} Assume that $G$ is symmetric and $v_0 \geq (1/h(0)) \vee [ (7/2) \max_{\g' \in \G} (1/ g(\g'))]$. The functions $U^{\mathrm{NE}}$, $U^{\mathrm{SP}}$, and $U^{\mathrm{E}}$ are all strictly convex and symmetric. Moreover,
\[
    U^{\mathrm{E}} \succ_{\mathrm{disp}} U^{\mathrm{NE}} \succ_{\mathrm{disp}} U^{\mathrm{SP}}.
\]

\end{prop}

In each setting, the consumer's utility function is single-dipped around $\g = 0$. Moving from spot pricing to non-exclusive contracting to exclusive contracting, the consumer's utility function becomes strictly more dispersed. Intuitively, subscription contracts, particularly exclusive contracts, benefit consumers with extreme preferences more than consumers with less extreme preferences. Mathematically, the utility difference between any two types is determined by incentive compatibility and the envelope theorem. By the envelope theorem, the slope of the consumer's utility function is the difference between interim demand at firm $B$ and firm $A$; see equation \eqref{eq:envelope_main}. Under non-exclusive contracting, types to the right of $0$ are partially locked into firm $B$, so relative to spot pricing, they purchase product $B$ more often and product $A$ less often. Under exclusive subscriptions, the lock-in becomes complete, so these consumers purchase product $B$ even more often and they never purchase product $A$.

\section{Conclusion} \label{sec:conclusion}

This paper introduces a model of competition between firms that offer menus of option contracts to sequentially screen consumers. We find that the effect of sequential screening, relative to spot pricing, is fundamentally different under competition. In our model, the consumer chooses a subscription contract from each firm, even though he knows that he will ultimately purchase only one product. Subscriptions induce allocative distortions by partially locking in consumers. Despite this inefficiency, we find that if firms can contract with consumers sufficiently early, before consumers have learned much about their preferences, then consumer surplus is higher than under spot pricing. In this limit, exclusive contracts further benefit consumers by stiffening competition. While a firm may unilaterally benefit by moving to subscription pricing, in the resulting equilibrium, consumers benefit at the expense of producers. 


We work within the canonical Hotelling model of horizontal differentiation. A natural direction for future research is to examine competition in dynamic contracts under alternative models of consumer heterogeneity and product differentiation. We expect that our solution technique---explicitly incorporating into each firm's auxiliary design problem the agent's choices at the competing firm---will be useful in these settings. 
 In particular, it would be interesting to analyze competition when consumers are differentiated both vertically and horizontally. This will likely be challenging, however, given the difficulty of multi-dimensional screening, even in the static monopoly case. 


\newpage

\appendix

\bibliography{References.bib}
\bibliographystyle{ecta.bst}

\newpage

\section{Proofs} \label{sec:proofs}


\subsection{Proof of Proposition~\ref{res:monopoly_benchmark}} \label{sec:proof_monopoly_benchmark}
We prove \cref{res:monopoly_benchmark}  under weaker assumptions. In place of \cref{as:regularity}, it suffices to assume that $(1 - G)/g$ is weakly decreasing\footnote{For the symmetric result with firm $A$ as the monopolist, we would instead assume that $G/g$ is weakly increasing.} and that the density $f$ is continuous and has \emph{monotone tails}, i.e., there exist $\ubar{\e}, \bar{\e} \in \R$ such that $f$ is weakly increasing over $(-\infty, \ubar{\e})$ and weakly decreasing over $(\bar{\e}, \infty)$. These technical assumptions on $f$ are only needed to differentiate under the integral sign.

The proof proceeds in three steps. First, we set up the monopoly program for firm $B$. Second, we solve a relaxation of this program. Third, we verify that our solution of the relaxed program is feasible, i.e., satisfies global incentive compatibility. 

\paragraph{Monopoly program} By the dynamic revelation principle, firm $B$ chooses an allocation rule $q_B \colon \G \times \Th \to [0,1]$ and a transfer rule $t_B \colon \G \times \Th \to \R$ to solve 
\begin{equation} \label{eq:firm_B_monopoly}
\begin{aligned}
    &\maz &&\E_{\g, \th} \Brac { t_B (\g, \th)} \\
    &\text{subject to} && \th \in \argmax_{\th' \in \Th}  \Set{ v_B (\th) q_B ( \g, \th') - t_B(\g, \th')}, \quad (\g, \th) \in \G \times \Th\\
    &&& \g \in \argmax_{\g'}\, \E_{\th | \g} \Brac{ \max_{\th' \in \Th}  \Set {v_B (\th) q_B ( \g', \th') - t_B(\g', \th') }}, \quad \g \in \G \\
    &&& \E_{\th| \g} \Brac{ v_B (\th) q_B ( \g, \th) - t_B(\g, \th)} \geq 0, \quad \g \in \G.
\end{aligned}
\end{equation}
The three constraints impose second-period incentive compatibility, first-period incentive compatibility, and participation.  Second-period incentive compatibility requires that after the consumer has reported truthfully in the first period, he always finds it optimal to report truthfully in the second period. First-period incentive compatibility requires that the consumer always finds it optimal to report truthfully in the first period, anticipating that he will subsequently report optimally in the second period. 

\paragraph{Solving the relaxation} To analyze the program in \eqref{eq:firm_B_monopoly}, we introduce additional notation. For any $\g' \in \G$ and $\th, \th' \in \Th$, let 
\[
    u(\g', \th'| \th) = v_B(\th) q_B(\g', \th') - t_B(\g', \th').
\]
With this notation, we consider the following relaxed problem
\begin{equation} \label{eq:firm_B_monopoly_relaxed}
\begin{aligned}
    &\maz &&\E_{\g, \th} \Brac { t_B (\g, \th)} \\
    &\text{subject to} && \th \in \argmax_{\th' \in \Th}  u (\g, \th'| \th), \quad (\g, \th) \in \G \times \Th \\
    &&& \g \in \argmax_{\g'}\, \E_{\th | \g} \Brac{ u (\g', \th | \th)}, \quad \g \in \G \\
    &&& \E_{\th| \ubar{\g}} [ u ( \ubar{\g}, \th | \th)] \geq 0.
\end{aligned}
\end{equation}
We have simplified the first-period incentive-compatibility constraint by applying the second-period incentive-compatibility constraint inside the expectation. We have also dropped the participation constraint for every type except $\ubar{\g}$. 

Next, we apply the envelope theorem of \cite{milgrom2002envelope} to the incentive constraints in \eqref{eq:firm_B_monopoly_relaxed}. We begin with second-period incentive compatibility. Fix $\th_0 \in \Th$. Since $v_B'(\th) = 1$, we have $u_3 ( \g, \th' | \th) = q_B (\g, \th')$, where $u_3$ denote the partial derivative of $u$ with respect to its third argument.  Note that this derivative has absolute value at most $1$. Applying the envelope theorem to the second-period incentive-compatibility constraint, we conclude that for each $\g \in \G$ and $\th \in \Th$, we have
\begin{equation} \label{eq:period_2_envelope}
\begin{aligned}
     u(\g, \th | \th) = u(\g, \th_0 |\th_0) + \int_{\th_0}^{\th} q_B(\g, \th') \de \th'.
\end{aligned}
\end{equation}

Now, we turn to first-period incentive compatibility. For all $\g, \g' \in \G$, let 
\[
    U(\g' | \g) = \E_{\th | \g} \Brac{ u(\g', \th | \th)}.
\]
Plug in \eqref{eq:period_2_envelope} to get
\begin{equation} \label{eq:U_integration_change}
\begin{aligned}
    U(\g' |\g) 
    = u(\g', \th_0 | \th_0) +  \int_{-\infty}^{\infty} \Brac{ \int_{\th_0}^{\th} q_B(\g', \th') \de \th'}  f (\th  -\g) \de \th.
\end{aligned}
\end{equation}
To evaluate the double integral, we separate the outer integral according to whether $\th \geq \th_0$. We have
\begin{equation} \label{eq:integral_interchange}
\begin{aligned}
   &\int_{-\infty}^{\th_0} \Brac{ \int_{\th_0}^{\th} q_B(\g', \th') \de \th'}  f (\th  -\g) \de \th  + \int_{\th_0}^{\infty} \Brac{ \int_{\th_0}^{\th} q_B(\g', \th') \de \th'}  f (\th  -\g) \de \th\\
   &=    - \int_{-\infty}^{\th_0} q_B(\g', \th')  F(\th' - \g) \de \th' +  \int_{\th_0}^{\infty}  q_B(\g', \th') (1 - F( \th' - \g)) \de \th',
\end{aligned}
\end{equation}
where the equality follows from interchanging the order of integration, by Fubini's theorem. Substitute \eqref{eq:integral_interchange} into \eqref{eq:U_integration_change} and simplify to get
\begin{equation} \label{eq:U_monopoly}
    U(\g' |\g) 
    = u(\g', \th_0 | \th_0) + \int_{-\infty}^{\infty}  q_B(\g', \th') ([\th' > \th_0] - F( \th' - \g)) \de \th'.
\end{equation}
Differentiating under the integral in \eqref{eq:U_monopoly},\footnote{\label{ft:under_integral}Since $q$ is bounded, this is justified by the dominated convergence theorem, together with the mean value theorem, provided that there exists $\d > 0$ such that the map $x \mapsto \hat{f} (x, \d) \coloneq \sup_{y \in [x-\d, x + \d]} f(y)$ is integrable. This integrability property follows from our assumption that $f$ is continuous and has monotone tails, i.e., there exist $\ubar{\e}, \bar{\e} \in \R$ such that $f$ is weakly increasing over $(-\infty, \ubar{\e})$ and weakly decreasing over $(\bar{\e}, \infty)$. Let $\bar{f} = \sup_{x \in [\ubar{\e} - 2\d , \bar{\e} + 2 \d]} f(x)$. For any $x \in [\ubar{\e} - \d, \bar{\e} + \d]$, we have $\hat{f}(x, \d)\leq \bar{f}$. For $x \leq \ubar{\e} - \d$, we have $\hat{f}(x, \d) \leq f(x + \d)$, and for $x \geq \bar{\e} + \d$, we have $\hat{f}(x, \d) \leq f(x - \d)$. Therefore,
\[
    \int_{-\infty}^{\infty} \hat{f}(x, \d) \de x \leq 1 - F(\bar{\e}) + F(\ubar{\e}) + (\bar{\e} - \ubar{\e} + 2 \d) \bar{f}< \infty. 
\]
} we obtain
\[
    U_2 (\g' | \g) = \int_{-\infty}^{\infty} q_B (\g', \th') f (\th' - \g) \de \th',
\]
where $U_2$ denotes the partial derivative of $U$ with respect to the second argument. Note that this derivative has absolute value at most $1$. Applying the envelope theorem to the first-period incentive-compatibility constraint, we conclude that for each $\g \in \G$, we have
\begin{equation} \label{eq:U}
\begin{aligned}
    U(\g | \g) 
    = U( \ubar{\g} | \ubar{\g}) + \int_{\ubar{\g}}^{\g} \Paren{\int_{-\infty}^{\infty}  q_B(\g', \th') f(\th' - \g') \de \th'}   \de \g'.
\end{aligned}
\end{equation}

Now, we decompose revenue as total surplus net of information rent. The objective in  \eqref{eq:firm_B_monopoly_relaxed} can be expressed as
\[
\E_{\g, \th} [ t_B(\g, \th)]  = \E_{\g, \th} [ v_B(\th) q_B(\g, \th)] - \E_{\g}[U( \g | \g)].
\]
From \eqref{eq:U}, we have
\begin{equation*}
\begin{aligned}
         \E_{\g} [ U(\g|\g)]  - U(\ubar{\g} | \ubar{\g})
         &= \int_{\ubar{\g}}^{\bar{\g}}  \Brac{ \int _{\ubar{\g}}^{\g} \Paren{\int_{-\infty}^{\infty}  q_B(\g', \th')  f(\th' - \g') \de \th'}   \de \g'} g(\g) \de \g \\
        &= \int_{\ubar{\g}}^{\bar{\g}} \Paren{\int_{-\infty}^{\infty}  q_B(\g', \th') (1 - G(\g')) f(\th' - \g') \de \th'} \de \g' \\
        &= \E_{\g, \th} \Brac{ \frac{1 - G(\g)}{g(\g)} q_B(\g, \th)},
\end{aligned}
\end{equation*}
where the second equality follows from interchanging the order of integration, by Fubini's theorem. Therefore, we obtain the following further relaxation of  \eqref{eq:firm_B_monopoly_relaxed}: choose $q_B$ and a nonnegative constant $U(\ubar{\g}| \ubar{\g})$ to maximize the objective
\[
    \E_{\g, \th} [ t_B(\g, \th)] = - U (\ubar{\g} | \ubar{\g}) + \E_{\g, \th} \Brac{ \Paren{ v_B(\th) -  \frac{1 - G(\g)}{g(\g)}} q_B(\g, \th)}.
\]
The expression inside the expectation is pointwise maximized by 
\[
    q_B^M(\g, \th) = [ v_B(\th) \geq (1 - G(\g))/g(\g) ].
\]
Clearly, it is optimal to set $U (\ubar{\g} | \ubar{\g}) = 0$. The transfers are thus pinned down. 

\paragraph{Global incentive compatibility} The profile $(s_B^M, p_B^M, q_B^M)$ induces the solution of the relaxed problem. It remains to check that global incentive compatibility is satisfied; then $U(\g |\g) \geq U(\ubar{\g} | \ubar{\g}) = 0$ for every type $\g$, so the participation constraints follow. Clearly, global incentive compatibility is satisfied in the second period. For first-period incentive compatibility, note that $s_B^M$ is constant over $[\bar{p}_B^M, \infty)$, so the consumer strictly prefers $\bar{p}_B^M$ to any strike price above $\bar{p}_B^M$. Thus, it suffices to consider deviations to strike prices in the interval $[0, \bar{p}_B^M] = p_B^M( \G)$. For any fixed type $\g$, the map $p \mapsto \E_{\th| \g} [(v_B(\th) - p)_+]$ has derivative $- Q_B^M (p | \g)$. Therefore, for all $\g, \g' \in \G$, we have
\begin{equation*}
\begin{aligned}
    &U(\g| \g) - U(\g' | \g) \\
    &= \E_{\th |\g} \Brac{ (v_B(\th) - p_B^M(\g))_+ -  (v_B(\th) - p_B^M(\g'))_+}  - \Paren{  s_B^M( p_B^M(\g)) - s_B^M (p_B^M(\g')) } \\
    &= \int_{p_B^M(\g')}^{p_B^M(\g)} [\hat{Q}_B^M (p_B') - Q_B^M ( p_B' | \g)] \de p_B'.
\end{aligned}
\end{equation*}
We verify that this integral is nonnegative. By assumption, $p_B^M = (1 - G)/g$ is weakly decreasing. If $\g' < \g$, then $p_B^M(\g') \geq p_B^M ( \g)$ and $\hat{Q}_B^M (p_B') \leq Q_B^M(p_B'| \g)$ for almost all $p_B' \in [p_B^M ( \g), p_B^M(\g')]$. If $\g' > \g$, then $p_B^M(\g') \leq p_B^M(\g)$ and  $\hat{Q}_B^M (p_B') \geq Q_B^M(p_B'| \g)$ for almost all $p_B' \in [p_B^M ( \g'), p_B^M(\g)]$. 

\subsection{Proof of Theorem~\ref{res:equilibrium}: Equilibrium} \label{sec:proof_equilibrium_existence}

The proof of \cref{res:equilibrium} is separated into two sections. In this section, we prove that $(s_A^\ast, s_B^\ast, p^\ast, q^\ast)$ is an equilibrium, under the assumption that $v_0 \geq \max_{\g' \in \G} (1/g(\g'))$. 
In \cref{sec:proof_equilibrium_uniqueness}, under the stronger assumption that $v_0 \geq (7/2) \max_{\g' \in \G} (1/g(\g'))$, we prove that all other equilibria are equivalent to $(s_A^\ast, s_B^\ast, p^\ast, q^\ast)$, thus establishing essential uniqueness. 

In fact, we prove here that $(s_A^\ast, s_B^\ast, p^\ast, q^\ast)$ is an equilibrium under a weakening of our standing assumptions. In place of \cref{as:regularity}, we assume only that (i) $G/g$ is weakly increasing and $(1 - G)/g$ is weakly decreasing, and (ii) the density $f$ is continuous and has \emph{monotone tails}, i.e., there exist $\ubar{\e}, \bar{\e} \in \R$ such that $f$ is weakly increasing over $(-\infty, \ubar{\e})$ and weakly decreasing over $(\bar{\e}, \infty)$. The technical assumptions on $f$ are only needed to differentiate under the integral sign.

In our proof, we check both conditions in the definition of equilibrium (\cref{def:equilibrium}). First, we check that the consumer is playing a best response. Then we check that each firm is playing a best response. 

\paragraph{Consumer best response} We check that the consumer strategy $(p^\ast, q^\ast)$ is a best response to $(s_A^\ast, s_B^\ast)$.

First, we check that the product-purchase strategy $q^\ast$ is optimal in the second period. From the definition of $q^\ast$, it is clear that in the second period the consumer chooses optimally between the two products; we check that it is always optimal for the consumer to purchase some product. For each type $\g \in \G$, we have
\[
    p_A^\ast (\g) + p_B^\ast (\g) = \frac{2}{g(\g)} \leq \max_{\g' \in \G} \frac{2}{g(\g')} \leq 2v_0,
\]
so for every position $\th$, we have
\begin{equation*}
\begin{aligned}
    \max \{ v_A( \th) - p_A^\ast(\g), v_B(\th) - p_B^\ast (\g) \} 
    &\geq \frac{ v_A (\th) - p_A^\ast (\g) + v_B (\th) - p_B^\ast (\g) }{2} \\
    &= \frac{ 2v_0 - p_A^\ast (\g) - p_B^\ast (\g)}{2} \\
    &\geq 0.
\end{aligned}
\end{equation*}

Next, we show that the contract selection strategy $p^\ast$ is optimal in the first period. For each type $\g \in \G$ and all prices $p_A, p_B \in [0,\infty]$, let 
\[
    u( p_A, p_B | \g)  =  - s_A^\ast (p_A) - s_B^\ast (p_B) +  \E_{\th | \g} \Brac{ \max \{ 0, v_A(\th)  - p_A, v_B (\th) - p_B \}}.
\]
Let
\[
    U_A ( p_A | \g) = \max_{p_{B}' \in [0, \infty] } u ( p_A, p_{B}' | \g), 
    \qquad
    U_B ( p_B | \g) = \max_{p_{A}' \in [0, \infty] } u ( p_A', p_{B} | \g).
\]
Each maximum is well-defined because the functions
$u (p_A, \cdot | \g)$ and $u( \cdot, p_B | \g)$ are both upper semicontinuous over the compact set $[0, \infty]$. 

The key step is the following lemma. 

\begin{lem}[Consumer response] \label{res:consumer_response} For each type $\g \in \G$, we have 
\[
    p_{A}^\ast (\g) \in \argmax_{ p_{A}' \in [0, \infty]} u ( p_A', p_{B}^\ast (\g) | \g ) 
    \quad
    \text{and}
    \quad
    p_B^\ast (\g) \in \argmax_{p_B' \in [0, \infty]} U_B ( p_B' | \g).
\]
\end{lem}
\Cref{res:consumer_response} implies that, for all prices $p_A, p_B \in [0,\infty]$, we have
\begin{equation*}
\begin{aligned}
    u (p_A, p_B | \g)  \leq 
    U_B (p_B | \g) 
    \leq  U_B (p_B^\ast (\g) | \g) = u (p_A^\ast (\g), p_B^\ast (\g) | \g),
\end{aligned}
\end{equation*}
where the first inequality follows from the definition of $U_B$, the second inequality follows from the second part of \cref{res:consumer_response}, and the equality follows from the first part of \cref{res:consumer_response}. 


\paragraph{Firm best response} We prove a stronger result, namely that neither firm can profit by deviating to a direct stochastic mechanism and then selecting the consumer's best response. Without loss of generality, we check that firm $B$ does not have such a profitable deviation. 

First, we introduce notation to represent a deviation by firm $B$. Let 
\[
    \mathcal{Q} = \{ (q_A, q_B) \in [0,1]^2 : q_A + q_B \leq 1\}.
\]
By our convention, $s_A^\ast (\infty) = 0$, so selecting the strike price $p_A = \infty$ from firm $A$ is equivalent to not contracting with firm $A$. Therefore, a deviation by firm $B$, together with the selected consumer best response, can be represented by functions 
\begin{equation} \label{eq:consumer_behavior}
    p_A \colon \G \to [0,\infty]
    \quad
    \text{and}
    \quad
    (q_A, q_B, t_B) \colon \G \times \Th \to \QQ \times \R.
\end{equation}
The interpretation is that firm $B$'s deviation induces type $\g$ to report type $\g$ to firm $B$ and to select from firm $A$ the contract $(p_A (\g), s_A^\ast ( p_A(\g)))$. In the second period, type $\g$ truthfully reports his position $\th$ to firm $B$. Firm $B$ charges the consumer $t_B( \g, \th)$, and with probability $q_B (\g, \th)$, allocates product $B$ to the consumer. With probability $q_A ( \g, \th)$, the consumer does not receive product $B$ and instead purchases product $A$ at strike price $p_A(\g)$. With probability $ 1 - q_B (\g, \th) - q_A ( \g, \th)$, the consumer receives neither product.\footnote{We are assuming that the consumer chooses whether to buy from firm $A$ after observing the realization of the lottery at firm $B$. Given this timing, there is no loss in assuming that the consumer is never allocated both products. If the consumer disposed of a product on-path, then firm $B$ could replicate this on-path outcome by specifying the post-disposal quantities; this modification makes deviations weakly less attractive.}

The plan of the proof is as follows. First, we establish necessary conditions that the tuple $(p_A, q_A, q_B, t_B)$ must satisfy if the consumer is indeed playing a best response. Second, we consider the relaxed problem of maximizing firm $B$'s expected revenue subject to these necessary conditions. We show that this problem is solved by the equilibrium profile. 

We begin by defining the consumer's utility under the profile $(p_A, q_A, q_B, t_B)$. For each $\g' \in \G$ and all positions $\th, \th' \in \Th$, let
\[
    u ( \g', \th' | \th) =  -s_A^\ast( p_A(\g')) + \Paren{ v_A(\th) - p_A(\g')} q_A ( \g', \th') + v_B(\th) q_B(\g', \th') - t_B (\g', \th').
\]
In words, $ u ( \g', \th' | \th)$ is the consumer's utility from selecting contract $(p_A(\g'), s_A^\ast(p_A(\g')))$ from firm $A$, reporting type $\g'$ to firm $B$, and then, after the realization $\th$, mimicking $(\g', \th')$ at both firms. 

The tuple $(p_A, q_A, q_B, t_B)$ and the associated function $u$ must satisfy the following conditions:
\begin{align}
\label{eq:IC_expost} &\th \in \argmax_{\th' \in \Th} u (\g, \th' | \th),  \quad (\g, \th) \in \G \times \Th \\
\label{eq:IC_interim} &\g \in \argmax_{\g' \in \G } \E_{\th | \g} \Brac{ u ( \g', \th | \th)},  \quad \g \in \G \\
\label{eq:outside_option} &\E_{\th| \ubar{\g}} [u ( \ubar{\g}, \th | \th)]  \geq \E_{\th | \ubar{\g}} \Brac{ -s_A^\ast( p_A^\ast(\ubar{\g})) + ( v_A (\th) - p_A^\ast (\ubar{\g}))_+}. 
\end{align}
Condition \eqref{eq:IC_expost} requires that, after following the strategy in the first-period, the consumer cannot profit in the second period by making the same ``position misreport'' to both firms. Condition~\eqref{eq:IC_interim} requires that the consumer cannot profit by making the same ``type misreport'' to both firms in the first period and then reporting his position truthfully in the second period. Condition~\eqref{eq:outside_option} requires that type $\ubar{\g}$ cannot profit by (i) not participating in firm $B$'s mechanism, (ii) selecting the equilibrium contract from firm $A$, and (iii) purchasing product $A$ optimally in the second period. To be sure, these conditions are \emph{necessary}, but not \emph{sufficient}. 

Define the interim utility function $U \colon \G \to \R$ by
\[
    U(\g)  = \E_{\th | \g}\Brac{ u ( \g, \th| \th)}. 
\]

\begin{lem}[Envelope formula] \label{res:necessary_implications} Conditions \eqref{eq:IC_expost}--\eqref{eq:IC_interim} imply that 
    \begin{equation} \label{eq:envelope}
            U(\g) - U(\ubar{\g}) =  \int_{\ubar{\g}}^{\g} \E_{\th | \g'} [q_B ( \g', \th) - q_A( \g', \th)]  \de \g', \qquad \g \in \G. 
    \end{equation}
\end{lem}

The total surplus is split between the three parties---the consumer, firm~$A$, and firm~$B$. Therefore, by the envelope formula \eqref{eq:envelope}, we can express firm $B$'s expected revenue as
\begin{equation} \label{eq:revenue_envelope}
\begin{aligned}
    \E_{\g, \th} [ t_B( \g, \th)] &= \E_{\g,\th} \Brac{-s_A^\ast ( p_A(\g)) + ( v_A(\th) - p_A (\g)) q_A ( \g, \th)  + v_B(\th) q_B(\g, \th) }  \\
   &\quad  - \E_\g \Brac{\int_{\ubar{\g}}^{\g} \E_{\th | \g'} [q_B ( \g', \th) - q_A( \g', \th)]  \de \g'} - U(\ubar{\g}).
\end{aligned}
\end{equation}
To simplify the expected information rent, change the order of integration (by Fubini's theorem) to get
\begin{equation} \label{eq:Fubini_U}
\begin{aligned}
   &\E_\g \Brac{\int_{\ubar{\g}}^{\g} \E_{\th | \g'} [q_B ( \g', \th) - q_A( \g', \th)]  \de \g'} \\
    &= \int_{\ubar{\g}}^{\bar{\g}} \Paren{\int_{\ubar{\g}}^{\g} \E_{\th | \g'} [q_B ( \g', \th) - q_A( \g', \th)]  \de \g'} g(\g) \de \g \\
    &= \int_{\ubar{\g}}^{\bar{\g}} \E_{\th | \g'} [q_B ( \g', \th) - q_A( \g', \th)] (1 - G(\g')) \de \g'.
\end{aligned}
\end{equation}
Write $\bar{G}(\g) = 1 - G(\g)$. Substitute \eqref{eq:Fubini_U} into \eqref{eq:revenue_envelope} and simplify to obtain 
\begin{equation} \label{eq:revenue_simplified}
    \E_{\g, \th} [ t_B (\g, \th)] = - U(\ubar{\g}) + \int_{\ubar{\g}}^{\bar{\g}} I_B(\g) g (\g) \de \g,
\end{equation}
where 
\begin{equation*}
\begin{aligned}
      I_B(\g) &= -s_A^\ast (p_A(\g)) \\
  &\quad + \E_{\th | \g} \Brac{ \Paren{ v_A( \th) -  p_A(\g) + \frac{\bar{G}(\g)}{g(\g)}} q_A(\g, \th)  + \Paren{ v_B(\th) - \frac{\bar{G}(\g)}{g(\g)}} q_B (\g, \th)}.
\end{aligned}
\end{equation*}
The integrand $I_B(\g)$ depends on the tuple $(s_A^\ast, p_A, q)$. Below, we write $I_B( \g ; s_A, p_A, q)$ to denote the analogous expression, with an arbitrary tuple $(s_A, p_A, q)$ in place of $(s_A^\ast, p_A, q)$.

Therefore, we obtain the following further relaxation of firm $B$'s best-response problem, given firm $A$'s schedule $s_A^\ast$: choose the functions $p_A$ and $q$ and the constant $U(\ubar{\g})$ to maximize \eqref{eq:revenue_simplified} subject to the constraint \eqref{eq:outside_option}. We claim that this program is solved by choosing $(p_A,q) = (p_A^\ast, q^\ast)$ and choosing $U(\ubar{\g})$ so that \eqref{eq:outside_option} holds with equality.

We show that $(p_A^\ast, q^\ast)$ maximizes the integrand $I_B$ pointwise. Fix $\g \in \G$. To analyze the integrand, it is convenient to subtract the constant $s_B^\ast ( p_B^\ast(\g)) +  \bar{G}(\g) /g(\g)$. For any pair $(p_A, q)$, we claim that
\begin{equation*}
\begin{aligned}
 & I_B(\g ; s_A^\ast, p_A, q)  -   s_B^\ast ( p_B^\ast(\g)) - \bar{G}(\g)/ g(\g) \\
 &\leq  I_B (\g ; s_A^\ast, p_A, q)  -    s_B^\ast ( p_B^\ast(\g)) - (\bar{G}(\g)/ g(\g)) \E_{\th |\g} \Brac{ q_A(\g, \th) + q_B( \g, \th) } \\
 &= -s_A^\ast (p_A(\g)) - s_B^\ast (p_B^\ast(\g))  + \E_{\th | \g}  \Brac{ \Paren{ v_A( \th) -  p_A(\g)} q_A(\g, \th)  + \Paren{ v_B(\th) - p_B^\ast(\g)} q_B(\g, \th)}  \\
&\leq - s_A^\ast(p_A^\ast(\g))  - s_B^\ast(p_B^\ast(\g)) + \E_{\th | \g}  \Brac{ \Paren{ v_A( \th) -  p_A^\ast(\g)} q_A^\ast(\g, \th)  + \Paren{ v_B(\th) - p_B^\ast(\g)} q_B^\ast(\g, \th)} \\
 &= I_B (\g ; s_A^\ast, p_A^\ast, q^\ast)  -   s_B^\ast ( p_B^\ast(\g)) - (\bar{G}(\g)/ g(\g) ) \E_{\th |\g} \Brac{ q_A^\ast(\g, \th) + q_B^\ast( \g, \th) }  \\
 &=  I_B (\g ; s_A^\ast, p_A^\ast, q^\ast) -  s_B^\ast ( p_B^\ast(\g)) - \bar{G}(\g)/ g(\g). 
\end{aligned}
\end{equation*}
The first inequality holds because $\E_ {\th | \g} [ q_A(\g, \th) + q_B( \g, \th)] \leq 1$ by the definition of $\QQ$; the second inequality holds because $(p^\ast, q^\ast)$ is a best response to $(s_A^\ast, s_B^\ast)$, as shown in the first part of the proof; and the final equality holds because $\E_ {\th | \g} [ q_A^\ast(\g, \th) + q_B^\ast( \g, \th)] = 1$, by the definition of $q^\ast$. 

Finally, to show that this solution of the relaxed problem induces the equilibrium transfer rule $t_B^\ast$, we check that under the equilibrium strategy, the lower bound on $U(\ubar{\g})$ from \eqref{eq:outside_option} holds with equality. We have
\begin{equation*}
\begin{aligned}
    &-s_A^\ast (p_A^\ast (\ubar{\g})) + \E_{\th | \ubar{\g}} \Brac{ ( v_A(\th) - p_A^\ast (\ubar{\g}))q_A^\ast (\ubar{\g}, \th) + v_B (\th) q_B^\ast (\ubar{\g}, \th) -  t_B^\ast(\ubar{\g}, \th)}\\
    &= -s_A^\ast (p_A^\ast (\ubar{\g}))  - s_B^\ast ( p_B^\ast(\ubar{\g})) +  \E_{\th | \ubar{\g}} \Brac{ ( v_A(\th) - p_A^\ast (\ubar{\g}))q_A^\ast (\ubar{\g}, \th) + (v_B (\th) - p_B^\ast (\ubar{\g})) q_B^\ast (\ubar{\g}, \th)}\\
    &= -s_A^\ast (p_A^\ast (\ubar{\g})) - s_B^\ast ( p_B^\ast (\ubar{\g})) + \E_{\th | \ubar{\g}} \Brac{ \max \{ 0, v_A(\th) - p_A^\ast (\ubar{\g}), v_B( \th) - p_B^\ast (\ubar{\g}) \} } \\
    &= -s_A^\ast (p_A^\ast (\ubar{\g})) + \E_{\th | \ubar{\g}} \Brac{ (v_A(\th) - p_A^\ast(\ubar{\g}))_+},
\end{aligned}
\end{equation*}
where the first equality follows from the definition of $t_B^\ast$, the second follows from the optimality of $q^\ast$ given $p^\ast$, and the third follows from the definition of $s_B^\ast ( p_B^\ast (\ubar{\g}))$, upon noting that $p_B^\ast (\ubar{\g}) = \bar{p}_B$ and $p_A^\ast (\ubar{\g}) = 0$.

\subsection{Proof of Lemma~\ref{res:consumer_response}} \label{sec:proof_consumer_response}

First, we check that for all types $\g \in \G$ and prices $p_B \in [0, \infty]$, we have $u ( \bar{p}_A, p_B | \g) \geq u( \infty, p_B | \g)$; a symmetric result holds with the roles of firm $A$ and $B$ reversed. Note that
\begin{equation*}
\begin{aligned}
    &\E_{\th| \g} \Brac{ \max \{ 0, v_A(\th) - \bar{p}_A, v_B(\th) - p_B \}} - \E_{\th |\g} \Brac{ (v_B(\th) - p_B)_+} \\
    &= \E_{\th | \g} \Brac{ \Paren{ v_A(\th) - \bar{p}_A - (v_B(\th) - p_B)_+ }_+} \\ 
    &\geq 
    \E_{\th | \g} \Brac{ \Paren{ v_A(\th) - \bar{p}_A - v_B(\th)_+}_+}  \\
    &\geq \E_{\th | \bar{\g}} \Brac{ \Paren{ v_A(\th) - \bar{p}_A - v_B(\th)_+}_+} \\
    &= s_A^\ast (\bar{p}_A),
\end{aligned}
\end{equation*}
where the last inequality holds because the conditional distribution of $v_A(\th) - \bar{p}_A - v_B(\th)_+$ is decreasing in $\g$ with respect to first-order stochastic dominance. Therefore, in our arguments below, it suffices to consider price vectors $(p_A, p_B) \in [0, \infty)^2$. 

First, we check that for every type $\g$, we have
\begin{equation} \label{eq:argmax}
    p_A^\ast (\g) \in \argmax_{p_A' \in [0, \infty)} u (p_A', p_B^\ast (\g) | \g).
\end{equation}
Fix $\g \in \G$. By the definition of $s_A^\ast$, we have $u (\bar{p}_A , p_B^\ast(\g) | \g) \geq u (p_A' , p_B^\ast(\g) | \g)$ for all $p_A' \in (\bar{p}_A, \infty)$. Therefore, it suffices to prove  \eqref{eq:argmax} with the maximization over $[0, \bar{p}_A]$ in place of $[0, \infty)$. The map 
\begin{equation} \label{eq:CS}
	(p_A, p_B)  \mapsto \E_{\th | \g}  \Brac{ \max \{0,  v_A (\th) - p_A, v_B( \th) - p_B \}}
\end{equation}
is differentiable; the partial derivatives are given by $-Q_A ( p_A, p_B | \g)$ and $-Q_B (p_B, p_A| \g)$, respectively.  Therefore, for any $p_A \in [0, \bar{p}_A]$, we have
\begin{equation} \label{eq:integral_comparison}
    u ( p_A^\ast(\g), p_B^\ast(\g) | \g) - u( p_A, p_B^\ast(\g) | \g)
    = \int_{p_A}^{p_A^\ast (\g)}  \Brac{ Q_A^\ast (p_A') - Q_A( p_A', p_B^\ast (\g) | \g)} \de p_A'.
\end{equation}
This integral in \eqref{eq:integral_comparison} is nonnegative because, by the definition of $Q_A^\ast$, for almost every price $p_A'$ we have
\[
    p_A' \mathrel{< (>)} p_A^\ast (\g) \implies Q_A^\ast (p_A') \mathrel{\geq (\leq)} Q_A( p_A', p_B^\ast(\g) | \g). 
\]

Second, we check that for every type $\g$, we have
\begin{equation} \label{eq:argmax_B}
    p_B^\ast (\g) \in \argmax_{p_B' \in [0, \infty)} U_B( p_B'
    | \g). 
\end{equation}
Since $s_B^\ast$ is flat over $[\bar{p}_B, \infty)$, it suffices to prove  \eqref{eq:argmax_B} with the maximization over $[0, \bar{p}_B]$ in place of $[0, \infty)$. For each fixed $p_B' = p_B^\ast(\g')$, we may select a weakly increasing function $\g \mapsto \hat{p}_A ( p_B' | \g)$ with $\hat{p}_A( p_B' | \g') = p_A^\ast(\g')$ such that 
\[
    \hat{p}_A ( p_B' | \g) \in \argmax_{p_A' \in [0, \infty]} u (p_A', p_B' | \g), \qquad \g \in \G.
\]
This follows from single-crossing; see \cref{res:BR_partial_monotone}.\ref{it:partial_strict_monotonicity} (\cref{sec:proof_equilibrium_uniqueness}) for a formal statement of a more general result. Then for any fixed $p_B \in [0, \bar{p}_B]$, the envelope theorem gives
\[
    U_B(p_B^\ast(\g) | \g) - U_B (p_B | \g) = \int_{ p_B}^{p_B^\ast(\g)} \Brac{ Q_B^\ast ( p_B') - Q_B( p_B', \hat{p}_A ( p_B' | \g) | \g) } \de p_B'.
\]
We claim that this integral is nonnegative. Fix  $p_B' < p_B^\ast (\g)$.  By assumption, $(1 - G)/g$ is weakly decreasing, so $p_B' = p_B^\ast ( \g')$ for some $\g' > \g$, and hence 
\[
  Q_B^\ast ( p_B') - Q_B( p_B', \hat{p}_A ( p_B'|\g) | \g) =  Q_B ( p_B', p_A^\ast (\g') | \g') - Q_B( p_B', \hat{p}_A ( p_B' | \g) | \g) \geq 0, 
\]
where the inequality follows because $\hat{p}_A (p_B' | \g) \leq  \hat{p}_A (p_B' | \g') = p_A^\ast (\g')$. Symmetrically, fix $p_B' > p_B^\ast(\g)$. By assumption, $(1 - G)/g$ is weakly decreasing, so $p_B' = p_B^\ast(\g')$ for some $\g' < \g$, and hence
\[
    Q_B^\ast ( p_B') - Q_B (p_B', \hat{p}_A(p_B'|\g) |\g)
    = Q_B ( p_B', p_A^\ast (\g') | \g') - Q_B( p_B', \hat{p}_A ( p_B' |\g) | \g) \leq 0,
\]
where the inequality follows because $\hat{p}_A (p_B' | \g) \ge \hat{p}_A (p_B' | \g') =  p_A^\ast (\g')$.

\subsection{Proof of Lemma~\ref{res:necessary_implications}} 

For all $\g' \in \G$ and $\th \in \Th$,  condition \eqref{eq:IC_expost} gives
\[
    u( \g', \th | \th) = \max_{\th' \in \Th} u (\g', \th' | \th).
\]
Let $u_3$ denotes the partial derivative of $u$ with respect to the third argument. Since $v_A'(\th) = -1$ and $v_B'(\th) = 1$, we have $u_3 ( \g', \th | \th)= q_B ( \g', \th) - q_A( \g', \th)$. Note that the absolute value of this derivative is at most $1$. Fix $\th_0 \in \Th$. By the envelope theorem \citep{milgrom2002envelope}, for each $\g' \in \G$ and $\th \in \Th$, we have
\begin{equation} \label{eq:envelope_shock} u( \g', \th | \th) =  u (\g', \th_0 | \th_0) +  \int_{\th_0}^{\th} (q_B (\g', \th') - q_A (\g', \th')) \de \th'.
\end{equation}
To simplify notation, let $u(\g' ,\th) = u( \g', \th| \th)$ and $u_0( \g') = u ( \g', \th_0)$. For all $\g, \g' \in \G$, let
\[
    U(\g' | \g) = \E_{\th | \g} \Brac{ u(\g',  \th) }.
\]
From \eqref{eq:envelope_shock}, we have
\begin{equation} \label{eq:U_def_duopoly}
\begin{aligned}
    U(\g' | \g) 
    &= u_0(\g') + \int_{-\infty}^{\infty} \Paren{ \int_{\th_0}^{\th} (q_B(\g', \th')  - q_A( \g', \th')) \de \th'}  f (\th  -\g) \de \th .
\end{aligned}
\end{equation}
To evaluate the integral we separate the outer integral according to whether $\th \geq \th_0$. We have 
\begin{equation} \label{eq:AB_interchange}
\begin{aligned}
    &\int_{-\infty}^{\th_0} \Paren{ \int_{\th_0}^{\th} (q_B(\g', \th')  - q_A( \g', \th')) \de \th'}  f (\th  -\g) \de \th\\
    &\quad+  \int_{\th_0}^{\infty} \Paren{ \int_{\th_0}^{\th} (q_B(\g', \th')  - q_A( \g', \th')) \de \th'}  f (\th  -\g) \de \th \\
    &= - \int_{-\infty}^{\th_0} (q_B(\g', \th')  - q_A( \g', \th'))  F (\th'  -\g) \de \th' \\
    &\quad +  \int_{\th_0}^{\infty} (q_B(\g', \th')  - q_A( \g', \th')) (1 - F (\th' - \g)) \de \th',
\end{aligned}
\end{equation}
where we have changed the order of integration by Fubini's theorem (since the density $f$ is integrable and $q$ is bounded). Substitute \eqref{eq:AB_interchange} into \eqref{eq:U_def_duopoly} and simplify to get
\begin{equation} \label{eq:U_AB_simplified}
   U(\g' | \g) = u_0( \g')
    +
    \int_{-\infty}^{\infty} (q_B (\g', \th') - q_A( \g', \th')) ([ \th' > \th_0] - F( \th'- \g)) \de \th'.
\end{equation}
Differentiating under the integral in \eqref{eq:U_AB_simplified},\footnote{This is justified by the monotone tails condition; see \cref{ft:under_integral}.} we obtain 
\begin{equation*}
\begin{aligned}
    U_2 (\g' | \g) 
    &= \int_{-\infty}^{\infty} ( q_B (\g',\th') - q_A (\g', \th')) f( \th' - \g) \de \th' \\
    &= \E_{\th | \g}  [q_B ( \g', \th) - q_A( \g', \th)].
\end{aligned}
\end{equation*}
Note that this derivative has absolute value at most $1$. For all $\g \in \G$, by \eqref{eq:IC_interim}, we have
\[
    U ( \g | \g) = \max_{\g' \in \G} U (\g' | \g). 
\]
By the envelope theorem, 
\begin{equation*}
\begin{aligned}
    U( \g | \g) = U( \ubar{\g} | \ubar{\g}) + \int_{\ubar{\g}}^{\g} \E_{\th | \g'}  [q_B ( \g', \th) - q_A( \g', \th)] \de \g'.
\end{aligned}
\end{equation*}

\subsection{Proof of Theorem~\ref{res:equilibrium}: Uniqueness} \label{sec:proof_equilibrium_uniqueness}

In this section, we prove essential uniqueness, under the assumption that $v_0 \geq (7/2) \max_{\g' \in \G} (1 / g(\g'))$. We begin with some preliminaries about the structure of the consumer's best response. For any subscription-schedule pair $(s_A, s_B)$, type $\g \in \G$, and prices $p_A, p_B \in [0, \infty]$, let
\begin{equation*} 
\begin{aligned}
    u(p_A, p_B, \g  ; s_A, s_B) = - s_A( p_A) - s_B (p_B) + \E_{\th| \g} \Brac{ \max\{ 0, v_A(\th) - p_A , v_B(\th) - p_B\}}.
\end{aligned}
\end{equation*}
Let
\begin{equation*}
    P(\g ; s_A, s_B) = \argmax_{(p_A', p_B')  \in [0,\infty]^2} u( p_A', p_B', \g ; s_A, s_B).
\end{equation*}
Thus, $P( \cdot ; s_A, s_B)$ is a correspondence from $\G$ to $[0, \infty]^2$ that assigns to each type $\g$ the set of all price vectors that are optimal for type $\g$, given the subscription-schedule pair $(s_A, s_B)$.  We establish order properties of this correspondence. Define the partial order $\succeq$ on $[0, \infty]^2$ to be the product order with the reversed order in the second component. Formally,  $(p_A, p_B) \succeq (p_A', p_B')$ if and only if $p_A \geq p_A'$ and $p_B \leq p_B'$. Hereafter, we assume that the space $[0, \infty]^2$ of price vectors is endowed with the order $\succeq$. With this order, $[0, \infty]^2$ is a complete lattice. Below, we write $p \succ p'$ if $p \succeq p'$ and $p \neq p'$.

\begin{lem}[Structure of best response] \label{res:BR_structure} Let $(s_A, s_B)$ be a profile of subscription schedules.  
\begin{enumerate}
    \item \label{it:complete} For each type $\g \in \G$, the set $P(\g ; s_A, s_B)$ is a nonempty, complete sublattice. 
    \item \label{it:strict_monotone} For all types $\g, \g' \in \G$, if $\g > \g'$, then $\inf P(\g ; s_A, s_B) \succeq \sup P (\g' ; s_A, s_B)$. 
\end{enumerate}
\end{lem}

Part \ref{it:complete} ensures that the infimum  and supremum in part~\ref{it:strict_monotone} exist. Part~\ref{it:strict_monotone} implies that every selection from the correspondence $P ( \cdot ; s_A, s_B)$ is weakly increasing (with respect to $\succeq$). Intuitively, more rightward types choose higher strike prices at firm $A$ and lower strike prices at firm $B$.

Building on \cref{res:BR_structure}, we next show that each firm's expected revenue from a given subscription schedule pair is independent of which best response the consumer plays. 

\begin{lem}[Revenue]  \label{res:revenue} Let $(s_A, s_B)$ be a profile of subscription schedules. For any best responses $(p,q)$ and $(p', q')$ to $(s_A, s_B)$, we have
\[
    \E_{\g, \th} [ s_i (p_i (\g)) + p_i (\g) q_i (\g, \th) ] = \E_{\g, \th} [ s_i (p_i' (\g)) + p_i' (\g) q_i'(\g, \th) ], \qquad i = A,B.
\]
\end{lem}

\cref{res:revenue} simplifies our equilibrium analysis below. To show that a strategy profile is \emph{not} an equilibrium, it suffices to identify a deviation by a firm and \emph{some} associated best response that makes this deviation profitable to the deviator. 

Next, we consider the consumer's \emph{partial} best response, given the price selected at one firm. Fix firm $i$. For any subscription-schedule $s_i$, type $\g \in \G$, and prices $p_i, p_{-i} \in [0, \infty]$, let
\begin{equation} \label{eq:ui_def}
    u_i ( p_i, p_{-i}, \g ; s_i) = - s_i( p_i) + \E_{\th| \g} \Brac{ \max\{ 0, v_i(\th) - p_i , v_{-i}(\th) - p_{-i} \}}. 
\end{equation} 
Let 
\[
    P_{i} (p_{-i}, \g ; s_i) = \argmax_{p_{i}' \in [0,\infty]} u_i( p_i', p_{-i}, \g ; s_i).
\]
Given the subscription-schedule $s_i$, the correspondence $P_i (\cdot ; s_i)$ from $[0, \infty] \times \G$ to $[0, \infty]$ assigns to each pair $(p_{-i}, \g)$ the set of prices at firm $i$ that are optimal for type $\g$ conditional on selecting price $p_{-i}$ at firm $-i$. The next result establishes monotonicity properties of $P_A (\cdot ;s_A)$; a symmetric result holds for $P_B (\cdot ; s_B)$. 

\begin{lem}[Monotonicity of partial best response] \label{res:BR_partial_monotone}
Let $s_A$ be a subscription schedule. 
\begin{enumerate}
    \item \label{it:nonempty_closed} For each type $\g \in \G$ and price $p_B \in [0, \infty]$, the set $P_A(p_B, \g ; s_A)$ is nonempty and closed.
    \item \label{it:partial_weak_monotonicity} For all types $\g \in \G$, the map $p_B \mapsto P_A( p_B, \g;  s_A)$ is weakly decreasing (in the strong set order). 
    \item \label{it:partial_strict_monotonicity} For all types $\g, \g' \in \G$ and prices $p_B, p_B' \in [0, \infty]$, if $\g > \g'$ and $p_B \leq p_B'$, then $\inf P_A( p_B, \g;  s_A) \geq \sup P_A (p_B', \g' ; s_A)$. 
\end{enumerate}
\end{lem}

Given an arbitrary subscription schedule $s_A$ offered by firm $A$, we next show that firm $B$ can, through its choice of subscription schedule, induce the consumer to follow any monotone price-selection rule $p_B \colon \G \to [0,\infty]$ at firm $B$. An analogous result holds for firm $A$.

\begin{lem}[Inducing prices] \label{res:subscription_construction} Fix a subscription schedule $s_A$.  Let $p_B \colon \G \to [0,\infty]$ be weakly decreasing. For every selection $p_A \colon \G \to [0, \infty]$ from the correspondence $\g \mapsto P_A ( p_B (\g), \g ; s_A)$, there exists a subscription schedule $s_B$ such that $(p_A(\g), p_B(\g))$ is in $P ( \g ; s_A, s_B)$ for each type $\g$, and
\begin{equation}\label{eq:lower_util}
\begin{aligned}
     &-  s_A( p_A (\ubar{\g})) - s_B (p_B( \ubar{\g})) + \E_{\th | \ubar{\g}} \Brac{ \max\{ v_A(\th) - p_A(\ubar{\g}), v_B(\th) - p_B(\ubar{\g}), 0 \}}  \\
     &= \max_{p_A' \in [0, \infty]} \Set{ -s_A( p_A') +  \E_{\th| \ubar{\g}} [(v_A(\th) - p_A')_+]  }.
\end{aligned}
\end{equation} 
\end{lem}

Condition \eqref{eq:lower_util} ensures that $s_B (p_B(\ubar{\g}))$ is chosen as large as possible in the sense that type $\ubar{\g}$ is indifferent between (i) selecting the contract $(p_A(\ubar{\g}), s_A (p_A(\ubar{\g})))$ from firm $A$ and the contract $(p_B( \ubar{\g}), s_B ( p_B( \ubar{\g})))$ from $B$, and (ii) selecting optimally from firm $A$ only. 

With these preliminaries in hand, we turn to the main proof. Let $(\hat{s}_A, \hat{s}_B, \hat{p}, \hat{q})$ 
be an equilibrium. By \cref{res:BR_structure}, the function $\hat{p}$ is weakly increasing (with respect to $\succeq$). We freely use this fact below. To prove that $(\hat{s}_A, \hat{s}_B, \hat{p}, \hat{q})$ is equivalent to $(s_A^\ast, s_B^\ast, p^\ast, q^\ast)$, we prove that for each firm $i$ and for each type $\g \in (\ubar{\g}, \bar{\g})$, we have $\hat{p}_i(\g) = p_i^\ast (\g)$ and $\hat{s}_i ( \hat{p}_i( \g)) =   s_i^\ast  (p_i^\ast (\g))$. First, recall our notation from the envelope formula \eqref{eq:revenue_simplified}. For each type $\g$, write $\bar{G}(\g)= 1 - G(\g)$. Given $p_A \colon \G \to [0, \infty]$ and $q  = (q_A, q_B) \colon \G \times \Th \to \QQ$, let 
\begin{equation*}
\begin{aligned}
         &I_B(\g ;s_A, p_A, q)\\
         &= -s_A (p_A(\g)) 
          + \E_{\th| \g} \Brac{ \Paren{ v_A( \th) -  p_A(\g) + \frac{\bar{G}(\g)}{g(\g)}} q_A(\g, \th)  + \Paren{ v_B(\th) - \frac{\bar{G}(\g)}{g(\g)}} q_B(\g, \th)}.
\end{aligned}
\end{equation*}
The next result is the key tool for narrowing down the profile $(\hat{s}_A, \hat{s}_B, \hat{p}, \hat{q})$.  A symmetric result holds for firm $A$, but we focus, without loss, on deviations by firm $B$. 

\begin{lem}[Equilibrium constraint] \label{res:profitable_deviation} Let $\tilde{p}_B \colon \G \to [0, \infty]$ be weakly decreasing. For every selection $\tilde{p}_A \colon \G \to [0, \infty]$ from the correspondence $\g \mapsto P_A ( \tilde{p}_B (\g), \g ; \hat{s}_A)$  and every product-purchase strategy $\tilde{q}$ that is optimal given $\tilde{p}$, we have
\begin{equation} \label{eq:I_bound}
    \E_\g [I_B (\g ; \hat{s}_A, \tilde{p}_A, \tilde{q})] \leq \E_\g [I_B(\g; \hat{s}_A, \hat{p}_A, \hat{q})].
\end{equation}
\end{lem}

\cref{res:profitable_deviation} is proven as follows. By \cref{res:subscription_construction}, we can construct a subscription schedule for firm $B$ that (i) induces the consumer to follow the contract-selection strategy $\tilde{p}$ and (ii) leaves type $\ubar{\g}$ with the same utility that he can obtain from firm $A$'s schedule $\hat{s}_A$ alone. Using the envelope theorem, we express firm $B$'s expected revenue from this deviation in terms of the function $I_B$. If inequality \eqref{eq:I_bound} is violated, then this deviation is profitable, contrary to the assumption that $(\hat{s}_A, \hat{s}_B, \hat{p}, \hat{q})$ is an equilibrium. 

Applying \cref{res:profitable_deviation}, we establish a sequence of properties of the profile $(\hat{s}_A, \hat{s}_B, \hat{p}, \hat{q})$. 

\begin{lem}[Full coverage] \label{res:full_coverage} For every type $\g$ in $(\ubar{\g}, \bar{\g})$ at which $\hat{p}$ is continuous, we have $\hat{p}_A(\g) + \hat{p}_B(\g) \leq 2v_0$.
\end{lem}

\begin{lem}[Price infimum] \label{res:price_lower_limit} For each firm $i$, we have $\inf_{ \g' \in (\ubar{\g}, \bar{\g})} \hat{p}_i (\g') = 0$. 
\end{lem}

Recall from the statement of \cref{res:equilibrium} that $\bar{p}_A = p_A^\ast ( \bar{\g})$ and $\bar{p}_B = p_B^\ast (\ubar{\g})$.

\begin{lem}[Equality under price bound] \label{res:prices} Fix $i \in \{A, B\}$. If\/ $\sup_{\g' \in (\ubar{\g}, \bar{\g})} \hat{p}_i(\g') \leq (5/2) \bar{p}_i$, then $\hat{p}_{-i}(\g) = p_{-i}^\ast( \g)$ for all $\g \in (\ubar{\g}, \bar{\g})$.
\end{lem}

By symmetry, it suffices to prove \cref{res:prices} with $i = A$. Here is a sketch of the proof. We apply \cref{res:profitable_deviation} with $\tilde{p}_{B} = p_B^\ast$. We show that if $\hat{p}_B$ disagrees with $p_B^\ast$ on a positive-measure set, then we obtain a contradiction with  \eqref{eq:I_bound}.\footnote{We then use the monotonicity of $\hat{p}_B$ to strengthen almost-everywhere equality to everywhere equality.} The key step is showing that if firm $B$ induces type $\g$ to select price $p_B^\ast(\g)$ at firm $B$, then full coverage is preserved. We show this by combining the assumed bound $\sup_{\g' \in (\ubar{\g}, \bar{\g})} \hat{p}_A(\g') \leq (5/2) \bar{p}_A$ with the fact that $\inf_{\g' \in (\ubar{\g}, \bar{\g})} \hat{p}_B (\g') = 0$ (from \cref{res:price_lower_limit}) and the assumption that $v_0 \geq (7/2) \max_{\g' \in \G} (1/ g(\g'))$.

\begin{lem}[Price upper bound] \label{res:price_bound}
For each firm $i$, we have $\sup_{\g' \in (\ubar{\g}, \bar{\g})} \hat{p}_i(\g') \leq (5/2) \bar{p}_i$.
\end{lem}

To prove \cref{res:price_bound}, we show that if the claimed upper bound is violated for some firm $i$, then the functions $\hat{p}_A$ and $\hat{p}_B$ must satisfy a sequence of restrictive properties that ultimately yield a contradiction. In particular, there exists a critical type $\g_0$ in $(\ubar{\g}, \bar{\g})$ at which (a) $\hat{p}_A$ jumps up from below $p_A^\ast (\g_0)$ to above $(5/2) \bar{p}_A$, and (b) $\hat{p}_B$ jumps down from above $(5/2) \bar{p}_B$ to below $p_B^\ast (\g_0)$. Using the fact that the density $f$ is symmetric and single-peaked around $0$ (by \cref{as:regularity}), we then show that at least one firm can profit by inducing types near $\g_0$ to select different prices.

Finally, we combine \cref{res:prices} and \cref{res:price_bound} to conclude that $\hat{p}(\g) = p^\ast(\g)$ for all types $\g$ in $(\ubar{\g}, \bar{\g})$. To complete the proof, we establish the following. 

\begin{lem}[Equality of subscription schedules] \label{res:schedules_equality} For each $i \in \{A, B\}$ and each $\g \in (\ubar{\g}, \bar{\g})$, we have $\hat{s}_i (\hat{p}_i(\g)) = s_i^\ast (p_i^\ast(\g))$. 
\end{lem}

We prove \cref{res:schedules_equality} as follows. Using the fact that $(\hat{p}, \hat{q})$ is a best response to $(\hat{s}_A, \hat{s}_B)$, we pin down each subscription schedule up to a constant, and we obtain an upper bound on the constant. Then, using firm optimality, we get a corresponding lower bound on each constant.

\subsection{Proof of Lemma~\ref{res:BR_structure}} \label{sec:BR_structure_proof}

Recall that the space $[0, \infty]^2$ of price vectors is endowed with the order $\succeq$. To simplify notation, write $u( p_A, p_B, \g) = u( p_A, p_B, \g ; s_A, s_B)$. We check that $u$ is supermodular in $p = (p_A, p_B)$ and has strictly increasing differences in $(p,\g)$.

First, we check that $u$ is supermodular in $(p_A, p_B)$ for each fixed type $\g$. The space of supermodular functions is closed under pointwise addition and averaging. Therefore, since $u$ is additively separable in $s_A(p_A)$ and $s_B(p_B)$, it suffices to check that for each fixed $\th \in \Th$, the map
\[
    (p_A, p_B) \mapsto \max\{ 0, v_A(\th) - p_A , v_B(\th) - p_B \}
\]
is supermodular. Equivalently, we prove that this function satisfies increasing differences (with respect to the usual order on $p_A$ and the reverse order on $p_B$). Fix $p_A, p_A' \in [0, \infty]$ with $p_A \leq p_A'$. For any $p_B \in [0, \infty]$, we have
\begin{equation*}
\begin{aligned}
&
\max\{ 0, v_A(\th) - p_A' , v_B(\th) - p_B \}
 - 
 \max\{ 0, v_A(\th) - p_A , v_B(\th) - p_B \}
 \\
 &= - \min \Set{ p_A' - p_A, \Paren{ v_A(\th) - p_A - (v_B(\th) - p_B)_+}_+ }.
\end{aligned}
\end{equation*}
This expression is weakly increasing in $v_B( \th) - p_B$, and hence weakly increasing, with respect to the reverse order, in $p_B$. 

Second, we check that $u$ has strictly increasing differences in $(p,\g)$. Fix $p, p' \in [0, \infty]^2$ and $\g ,\g' \in \G$ with $p \prec p'$ and $\g < \g'$. We claim that 
\[
    u( p,\g') + u(p',\g) < u ( p,\g) + u(p',\g').
\]
For each fixed $\th$, we have 
\begin{equation*}
\begin{aligned}
    &\max\{ 0, v_A(\th) - p_A , v_B(\th) - p_B \} - \max\{ 0, v_A(\th) - p_A' , v_B(\th) - p_B' \} \\
    &= \max\{ 0, v_A(\th) - p_A , v_B(\th) - p_B \} - \max\{ 0, v_A(\th) - p_A' , v_B(\th) - p_B \} \\
    &\quad + \max\{ 0, v_A(\th) - p_A' , v_B(\th) - p_B \} - \max\{ 0, v_A(\th) - p_A' , v_B(\th) - p_B' \} \\
    &= -\int_{p_A'}^{p_A} \Brac{ v_A(\th) - \tilde{p}_A > ( v_B( \th) - p_B)_+ } \de \tilde{p}_A 
    +  \int_{p_B}^{p_B'} \Brac{ v_B(\th) - \tilde{p}_B > ( v_A( \th) - p_A')_+ } \de \tilde{p}_B. 
\end{aligned}
\end{equation*}
Taking expectations and simplifying, we conclude that 
\begin{equation*}
\begin{aligned}
    &( u (p,\g) + u(p', \g')) - ( u(p,\g') + u(p', \g)) \\
    &= (u(p' , \g') - u (p, \g')) - (u(p', \g) - u (p, \g)) \\
    &= \int_{p_A}^{p_A'} [ Q_A ( \tilde{p}_A, p_B | \g) - Q_A ( \tilde{p}_A, p_B | \g')]
\de \tilde{p}_A +  \int_{p_B'}^{p_B} [Q_B( \tilde{p}_B, p_A' | \g') - Q_B( \tilde{p}_B, p_A' | \g)]
\de \tilde{p}_B.
\end{aligned}
\end{equation*}
This expression is strictly positive because $Q_A$ is strictly decreasing in the consumer's type and $Q_B$ is strictly increasing in the consumer's type. 

With supermodularity in $(p_A,p_B)$ and increasing differences in $(p,\g)$ established for $u$, we now prove each part of \cref{res:BR_structure}. To simplify notation, write $P(\g) = P(\g ; s_A, s_B)$. 

\begin{enumerate}
\item Fix $\g \in \G$. Since $u ( \cdot, \g)$ is supermodular, the set $P(\g)$ is a sublattice of $[0, \infty]^2$. Since $u(\cdot, \g)$ is upper semicontinuous and $[0, \infty]^2$ is complete, we conclude that $P(\g)$ is complete. 
\item This part essentially follows from the monotone selection theorem \citep[Theorem 4', p.~163]{MilgromShannon1994}, but we include the proof for completeness. Fix types $\g,\g'$ with $\g > \g'$. Write $\ubar{p}(\g) = \inf P( \g )$ and $\bar{p}(\g') = \sup P(\g')$. Since $\bar{p}(\g')$ is in $P(\g')$, we have $u(\bar{p}(\g'), \g' )\geq u ( \ubar{p}(\g) \wedge \bar{p}(\g'), \g')$. Since $u(\cdot, \g')$ is supermodular, it follows that $u(\ubar{p}(\g) \vee \bar{p}(\g'), \g') \geq u(\ubar{p}(\g), \g')$. Suppose, for a contradiction, that $\ubar{p}(\g) \not\succeq \bar{p}(\g')$. Then $\ubar{p}(\g) \vee \bar{p}(\g') \neq \ubar{p}(\g)$, so $\ubar{p}(\g) \vee \bar{p}(\g') \succ \ubar{p}(\g)$. Since $u(\ubar{p}(\g) \vee \bar{p}(\g'), \g') \geq u(\ubar{p}(\g), \g')$, strictly increasing differences imply that $u( \ubar{p}(\g) \vee \bar{p}(\g'), \g) > u(\ubar{p}(\g), \g)$, contrary to the fact that $\ubar{p}(\g)$ is in $P(\g)$. 
\end{enumerate}

\subsection{Proof of Lemma~\ref{res:revenue}}
To simplify notation, write $P(\g) = P(\g ; s_A, s_B)$. Let $P_i(\g)$ denote the projection of the set $P(\g)$ onto its $i$-th component. Let $\bar{p}(\g) = \sup P(\g)$ and $\ubar{p}(\g) = \inf P(\g)$; these vectors exist by \cref{res:BR_structure}.\ref{it:complete}. The $i$-th components of $\bar{p}(\g)$ and $\ubar{p}(\g)$ are denoted by $\bar{p}_i(\g)$ and $\ubar{p}_i(\g)$, respectively. 

By symmetry, it suffices to establish the equality for $i = A$. Let $\G_A$ be the set of all types $\g$ such that $P_A(\g)$ is nonsingleton. It suffices to prove that $\G_A$ is countable. For each $\g \in \G_A$, let $I(\g) = (\ubar{p}_A(\g), \bar{p}_A (\g))$. By the definition of $\G_A$, the interval $I(\g)$ is nonempty for each $\g$ in $\G_A$. By \cref{res:BR_structure}.\ref{it:strict_monotone}, the intervals $I(\g)$ and $I(\g')$ are disjoint for all distinct types $\g, \g' \in \G_A$. Thus, there is an injection from $\G_A$ to the rationals. Hence, $\G_A$ is countable. 

\subsection{Proof of Lemma~\ref{res:BR_partial_monotone}}

Part~\ref{it:nonempty_closed} is immediate since the utility function $u_A ( \cdot, p_B, \g ; s_A)$ is upper semicontinuous over the compact set $[0, \infty]$.  Part~\ref{it:partial_weak_monotonicity} holds because $u( p_A, p_B, \g; s_A, s_B)$ is supermodular in $(p_A, p_B)$, with respect to the order $\succeq$ (the product order with the reversed order in $p_B$), as shown in the proof of \cref{res:BR_structure} (\cref{sec:BR_structure_proof}). Part~\ref{it:partial_strict_monotonicity} follows from the proof of \cref{res:BR_structure} (\cref{sec:BR_structure_proof}). 


\subsection{Proof of Lemma~\ref{res:subscription_construction}} \label{sec:proof_subscription_construction}

If $p_B (\bar{\g}) = \infty$, then we may take $s_B(p_B) = \infty$ for all $p_B \in [0, \infty)$. Therefore, we may assume that $p_B( \bar{\g}) < \infty$. 

Let $p_A \colon \G \to [0, \infty]$ be a selection from the correspondence $\g \mapsto P_A ( p_B (\g), \g ; s_A)$. Since $p_B \colon \G \to [0, \infty]$ is weakly decreasing, \cref{res:BR_partial_monotone}.\ref{it:partial_strict_monotonicity} implies that $p_A$ is weakly increasing. For each $\g \in \G$ and $p_B' \in   [  p_B(\bar{\g}),p_B(\ubar{\g})]$, choose $\hat{p}_A( p_B', \g)$ from $P_A (p_B', \g; s_A)$ so that, for all types $\g$, we have $\hat{p}_A ( p_B (\g), \g) = p_A(\g)$. By \cref{res:BR_partial_monotone}.\ref{it:partial_strict_monotonicity}, $\hat{p}_A( p_B' ,\cdot)$ is weakly increasing. 

Let $p_B^{-1} \colon  [  p_B(\bar{\g}),p_B(\ubar{\g})] \to [\ubar{\g}, \bar{\g}]$ be a generalized inverse of $p_B$.  Define $\hat{Q}_B \colon [0, \infty] \to [0,1]$ by 
\[
\hat{Q}_B( p_B')
=
\begin{cases}
    1 &\text{if}~ p_B' < p_B (\bar{\g}), \\
 Q_B \bigl( p_B', \hat{p}_A(p_B', p_B^{-1}(p_B')) | p_B^{-1}(p_B') \bigr)
&\text{if}~p_B( \bar{\g}) \leq p_B' \leq p_B( \ubar{\g}), \\
0 &\text{if}~p_B' > p_B( \ubar{\g}).
\end{cases}
\]
Define the subscription schedule $s_B \colon [0, \infty) \to [0, \infty]$ by
\begin{equation} \label{eq:s_B_def}
\begin{aligned}
  s_B( p_B) &= \int_{p_B}^{\infty} \hat{Q}_B (p_B') \de p_B' + 
       \E_{\th | \ubar{\g}} \Brac{ \max\{ v_A(\th) - p_A(\ubar{\g}), v_B(\th) - p_B(\ubar{\g}), 0 \}} \\
       &\quad - s_A( p_A (\ubar{\g})) - \max_{p_A' \in [0, \infty]} \Set{ -s_A( p_A') +  \E_{\th| \ubar{\g}} [(v_A(\th) - p_A')_+]  }.
\end{aligned}
\end{equation}
By construction, $s_B$ is finite-valued.\footnote{We may assume $p_B (\ubar{\g}) = \infty$; otherwise this is immediate. For all prices $p_B' \in [p_B( \bar{\g}), \infty)$, we have $\hat{Q}_B ( p_B') \leq Q_B^M ( p_B' | \bar{\g})$, so
\[
   \int_{p_B(\bar{\g})}^{\infty} \hat{Q}_B ( p_B') \de p_B'  \leq \int_{p_B(\bar{\g})}^{\infty} Q_B^M ( p_B' | \bar{\g}) = \int_{p_B(\bar{\g})}^{\infty} \P_{\th | \bar{\g}} ( v_B( \th) \geq p_B' ) \de p_B'  \leq \E_{\th | \bar{\g}} [v_B(\th)_+] < \infty.
\]} Since $p_A( \ubar{\g})$ is in $P_A( p_B (\ubar{\g}), \ubar{\g}; s_A)$, it follows that $s_B$ is nonnegative-valued.\footnote{Here are the details. Choose $p_A^0 \in \argmax_{p_A' \in [0, \infty]} \Set{ - s_A( p_A') + \E_{\th | \ubar{\g}}[ (v_A(\th) - p_A')_+]}$. Since $p_A( \ubar{\g})$ is in $P_A( p_B (\ubar{\g}), \ubar{\g}; s_A)$, we have
\begin{equation*}
\begin{aligned}
    &- s_A( p_A (\ubar{\g}))  + \E_{\th | \ubar{\g}} \Brac{ \max\{ v_A(\th) - p_A(\ubar{\g}), v_B(\th) - p_B(\ubar{\g}), 0 \}} \\
    &\geq - s_A( p_A^0)  + \E_{\th | \ubar{\g}} \Brac{ \max\{ v_A(\th) - p_A^0, v_B(\th) - p_B(\ubar{\g}), 0 \}} \\
    &\geq  - s_A( p_A^0)  + \E_{\th | \ubar{\g}} [ (v_A(\th) - p_A^0)_+] \\
    &= \max_{p_A' \in [0, \infty]} \Set{ - s_A( p_A') + \E_{\th | \ubar{\g}}[ (v_A(\th) - p_A')_+]}.
\end{aligned}
\end{equation*}}

First, we check that \eqref{eq:lower_util} is satisfied. If $p_B( \ubar{\g}) < \infty$, then 
\eqref{eq:lower_util} follows from \eqref{eq:s_B_def} since $\hat{Q}_B (p_B') = 0$ for $p_B' > p_B( \ubar{\g})$. If $p_B( \ubar{\g}) = \infty$, then $p_A( \ubar{\g})$ is in $P_A( \infty, \ubar{\g}; s_A)$, so \eqref{eq:lower_util} is immediate.

It remains to prove that $(p_A(\g), p_B(\g))$ is in $P(\g; s_A, s_B)$ for every type $\g$. To simplify notation, write $u(p_A, p_B, \g) = u( p_A, p_B, \g ; s_A, s_B)$. Similarly, write $P_A ( p_B, \g) = P_A (p_B,\g ;s_A)$ and $P(\g) = P( \g ; s_A, s_B)$. Our argument is similar to the proof of \cref{res:consumer_response}. For each type $\g \in \G$ and each price $p_B \in [0, \infty]$, let 
\[
    U_B(p_B, \g) = \max_{p_A' \in [0, \infty]} u (p_A', p_B, \g).
\]
By assumption, $p_A(\g)$ is in $P_A (p_B (\g), \g)$ for each type $\g$. Therefore, it suffices to show that for every type $\g \in \G$, we have
\[
    U_B( p_B(\g), \g) \geq U_B( p_B', \g) \quad \text{for all}~ p_B' \in [0, \infty]. 
\]
In fact, we claim that it suffices to prove this inequality for all $p_B' \in [0, p_B (\ubar{\g})] \cap [0, \infty)$.  We verify the claim by cases. \begin{itemize}
    \item Suppose $p_B( \ubar{\g}) < \infty$.  Then the definition of $s_B$ ensures that the strike price $p_B( \ubar{\g})$ strictly dominates every strike price in $(p_B( \ubar{\g}), \infty)$. And \eqref{eq:lower_util} implies that $U_B( p_B (\ubar{\g}), \ubar{\g}) \geq U_B( \infty, \ubar{\g})$, so by \cref{res:BR_structure}, we have $U_B( p_B (\ubar{\g}), \g) \geq U_B( \infty, \g)$ for all types $\g$. 
    \item Suppose $p_B( \ubar{\g}) = \infty$. In this case, it follows from \eqref{eq:s_B_def} that $\lim_{p_B \to \infty} s_B(p_B) = 0$. For every price $p_B \in [0,\infty)$, we have $U_B(p_B, \g) + s_B ( p_B) \geq  U_B(\infty, \g)$. Passing to the limit yields $\lim_{p_B \to \infty} U_B(p_B, \g) \geq U_B(\infty, \g)$. 
\end{itemize}

With the claim established, we now complete the proof. Fix $\g \in \G$. For any fixed $\tilde{p}_B \in [0, p_B( \ubar{\g})] \cap [0, \infty)$, the envelope theorem gives
\[
    U_B(p_B(\g) , \g) - U_B(\tilde{p}_B , \g)
    = \int_{\tilde{p}_B}^{p_B(\g)} 
        \Brac{ \hat{Q}_B(p_B') - Q_B(p_B', \hat{p}_A(p_B' , \g) | \g) } 
    \de p_B'.
\]
We claim that this integral is nonnegative. There are two cases. Recall that $p_B^{-1} \colon  [  p_B(\bar{\g}),p_B(\ubar{\g})] \to [\ubar{\g}, \bar{\g}]$ is a generalized inverse of $p_B$. 
\begin{itemize}
    \item Suppose $p_B(\g) > \tilde{p}_B$. 
    Fix $p_B' \in (\tilde{p}_B,\, p_B(\g))$. 
    Since the function $p_B(\cdot)$ is weakly decreasing, we have 
    $p_B' < p_B(\g) \le p_B(\ubar{\g})$. 
    If $p_B' < p_B(\bar{\g})$, then $\hat{Q}_B(p_B') = 1$, 
    so the integrand is clearly nonnegative. 
    Therefore, we may assume that $p_B'$ is in $ [  p_B(\bar{\g}),p_B(\ubar{\g})]$. Let $\g' = p_B^{-1}(p_B')$. Since $p_B' < p_B(\g)$, we conclude that $\g' \ge \g$. 
    Therefore,
    \begin{equation*}
    \begin{aligned}
        &\hat{Q}_B(p_B') - Q_B(p_B', \hat{p}_A(p_B' , \g) | \g) \\
        &=
        Q_B(p_B', \hat{p}_A(p_B' , \g') | \g') 
        - Q_B(p_B', \hat{p}_A(p_B' , \g) | \g) \\
        &\ge 0,
    \end{aligned}
    \end{equation*}
    where the inequality holds because $\g' \ge \g$ and 
    $\hat{p}_A(p_B' , \g') \ge \hat{p}_A(p_B' , \g)$ 
    because $\hat{p}_A(p_B' , \cdot)$ is weakly increasing. 

    \item Suppose $p_B(\g) < \tilde{p}_B$. 
    Fix $p_B' \in (p_B(\g),\, \tilde{p}_B)$. Hence, $p_B'$ is in $[ p_B( \bar{\g}), p_B ( \ubar{\g})]$. 
    Let $\g' = p_B^{-1}(p_B')$ and note that $\g' \le \g$. 
    By construction, we have
    \begin{equation*}
    \begin{aligned}
        &\hat{Q}_B(p_B') - Q_B(p_B', \hat{p}_A(p_B' , \g) | \g) \\
        &=
        Q_B(p_B', \hat{p}_A(p_B', \g') | \g') 
        - Q_B(p_B', \hat{p}_A(p_B' , \g) | \g) \\
        &\le 0,
    \end{aligned}
    \end{equation*}
    where the inequality holds because $\g' \le \g$ and 
    $\hat{p}_A(p_B' ,\g') \le \hat{p}_A(p_B' ,\g)$ 
    because $\hat{p}_A(p_B' , \cdot)$ is weakly increasing.
\end{itemize}

\subsection{Proof of Lemma~\ref{res:profitable_deviation}} \label{sec:proof_profitable_deviation}

For any strategy profile $(s_A, s_B, p, q)$, let
\begin{equation*}
\begin{aligned}
    U( \ubar{\g} ; s_A, s_B, p,q) &= - s_A( p_A (\ubar{\g}))  -s_B( p_B( \ubar{\g})) \\
    &\quad + \E_{\th | \ubar{\g}} \Brac{ (v_A( \th) - p_A (\ubar{\g}))q_A (\ubar{\g}, \th) + (v_B( \th) - p_B( \ubar{\g})) q_B( \ubar{\g}, \th) }.
\end{aligned}
\end{equation*}
 Let $\tilde{p}_B \colon \G \to [0, \infty]$ be weakly decreasing. Let $\tilde{p}_A \colon \G \to [0, \infty]$ be a selection from the correspondence $\g \mapsto P_A ( \tilde{p}_B (\g), \g ; \hat{s}_A)$. Let $\tilde{q}$ be a product-purchase strategy that is optimal given $\tilde{p}$. By \cref{res:subscription_construction}, there exists a subscription schedule $\tilde{s}_B \colon [0, \infty) \to  [0, \infty]$ such that $\tilde{p} (\g)$ is in $P (\g ; \hat{s}_A, \tilde{s}_B)$ for each type $\g$, and
\begin{equation} \label{eq:tilde_extraction}
\begin{aligned}
      & - \hat{s}_A( \tilde{p}_A (\ubar{\g})) - \tilde{s}_B (\tilde{p}_B( \ubar{\g})) + \E_{\th | \ubar{\g}} \Brac{ \max\{0, v_A(\th) - \tilde{p}_A(\ubar{\g}), v_B(\th) - \tilde{p}_B(\ubar{\g}) \}}  \\
     &= \max_{p_A' \in [0, \infty]} \Set{ -\hat{s}_A( p_A') +  \E_{\th| \ubar{\g}} [(v_A(\th) - p_A')_+]  }.
\end{aligned}
\end{equation}
By the envelope formula, which follows from the derivation of \eqref{eq:revenue_simplified} in the proof of \cref{res:equilibrium} (\cref{sec:proof_equilibrium_existence}), firm $B$'s equilibrium expected revenue is 
\[
    -U( \ubar{\g} ; \hat{s}_A, \hat{s}_B, \hat{p}, \hat{q}) + \E_{\g} [ I_B ( \g ; \hat{s}_A, \hat{p}_A, \hat{q})].
\]
Firm $B$'s expected revenue from deviating to $\tilde{s}_B$ and inducing the best response $(\tilde{p}, \tilde{q})$ is 
\[
    -U( \ubar{\g} ; \hat{s}_A, \tilde{s}_B, \tilde{p}, \tilde{q}) + \E_{\g} [ I_B ( \g ; \hat{s}_A, \tilde{p}_A, \tilde{q})].
\]
By \cref{res:revenue}, firm $B$ gets this same revenue whichever best response is selected. Since $(\hat{s}_A, \hat{s}_B, \hat{p}, \hat{q})$ is an equilibrium, we must have 
\[
     -U( \ubar{\g} ; \hat{s}_A, \hat{s}_B, \hat{p}, \hat{q}) + \E_{\g} [ I_B ( \g ; \hat{s}_A, \hat{p}_A, \hat{q})] \geq    -U( \ubar{\g} ; \hat{s}_A, \tilde{s}_B, \tilde{p}, \tilde{q}) + \E_{\g} [ I_B ( \g ; \hat{s}_A, \tilde{p}_A, \tilde{q})]. 
\]
To conclude that $\E_\g [I_B(\g; \hat{s}_A, \hat{p}_A, \hat{q})] \geq \E_\g [I_B (\g ; \hat{s}_A, \tilde{p}_A, \tilde{q})]$, it suffices to show that $U ( \ubar{\g};  \hat{s}_A, \hat{s}_B, \hat{p}, \hat{q}) \geq U( \ubar{\g} ; \hat{s}_A, \tilde{s}_B, \tilde{p}, \tilde{q})$. We claim that
\begin{equation*}
\begin{aligned}
    & U( \ubar{\g} ; \hat{s}_A, \tilde{s}_B, \tilde{p}, \tilde{q}) \\
    &= - \hat{s}_A( \tilde{p}_A(\ubar{\g})) - \tilde{s}_B( \tilde{p}_B( \ubar{\g})) + \E_{\th | \ubar{\g}} \Brac{ (v_A(\th) - \tilde{p}_A(\ubar{\g})) \tilde{q}_A( \ubar{\g}, \th) + ( v_B( \th)  -\tilde{p}_B(\ubar{\g})) \tilde{q}_B(\ubar{\g}, \th)} \\
    &=  - \hat{s}_A( \tilde{p}_A(\ubar{\g})) - \tilde{s}_B( \tilde{p}_B( \ubar{\g})) +  \E_{\th | \ubar{\g}} \Brac{ \max\{0, v_A( \th) - \tilde{p}_A(\ubar{\g}), v_B(\th) - \tilde{p}_B(\ubar{\g}) \} } \\
    &= \max_{p_A' \in [0, \infty]} \Set{ -\hat{s}_A( p_A') +  \E_{\th| \ubar{\g}} [(v_A(\th) - p_A')_+]  } \\
    &\leq  - \hat{s}_A( \hat{p}_A(\ubar{\g})) - \hat{s}_B( \hat{p}_B( \ubar{\g})) + \E_{\th | \ubar{\g}} \Brac{ (v_A(\th) - \hat{p}_A(\ubar{\g})) \hat{q}_A( \ubar{\g}, \th) + ( v_B( \th)  -\hat{p}_B(\ubar{\g})) \hat{q}_B(\ubar{\g}, \th)} \\
    &= U ( \ubar{\g};  \hat{s}_A, \hat{s}_B, \hat{p}, \hat{q}).
\end{aligned}
\end{equation*}
The second equality holds because $\tilde{q}$ is optimal given $\tilde{p}$; the third equality follows from \eqref{eq:tilde_extraction}; and the inequality holds because $(\hat{p}, \hat{q})$ is a best response to $(\hat{s}_A, \hat{s}_B)$.

\subsection{Proof of Lemma~\ref{res:full_coverage}} \label{sec:proof_full_coverage}

Suppose, for a contradiction, that there is a $\hat{p}$-continuity point $\g_0$ in $(\ubar{\g}, \bar{\g})$ such that $\hat{p}_A (\g_0) + \hat{p}_B(\g_0) > 2v_0$. Since $v_0 \geq \max_{\g' \in \G} (1 / g(\g'))$, we have
\[
    p_A^\ast (\g_0) + p_B^\ast (\g_0)  = 2 / g(\g_0) \leq 2v_0.
\]
Thus, $\hat{p}_i (\g_0) > p_i^\ast (\g_0)$ for some firm $i$. Without loss, we may assume that $\hat{p}_B( \g_0) > p_B^\ast(\g_0)$. We derive a contradiction with \cref{res:profitable_deviation}. Define $\tilde{p}_B \colon \G \to [0, \infty]$ by 
\[
    \tilde{p}_B(\g) 
    =
    \begin{cases}
        \hat{p}_B (\g) &\text{if}~\g < \g_0, \\
    \max \{ p_B^\ast (\g_0), 2 v_0 - \hat{p}_A( \g_0 )\} \wedge \hat{p}_B(\g) &\text{if}~ \g \geq \g_0.
    \end{cases}
\]
Since $\hat{p}_B$ is weakly decreasing, so is the function $\tilde{p}_B$. Note that $\hat{p}_B(\g_0) >  \max \{ p_B^\ast (\g_0), 2 v_0 - \hat{p}_A( \g_0 )\}$. Since $\g_0$ is a continuity point of $\hat{p}_B$, there exists 
$\d > 0$ such that $\hat{p}_B(\g) >  \max \{ p_B^\ast (\g_0), 2 v_0 - \hat{p}_A( \g_0 )\}$ for all $\g \in [\g_0, \g_0 + \d]$. 

Define $\tilde{p}_A \colon \G \to [0, \infty]$ by $\tilde{p}_A (\g) = \hat{p}_A(\g)$ for all types $\g$. To apply \cref{res:profitable_deviation}, we check that for each type $\g$, we have $\hat{p}_A (\g) \in P_A  (\tilde{p}_B (\g), \g ; \hat{s}_A)$. This is immediate if $\tilde{p}_B(\g) = \hat{p}_B(\g)$ since $\hat{p}(\g)$ is in $P(\g ; \hat{s}_A, \hat{s}_B)$. Fix a type $\g \in \G$ with $\tilde{p}_B( \g) \neq \hat{p}_B (\g)$. We must have $\g \geq \g_0$. By construction, we have $\tilde{p}_B(\g) \geq 2 v_0 - \hat{p}_A( \g_0)$. Since $\hat{p}_A$ is weakly increasing, we have $\hat{p}_A (\g) \geq \hat{p}_A(\g_0)$. It follows that $\hat{p}_A(\g) + \tilde{p}_B(\g) \geq 2v_0$.  Since $\tilde{p}_B(\g) \leq \hat{p}_B (\g)$, we have $Q_A ( p_A', \tilde{p}_B(\g) | \g) \leq Q_A (p_A', \hat{p}_B( \g) | \g)$ for all prices $p_A' \in [0, \infty]$. We conclude that this inequality holds with equality whenever $p_A' \geq \hat{p}_A(\g)$. Since $\hat{p}_A( \g)$ is in $P_A (\hat{p}_B( \g), \g ; \hat{s}_A)$, it follows that $\hat{p}_A( \g)$ is in $P_A ( \tilde{p}_B( \g), \g ; \hat{s}_A)$ as well. 

Finally, to reach a contradiction, observe that for all types $\g$ with $\tilde{p}_B( \g) = \hat{p}_B (\g)$, we immediately have $ I_B(\g ; \hat{s}_A, \tilde{p}_A, \tilde{q}) = I_B(\g; \hat{s}_A, \hat{p}_A, \hat{q})$. For all types $\g$ with $\tilde{p}_B( \g) \neq \hat{p}_B (\g)$, we have $\hat{p}_A(\g) + \tilde{p}_B(\g) \geq 2v_0$, as shown above, so $\tilde{q}_A(\g, \cdot) = \hat{q}_A (\g, \cdot)$. Therefore, 
\begin{equation*}
\begin{aligned}
&
 I_B(\g ; \hat{s}_A, \tilde{p}_A, \tilde{q}) - I_B(\g; \hat{s}_A, \hat{p}_A, \hat{q})\\
 &= \E_{\th| \g} \Brac{ (v_B(\th) - \bar{G}(\g) / g(\g)) \Paren{ \tilde{q}_B(\g, \th) - \hat{q}_B(\g, \th)}} \\
 &> 0,
\end{aligned}
\end{equation*}
where the inequality is strict because $\tilde{p}_B (\g) < \hat{p}_B( \g)$. Since $\tilde{p}_B( \g) < \hat{p}_B (\g)$ for all types $\g$ in $[\g_0, \g_0 + \d]$, we conclude that 
\[
\E_{\g} \Brac{ I_B(\g ; \hat{s}_A, \tilde{p}_A, \tilde{q})} > \E_{\g} \Brac{ I_B(\g; \hat{s}_A, \hat{p}_A, \hat{q})},
\]
contrary to \cref{res:profitable_deviation}.

\subsection{Proof of Lemma~\ref{res:price_lower_limit}}
Without loss, we may assume $i = B$. Let $\ubar{p}_B = \inf_{\g' \in (\ubar{\g}, \bar{\g})} \hat{p}_B( \g')$. Suppose for a contradiction that $\ubar{p}_B > 0$.  For each type $\g$, let
\begin{equation} \label{eq:delta_def_first}
\begin{aligned}
    \D( \g) 
    &= \E_{\th | \g} \Brac{ \max \{ 0, v_A(\th) - \hat{p}_A ( \g), v_B(\th) - p_B^\ast (\g) \} }  \\
    &\quad- \E_{\th | \g} \Brac{ (v_A(\th) - \hat{p}_A(\g)) \hat{q}_A( \g, \th) + (v_B( \th) - p_B^\ast (\g)) \hat{q}_B( \g, \th)}.
\end{aligned}
\end{equation}
We claim that there exists $\d > 0$ such that for all $\g \geq \bar{\g} - \d$, we have
\[
    p_B^\ast (\g) \leq \ubar{p}_B
    \quad
    \text{and}
    \quad
    \D(\g) > \bar{G}(\g) / g(\g).
\]
This claim is verified at the end of the proof. Here, we complete the proof, assuming the claim. Define $\tilde{p}_B \colon \G \to [0, \infty]$ by 
\[
    \tilde{p}_B(\g) =
    \begin{cases}
        \hat{p}_B (\g) & \text{if}~ \g < \bar{\g} - \d, \\
        p_B^\ast (\g) &\text{if}~ \g \geq \bar{\g} - \d.
    \end{cases}
\]
By construction, $\tilde{p}_B$ is weakly decreasing. 
Define $\tilde{p}_A \colon \G \to [0, \infty]$ as follows. For $\g < \bar{\g} - \d$, let $\tilde{p}_A(\g) = \hat{p}_A( \g)$. For $\g \geq \bar{\g} - \d$, choose $\tilde{p}_A(\g)$ from $P_A ( \tilde{p}_B(\g), \g ; \hat{s}_A)$.  By construction, $\tilde{p}_A (\g)$ is in $P_A( \tilde{p}_B( \g), \g ; \hat{s}_A)$ for each type $\g$.  Let $\tilde{q}$ be a product-purchase strategy that is optimal given $\tilde{p}$. We now seek a contradiction with \cref{res:profitable_deviation}. For each $\g < \bar{\g} - \d$, we have $I_B( \g; \hat{s}_A, \tilde{p}_A, \tilde{q}) = I_B( \g; \hat{s}_A, \hat{p}_A, \hat{q})$. For each $\g \geq \bar{\g} - \d$, we have
\begin{equation*}
\begin{aligned}
    & I_B(\g ; \hat{s}_A, \tilde{p}_A, \tilde{q}) - ( \bar{G}(\g) / g(\g)) \E_{\th| \g} [ \tilde{q}_A (\g, \th) + \tilde{q}_B(\g,\th)] \\
    &= - \hat{s}_A ( \tilde{p}_A(\g))  + \E_{\th | \g} \Brac{ (v_A( \th) - \tilde{p}_A(\g)) \tilde{q}_A (\g, \th) +  (v_B(\th) - p_B^\ast (\g)) \tilde{q}_B( \g, \th)}  \\
    &= - \hat{s}_A ( \tilde{p}_A(\g))  + \E_{\th | \g} \Brac{ \max \{0, v_A( \th) - \tilde{p}_A(\g), v_B(\th) - p_B^\ast (\g) \} } \\
    &\geq - \hat{s}_A ( \hat{p}_A(\g))  + \E_{\th | \g} \Brac{ \max \{0, v_A( \th) - \hat{p}_A(\g), v_B(\th) - p_B^\ast (\g) \} } \\
    &= - \hat{s}_A ( \hat{p}_A(\g)) + \E_{\th | \g} \Brac{ (v_A(\th) - \hat{p}_A(\g)) \hat{q}_A(\g, \th) + ( v_B(\th) - p_B^\ast ( \g)) \hat{q}_B(\g, \th) } + \D(\g) \\
    &= I_B( \g; \hat{s}_A, \hat{p}_A, \hat{q}) - ( \bar{G}(\g) / g(\g)) \E_{\th| \g} [ \hat{q}_A (\g, \th) + \hat{q}_B(\g,\th)] + \D(\g),
\end{aligned}
\end{equation*}
where we are using the facts that $\tilde{q}$ is optimal given $\tilde{p}$ and that $\tilde{p}_A(\g)$ is in $P_A (p_B^\ast ( \g), \g ; \hat{s}_A)$, together with the definition of $\D(\g)$ from \eqref{eq:delta_def_first}. Subtracting, we conclude that
\begin{equation*}
\begin{aligned}
    &I_B (\g ; \hat{s}_A, \tilde{p}_A, \tilde{q}) - I_B ( \g ; \hat{s}_A, \hat{p}_A, \hat{q})  \\
    &\geq \D (\g) - \bar{G}(\g)/g(\g) \Paren{\E_{\th| \g} [ \hat{q}_A (\g, \th) + \hat{q}_B(\g,\th)] - \E_{\th |\g} \Brac{ \tilde{q}_A (\g, \th) + \tilde{q}_B( \g, \th)}} \\
    &\geq \D (\g) - \bar{G}(\g)/g(\g) \\ 
    &> 0,
\end{aligned}
\end{equation*}
where the last inequality follows from the claim. Thus, we have a contradiction with \cref{res:profitable_deviation}. 

\paragraph{Proof of claim} Since $p_B^\ast (\bar{\g}) = 0$ and $p_B^\ast$ is continuous, we may select $\d' > 0$ such that $p_B^\ast (\g) < \ubar{p}_B/2$ for all $\g \in [\bar{\g} - \d', \bar{\g}]$. Therefore, for each  $\g \in (\bar{\g} - \d', \bar{\g}]$, we have $\hat{p}_B(\g) - p_B^\ast (\g) \geq \ubar{p}_B - \ubar{p}_B/2 = \ubar{p}_B/2$, and by \cref{res:full_coverage}, $\hat{p}_A(\g) + p_B^\ast (\g) < 2 v_0$, so
\begin{equation*}
\begin{aligned}
    \D(\g) 
    &\geq \int_{ (p_B^\ast (\g) - \hat{p}_A (\g))/2}^{(\hat{p}_B (\g) - \hat{p}_A (\g))/2} \bigl( 2\th -  (p_B^\ast (\g) - \hat{p}_A (\g)) \bigr) f(\th - \g) \de \th \\
    &\geq  \int_{ (p_B^\ast (\g) - \hat{p}_A (\g))/2}^{(p_B^\ast (\g) - \hat{p}_A (\g))/2 + \ubar{p}_B/4} \bigl( 2\th -  (p_B^\ast (\g) - \hat{p}_A (\g)) \bigr) f(\th - \g) \de \th \\
    &\geq (\ubar{p}_B/4)^2 \ubar{f},
\end{aligned}
\end{equation*}
where $\ubar{f} = \min_{x \in J} f(x)$ with $J$ defined to be the compact interval equal to the closure of the union 
\[
    \bigcup_{ \bar{\g} - \d' < \g < \bar{\g}} [ (p_B^\ast (\g) - \hat{p}_A (\g))/2 - \g , (p_B^\ast (\g) - \hat{p}_A (\g))/2 -\g + \ubar{p}_B/4 ].
\]
To complete the proof, choose $\d \in (0, \d')$ such that for all $\g \geq \bar{\g} - \d$, we have
\[
    \frac{\bar{G}(\g)}{g(\g)} < (\ubar{p}_B/4)^2 \ubar{f}.
\]

\subsection{Proof of Lemma~\ref{res:prices}}

Without loss, we may assume that $i = A$. Suppose that $\sup_{\g' \in (\ubar{\g}, \bar{\g})} \hat{p}_A( \g') \leq (5/2) \bar{p}_A$.  By \cref{res:price_lower_limit}, we have $\inf_{\g' \in (\ubar{\g}, \bar{\g})} \hat{p}_B (\g') = 0$. Let $\tilde{p}_B = p_B^\ast$. Thus, $\tilde{p}_B$ is weakly decreasing. For each type $\g$, choose $\tilde{p}_A(\g)$ from $P_A ( p_B^\ast (\g) , \g ; \hat{s}_A)$. Let $\tilde{q}$ be a product-purchase strategy that is optimal given $\tilde{p}$. We seek a contradiction with \cref{res:profitable_deviation}. Fix $\g \in (\ubar{\g}, \bar{\g})$. We have $p_B^\ast (\g) > 0$. Since $\hat{p}_B$ is weakly decreasing and $\inf_{\g' \in (\ubar{\g}, \bar{\g})} \hat{p}_B (\g') = 0$, we can find $\g' \in (\g, \bar{\g})$ such that $p_B^\ast (\g) > \hat{p}_B (\g')$. By \cref{res:BR_partial_monotone}.\ref{it:partial_strict_monotonicity}
\[
    \tilde{p}_A (\g)  \leq  \sup  P_A ( p_B^\ast (\g), \g ; \hat{s}_A) \leq \inf P_A ( \hat{p}_B (\g'), \g' ; \hat{s}_A) \leq \hat{p}_A ( \g') \leq (5/2) \bar{p}_A.
\]
Thus, 
\[
    \tilde{p}_A(\g) + \tilde{p}_B(\g) = \tilde{p}_A(\g) + p_B^\ast (\g) \leq (5/2) \bar{p}_A  + \bar{p}_B \leq \max_{\g' \in \G} \frac{7}{g(\g')} \leq 2 v_0. 
\]
Therefore, $\E_{\th | \g} \Brac{ \tilde{q}_A( \g, \th) + \tilde{q}_B(\g, \th)} = 1$, so we conclude that 
\begin{equation*}
\begin{aligned}
    & I_B( \g ; \hat{s}_A, \tilde{p}_A, \tilde{q} ) - \bar{G}(\g)/ g(\g) \\
    &=  I_B ( \g ; \hat{s}_A, \tilde{p}_A, \tilde{q} ) - ( \bar{G}(\g)/ g(\g)) \E_{\th | \g} \Brac{ \tilde{q}_A( \g, \th) + \tilde{q}_B(\g, \th)}  \\
    &=  - \hat{s}_A( \tilde{p}_A(\g))  + \E_{\th | \g} \Brac{ (v_A(\th) - \tilde{p}_A(\g)) \tilde{q}_A (\g, \th) + (v_B( \th) - p_B^\ast (\g)) \tilde{q}_B (\g, \th)} \\
    &=  - \hat{s}_A( \tilde{p}_A(\g))  + \E_{\th | \g} \Brac{ \max \{0, v_A(\th) - \tilde{p}_A(\g), v_B( \th) - p_B^\ast (\g)\}} \\
    &\geq - \hat{s}_A( \hat{p}_A(\g))  + \E_{\th | \g} \Brac{ \max \{0, v_A(\th) - \hat{p}_A(\g), v_B( \th) - p_B^\ast (\g)\}} \\
    &\geq  - \hat{s}_A( \hat{p}_A(\g))  + \E_{\th | \g} \Brac{ (v_A(\th) - \hat{p}_A(\g)) \hat{q}_A(\g, \th) + (v_B( \th) - p_B^\ast (\g)) \hat{q}_B (\g, \th)} \\
    &= I_B ( \g ; \hat{s}_A, \hat{p}_A, \hat{q} ) - ( \bar{G}(\g)/ g(\g)) \E_{\th | \g} \Brac{ \hat{q}_A( \g, \th) + \hat{q}_B(\g, \th)} \\
    &\geq I_B ( \g ; \hat{s}_A, \hat{p}_A, \hat{q} ) - \bar{G}(\g)/ g(\g),
\end{aligned}
\end{equation*}
where we are using the facts that $\tilde{q}$ is optimal given $\tilde{p}$ and that $\tilde{p}_A(\g)$ is in  $P_A ( p_B^\ast (\g) , \g ; \hat{s}_A)$. Moreover, since $\hat{q}$ is optimal given $\hat{p}$, the second inequality is strict if $\hat{p}_B(\g) \neq p_B^\ast (\g)$. By \cref{res:profitable_deviation}, we conclude that $\hat{p}_B(\g) = p_B^\ast (\g)$ for almost every type $\g$. Since $\hat{p}_B$ is weakly decreasing and 
$p_B^\ast$ is continuous, it follows that $\hat{p}_B$ and $p_B^\ast$ agree at \emph{every} type in $(\ubar{\g}, \bar{\g})$.\footnote{Let $\G_0$ denote the set of types in $(\ubar{\g}, \bar{\g})$ at which $\hat{p}_B$ and $p_B^\ast$ agree. For any type $\g \in (\ubar{\g}, \bar{\g})$, choose sequences $(\g_n)$ and $(\g_n')$ in $\G_0$ such that $\g_n \uparrow \g$ and $\g_n' \downarrow \g$.}

\subsection{Proof of Lemma~\ref{res:price_bound}} \label{sec:proof_price_bound}

Suppose, for a contradiction, that $\sup_{\g' \in (\ubar{\g}, \bar{\g})} \hat{p}_i(\g') > (5/2) \bar{p}_i$ for some firm $i$. By \cref{res:prices}, it follows that $\sup_{\g' \in (\ubar{\g}, \bar{\g})} \hat{p}_{-i}(\g') > (5/2) \bar{p}_{-i}$. Let 
\[
    \g_0 = \sup \{ \g \in (\ubar{\g}, \bar{\g}): \hat{p}_B( \g) > (5/2) \bar{p}_B \}.
\] 
By \cref{res:price_lower_limit}, we have $\inf_{\g' \in (\ubar{\g}, \bar{\g})} \hat{p}_B(\g') = 0$. Since $\hat{p}_B$ is weakly decreasing, we know that $\g_0$ is in $(\ubar{\g}, \bar{\g})$. Define the following price limits:
\[
   \ubar{p}_A^0 = \lim_{\g \uparrow \g_0} \hat{p}_A(\g), \qquad  \bar{p}_A^0 = \lim_{\g \downarrow \g_0} \hat{p}_A(\g), \qquad 
    \ubar{p}_B^0 = \lim_{\g \downarrow \g_0} \hat{p}_B(\g), \qquad
    \bar{p}_B^0 = \lim_{\g \uparrow \g_0} \hat{p}_B(\g).     
\]
Since $\hat{p}$ is monotone, these limits are well-defined and we have $\bar{p}_i^0 \geq \ubar{p}_i^0$ for each $i = A, B$. From the definition of $\g_0$, we have $\bar{p}_B^0 \geq (5/2) \bar{p}_B$.  We complete the proof, using two claims, which we prove at the end. 

\emph{Claim 1}: We have
\begin{equation}
    \bar{p}_A^0 + \ubar{p}_B^0 = \ubar{p}_A^0 + \bar{p}_B^0 = 2v_0,
\end{equation}
and 
\begin{equation}
   p_i^M(\g_0) \leq \ubar{p}_i^0 < p_i^\ast (\g_0) < \bar{p}_i < (5/2) \bar{p}_i \leq  \bar{p}_i^0, \qquad i = A, B.
\end{equation}

Without loss, we may assume hereafter that $v_B ( \g_0) - \ubar{p}_B^0 \geq v_A( \g_0) - \ubar{p}_A^0$; otherwise, we can follow a symmetric argument with firm $A$ in place of firm $B$. For each type $\g$, let 
\begin{equation} \label{eq:delta_def}
\begin{aligned}
    \D( \g) 
    &= \E_{\th | \g} \Brac { \max\{ 0, v_A(\th) - \hat{p}_A( \g), v_B(\th) - p_B^\ast (\g) \}} \\
    &\quad- \E_{\th | \g} \Brac{ (v_A(\th) - \hat{p}_A ( \g)) \hat{q}_A (\g,\th) + ( v_B(\th) - p_B^\ast (\g)) \hat{q}_B( \g, \th) }. 
\end{aligned}
\end{equation}

\emph{Claim 2:} There exists $\d > 0$ such that for all $\g$ in $(\g_0 - \d, \g_0)$, we have 
\begin{equation} \label{eq:ineq_claim2}
    \D (\g) > \frac{ \bar{G}(\g)}{g(\g)} \P_{\th | \g} \Paren{ \max \Set{ v_A(\th) - \inf P_A (p_B^\ast(\g), \g ; \hat{s}_A), v_B(\th) - p_B^\ast (\g) } < 0 }.
\end{equation}
Now we complete the proof, assuming the claims. Define $\tilde{p}_B \colon \G \to [0, \infty]$ by 
\[
    \tilde{p}_B (\g) 
    = \begin{cases} 
    p_B^\ast (\g) &\text{if}~ \g \in (\g_0 - \d, \g_0], \\
    \hat{p}_B (\g) &\text{otherwise}. 
    \end{cases}
\]
It is easily verified that $\tilde{p}_B$ is weakly decreasing.\footnote{The functions $p_B^\ast$ and $\hat{p}_B$ are weakly decreasing. Moreover, for all $\g \leq \g_0 - \d$, we have $\hat{p}_B( \g) \geq (5/2) \bar{p}_B \geq p_B^\ast (\g_0 - \d)$. For all $\g > \g_0$, we have $p_B^\ast (\g_0) \geq \ubar{p}_B^0 \geq \hat{p}_B (\g)$.} Define $\tilde{p}_A \colon \G \to [0, \infty]$ by 
\[
\tilde{p}_A(\g) 
= 
 \begin{cases} 
   \inf P_A (p_B^\ast (\g), \g ; \hat{s}_A) &\text{if}~ \g \in (\g_0 - \d, \g_0], \\
    \hat{p}_A (\g) &\text{otherwise}. 
    \end{cases}
\]
By construction, $\tilde{p}_A (\g)$ is in $P_A( \tilde{p}_B( \g), \g ; \hat{s}_A)$ for each type $\g$.  Let $\tilde{q}$ be a product-purchase strategy that is optimal given $\tilde{p}$. We now seek a contradiction with \cref{res:profitable_deviation}.  For each $\g \not\in (\g_0 - \d, \g_0]$, we have $I_B( \g; \hat{s}_A, \tilde{p}_A, \tilde{q}) = I_B (\g; \hat{s}_A, \hat{p}_A, \hat{q})$.  For each $\g \in (\g_0 - \d,\g_0)$,\footnote{We can ignore the point $\g_0$ since it has measure zero.} we have
\begin{equation*}
\begin{aligned}
& I_B( \g ; \hat{s}_A, \tilde{p}_A, \tilde{q}) - ( \bar{G}(\g) / g(\g)) \E_{\th| \g} \Brac{ \tilde{q}_A (\g, \th) + \tilde{q}_B( \g, \th) } \\
&= -\hat{s}_A (\tilde{p}_A (\g)) + \E_{\th | \g} \Brac{ \Paren{ v_A( \th) - \tilde{p}_A ( \g)} \tilde{q}_A (\g, \th) + \Paren{ v_B( \th) - p_B^\ast (\g)} \tilde{q}_B( \g, \th) }  \\ 
&= - \hat{s}_A ( \tilde{p}_A ( \g)) 
   + \E_{\th | \g} \Brac{ \max \{0, v_A( \th) - \tilde{p}_A ( \g), v_B( \th) - p_B^\ast (\g)\}} \\
&\geq  - \hat{s}_A ( \hat{p}_A ( \g)) 
   + \E_{\th | \g} \Brac{ \max \{0, v_A( \th) - \hat{p}_A ( \g), v_B( \th) - p_B^\ast (\g)\}} \\
&= - \hat{s}_A ( \hat{p}_A ( \g)) 
+  \E_{\th | \g} \Brac{ ( v_A(\th) - \hat{p}_A(\g)) \hat{q}_A (\g, \th) + \Paren{ v_B( \th) - p_B^\ast (\g)} \hat{q}_B (\g, \th) } + \D(\g) \\ 
&= I_B (\g ; \hat{s}_A, \hat{p}_A, \hat{q}) - (\bar{G}(\g) / g(\g)) \E_{\th| \g} \Brac{ \hat{q}_A ( \g, \th) + \hat{q}_B( \g, \th) } + \D(\g),
\end{aligned}
\end{equation*}
where we are using the facts that $\tilde{q}$ is optimal given $\tilde{p}$ and that $\tilde{p}_A (\g)$ is in $P_A (p_B^\ast (\g),\g ; \hat{s}_A)$, together with the definition of $\D(\g)$ from \eqref{eq:delta_def}. Subtracting gives
\begin{equation*}
\begin{aligned}
    &I_B( \g ; \hat{s}_A, \tilde{p}_A, \tilde{q}) - I_B (\g ; \hat{s}_A, \hat{p}_A, \hat{q}) \\
    &\geq \D(\g) - (\bar{G}(\g) / g(\g))\Paren{\E_{\th| \g} \Brac{ \hat{q}_A (\g, \th) + \hat{q}_B( \g, \th) } - \E_{\th| \g} \Brac{ \tilde{q}_A (\g, \th) + \tilde{q}_B( \g, \th) }} \\
    &\geq  \D(\g) - (\bar{G}(\g) / g(\g)) \Paren{1 - \E_{\th| \g} \Brac{ \tilde{q}_A (\g, \th) + \tilde{q}_B( \g, \th) }}  \\
    &= \D (\g) -  (\bar{G}(\g) / g(\g))\P_{\th | \g} \Paren{ \max\{ v_A(\th) - \tilde{p}_A(\g), v_B(\th) - p_B^\ast (\g) \} < 0 }\\
    &>0,
\end{aligned}
\end{equation*}
where the last inequality follows from Claim 2 and the definition of $\tilde{q}$. 

\paragraph{Proof of Claim 1} From the definition of $\g_0$, we have $\bar{p}_B^0 \geq (5/2) \bar{p}_B$. It suffices to prove the following:
\begin{enumerate}[label = \roman*., ref = \roman*]
    \item \label{it:low} $\ubar{p}_B^0 < p_B^\ast (\g_0)$;
    \item \label{it:sum} $\ubar{p}_B^0 + \bar{p}_A^0 = 2v_0$;
    \item \label{it:high} $\ubar{p}_B^0 \geq p_B^M (\g_0)$.
\end{enumerate}
Indeed, applying part~\ref{it:sum} and then part~\ref{it:low} gives
\begin{equation*}
\begin{aligned}
    \bar{p}_A^0 &= 2 v_0 - \ubar{p}_B^0\\
    &> 2v_0 - p_B^\ast (\g_0) \\
    &\geq 7 \max_{\g' \in \G} (1/ g(\g')) - 2/g(\g_0) \\
    &\geq 5 \max_{\g' \in \G} (1/ g(\g'))\\
    &\geq (5/2) \bar{p}_A,
\end{aligned}
\end{equation*}
where the second inequality follows from the assumption that $v_0 \geq (7/2) \max_{\g' \in \G} (1 /g(\g'))$. Thus, we can reason symmetrically for firm $A$ to establish the claim. We now prove parts~\ref{it:low}--\ref{it:high}.

\paragraph{Part i} We check that $\ubar{p}_B^0 < p_B^\ast (\g_0)$. Suppose for a contradiction that $\ubar{p}_B^0 \geq p_B^\ast (\g_0)$. To reach a contradiction, we show that firm $B$ can profit by inducing lower strike prices for types just left of $\g_0$. We have $\lim_{\g \uparrow \g_0} \hat{p}_A(\g) = \ubar{p}_A^0$ and $\lim_{\g \uparrow \g_0} \hat{p}_B(\g) = \bar{p}_B^0 \geq (5/2) \bar{p}_B$. Since $p_B^\ast$ is continuous, there exists $\d > 0$ such that for all $\g \in (\g_0 - \d, \g_0)$, we have 
\begin{equation} \label{eq:comparison}
\begin{aligned}
    &\E_{\th | \g} \Brac{ \max\{0, v_A(\th) - \hat{p}_A(\g), v_B( \th) - \ubar{p}_B^0\} }\\
    &\quad + ( \ubar{p}_B^0  - p_B^\ast (\g) ) \P_{\th | \g} \Paren{  v_B(\th) - \ubar{p}_B^0 > \Paren{ v_A(\th) - \inf P_A (\ubar{p}_B^0, \g ; \hat{s}_A)}_+} \\
&> \E_{\th |\g} \Brac{ (v_A(\th) - \hat{p}_A (\g) ) \hat{q}_A (\g, \th) + (v_B(\th) - \ubar{p}_B^0) \hat{q}_B (\g, \th) } \\
&\quad +  ( \ubar{p}_B^0  - p_B^\ast (\g) ) \E_{\th |\g} \Brac{ \hat{q}_B( \g, \th)} \\
&= \E_{\th |\g} \Brac{ (v_A(\th) - \hat{p}_A (\g) ) \hat{q}_A (\g, \th) + (v_B(\th) - p_B^\ast (\g) ) \hat{q}_B (\g, \th) }.
\end{aligned}
\end{equation}
Define $\tilde{p}_B \colon \G \to [0, \infty]$ by 
\[
    \tilde{p}_B (\g) = 
    \begin{cases}
        \ubar{p}_B^0 &\text{if}~ \g \in (\g_0 - \d, \g_0], \\
    \hat{p}_B(\g) &\text{otherwise}.
    \end{cases}
\]
It is easily verified that $\tilde{p}_B$ is weakly decreasing. Define $\tilde{p}_A \colon \G \to [0,\infty]$ by 
\[
    \tilde{p}_A (\g) = 
    \begin{cases}
        \inf P_A ( \ubar{p}_B^0, \g; \hat{s}_A) &\text{if}~ \g \in (\g_0 - \d, \g_0], \\
    \hat{p}_A(\g) &\text{otherwise}.
    \end{cases}
\]
Let $\tilde{q}$ be a product-purchase strategy that is optimal given $\tilde{p}$.  We seek a contradiction with \cref{res:profitable_deviation}. For each $\g' > \g_0$, we have $\hat{p}_B(\g') \leq \ubar{p}_B^0$, so for each $\g \in (\g_0 - \d, \g_0]$, \cref{res:BR_partial_monotone}.\ref{it:partial_weak_monotonicity} gives
\[
    \tilde{p}_A (\g) = \inf P_A ( \ubar{p}_B^0, \g ; \hat{s}_A) \leq \inf P_A (  \hat{p}_B( \g'), \g'; \hat{s}_A) \leq \hat{p}_A( \g').
\]
Passing to the limit as $\g' \downarrow \g_0$ gives $  \tilde{p}_A (\g) \leq \bar{p}_A^0$. Therefore, 
\[ 
    \tilde{p}_A (\g) + \tilde{p}_B(\g) \leq \bar{p}_A^0 + \ubar{p}_B^0 \leq 2v_0, 
\]
where the last inequality follows from \cref{res:full_coverage}. Therefore, for each $\g \in (\g_0 - \d, \g_0)$, we have $\E_{\th | \g} \Brac{ \tilde{q}_A(\g, \th) + \tilde{q}_B( \g, \th)} = 1$, so
\begin{equation*}
\begin{aligned}
    &  I_B (\g; \hat{s}_A, \tilde{p}_A, \tilde{q}) - \bar{G}(\g)/ g(\g) \\
    &=I_B (\g; \hat{s}_A, \tilde{p}_A, \tilde{q}) - (\bar{G}(\g)/ g(\g)) \E_{\th | \g} \Brac{ \tilde{q}_A(\g, \th) + \tilde{q}_B( \g, \th)} \\
    &= - \hat{s}_A ( \tilde{p}_A (\g)) + \E_{\th | \g} \Brac{ ( v_A(\th) - \tilde{p}_A(\g)) \tilde{q}_A(\g, \th) + (v_B( \th) - p_B^\ast (\g)) \tilde{q}_B(\g, \th) } \\
    &= - \hat{s}_A ( \tilde{p}_A (\g)) + \E_{\th | \g} \Brac{ \max\{ 0,v_A(\th) - \tilde{p}_A(\g), v_B( \th) - \ubar{p}_B^0 \} } + ( \ubar{p}_B^0  - p_B^\ast (\g) ) \E_{\th |\g} \Brac{ \tilde{q}_B( \g, \th)} \\
    &\geq - \hat{s}_A ( \hat{p}_A (\g)) + \E_{\th | \g} \Brac{ \max\{0, v_A(\th) - \hat{p}_A(\g), v_B( \th) - \ubar{p}_B^0\} } + ( \ubar{p}_B^0  - p_B^\ast (\g) ) \E_{\th |\g} \Brac{ \tilde{q}_B( \g, \th)} \\
       &>  - \hat{s}_A ( \hat{p}_A (\g)) +  \E_{\th |\g} \Brac{ (v_A(\th) - \hat{p}_A (\g) ) \hat{q}_A (\g, \th) + (v_B(\th) - p_B^\ast (\g)) \hat{q}_B (\g, \th) }\\
    &=I_B ( \g ; \hat{s}_A, \hat{p}_A, \hat{q}) - (\bar{G}(\g)/ g(\g)) \E_{\th | \g} [ \hat{q}_A (\g, \th) + \hat{q}_B( \g, \th)] \\
    &\geq I_B ( \g ; \hat{s}_A, \hat{p}_A, \hat{q}) - \bar{G}(\g)/ g(\g),
\end{aligned}
\end{equation*}
where we are using the facts that $\tilde{q}$ is optimal given $\tilde{p}$ and that $\tilde{p}_A(\g)$ is in  $P_A ( \ubar{p}_B^0, \g ; \hat{s}_A)$, together with \eqref{eq:comparison}.

\paragraph{Part ii} We prove that $\ubar{p}_B^0 + \bar{p}_A^0 = 2v_0$.  It suffices to check that $\ubar{p}_B^0 + \bar{p}_A^0 \geq 2v_0$; then \cref{res:full_coverage} implies that we must have equality.\footnote{If $\ubar{p}_B^0 + \bar{p}_A^0 > 2v_0$, then we would obtain a contradiction with \cref{res:full_coverage} by considering a sequence of $\hat{p}$-continuity points converging downward to $\g_0$.} Suppose, for a contradiction, that  $\ubar{p}_B^0 + \bar{p}_A^0 < 2v_0$. To reach a contradiction, we show that firm $B$ can profit by inducing higher strike prices for types just right of $\g_0$. We showed above that $\ubar{p}_B^0 < p_B^\ast (\g_0)$. Let
\[
\k = \min \{ 2v_0 - \ubar{p}_B^0 - \bar{p}_A^0, p_B^\ast (\g_0) - \ubar{p}_B^0 \}.
\]
Note that $\k > 0$. By the definition of $\bar{p}_A^0$, we may choose $\d' > 0$ such that $\hat{p}_A( \g) < \bar{p}_A^0 + \k/2$ for all $\g \in (\g_0, \g_0 + \d')$. Since $p_B^\ast(\g_0) \geq \ubar{p}_B^0 + \k$ and $p_B^\ast$ is continuous, we may choose $\d'' > 0$ such that $p_B^\ast (\g) > \ubar{p}_B^0 + \k/2$ for all $\g \in (\g_0, \g_0 + \d'')$. Let $\d = \min \{ \d', \d'' \}$. Define $\tilde{p}_B \colon \G \to [0, \infty]$ by 
\[
    \tilde{p}_B (\g) = 
    \begin{cases}
        \ubar{p}_B^0 + \k/2 &\text{if}~ \g \in [\g_0,\g_0 + \d),\\
    \hat{p}_B(\g) &\text{otherwise}.
    \end{cases}
\]
Since $\bar{p}_B^0 \geq (5/2) \bar{p}_B$, it is easily verified that $\tilde{p}_B$ is weakly decreasing. Define $\tilde{p}_A \colon \G \to [0,\infty]$ by
\[
    \tilde{p}_A (\g) = 
    \begin{cases}
        \inf P_A ( \ubar{p}_B^0 + 
\k/2, \g;  \hat{s}_A) &\text{if}~ \g \in [\g_0,\g_0 + \d),\\
    \hat{p}_A(\g) &\text{otherwise}.
    \end{cases}
\]
By construction, $\tilde{p}_A (\g)$ is in $P_A (\tilde{p}_B(\g), \g; \hat{s}_A)$ for every type $\g$. Let $\tilde{q}$ be an optimal product-purchase strategy given $\tilde{p}$.  For each $\g \not\in [ \g_0, \g_0 + \d)$, we have $\tilde{p} = \hat{p}$, hence $I_B( \g ; \hat{s}_A, \tilde{p}_A, \tilde{q})  = I_B( \g ; \hat{s}_A, \hat{p}_A, \hat{q})$. For $\g \in (\g_0, \g_0 + \d)$, we have $\ubar{p}_B^0 + \k/2 \geq \hat{p}_B (\g)$, so \cref{res:BR_partial_monotone}.\ref{it:partial_weak_monotonicity} gives
\[
    \tilde{p}_A(\g) = \inf P_A(\ubar{p}_B^0 + \k/2, \g ; \hat{s}_A) \leq \inf P_A(\hat{p}_B(\g), \g ; \hat{s}_A) \leq \hat{p}_A(\g). 
\]
So, by the definitions of $\d$ and $\k$, we have
\[
    \tilde{p}_A(\g) + \tilde{p}_B(\g) \leq \hat{p}_A(\g) + \ubar{p}_B^0 + \k/2 \leq \bar{p}_A^0 + \ubar{p}_B^0 + \k \leq 2v_0.
\]
Thus, $\E_{\th | \g} \Brac{ \tilde{q}_A(\g, \th) + \tilde{q}_B(\g, \th)} = 1$, so we have
\begin{equation*}
\begin{aligned}
    &I_B ( \g ; \hat{s}_A, \tilde{p}_A, \tilde{q}) - \bar{G}(\g)/ g(\g) \\
    &= I_B ( \g ; \hat{s}_A, \tilde{p}_A, \tilde{q}) - (\bar{G}(\g)/ g(\g))\E_{\th | \g} \Brac{ \tilde{q}_A(\g, \th) + \tilde{q}_B(\g, \th)} \\
    &= - \hat{s}_A( \tilde{p}_A(\g)) + \E_{\th | \g} \Brac{  ( v_A( \th) - \tilde{p}_A(\g)) \tilde{q}_A (\g, \th) + ( v_B( \th) - p_B^\ast (\g)) \tilde{q}_B( \g, \th)} \\
    & = - \hat{s}_A( \tilde{p}_A(\g)) + \E_{\th | \g} \Brac{  ( v_A( \th) - \tilde{p}_A(\g)) \tilde{q}_A (\g, \th) + ( v_B( \th) - \tilde{p}_B (\g)) \tilde{q}_B( \g, \th)} \\
    &\quad+ (\tilde{p}_B(\g) - p_B^\ast(\g)) \E_{\th | \g} \Brac{   \tilde{q}_B(\g, \th)}  \\
    &= - \hat{s}_A( \tilde{p}_A(\g)) + \E_{\th | \g} \Brac{  \max \{ 0, v_A(\th) - \tilde{p}_A(\g), v_B(\th) - \tilde{p}_B(\g) \} }  \\
    &\quad +   (\tilde{p}_B(\g) - p_B^\ast(\g)) \E_{\th | \g} \Brac{  \tilde{q}_B(\g, \th)} \\
    &\geq - \hat{s}_A( \hat{p}_A(\g)) + \E_{\th | \g} \Brac{  \max \{ 0, v_A(\th) - \hat{p}_A(\g), v_B(\th) - \tilde{p}_B(\g) \} }  \\
    &\quad +    (\tilde{p}_B(\g) - p_B^\ast(\g)) \E_{\th | \g} \Brac{  \tilde{q}_B(\g, \th)},
\end{aligned}
\end{equation*}
where we have used the facts that $\tilde{q}$ is optimal given $\tilde{p}$ and that $\tilde{p}_A(\g)$ is in $P_A( \tilde{p}_B(\g), \g ; \hat{s}_A)$. Note that $p_B^\ast (\g) > \tilde{p}_B(\g) > \hat{p}_B(\g)$ and $\E_{\th |\g}[\tilde{q}_B (\g, \th)] < \E_{\th |\g} [\hat{q}_B (\g, \th)]$. Thus,
\[
     (\tilde{p}_B(\g) - p_B^\ast(\g)) \E_{\th | \g} \Brac{  \tilde{q}_B(\g, \th)} >  (\tilde{p}_B(\g) - p_B^\ast(\g)) \E_{\th | \g} \Brac{  \hat{q}_B(\g, \th)}.
\]
Applying this fact, we conclude that 
\begin{equation*}
\begin{aligned}
    &I_B ( \g ; \hat{s}_A, \tilde{p}_A, \tilde{q}) - \bar{G}(\g)/ g(\g) \\
    &>- \hat{s}_A (\hat{p}_A (\g)) + \E_{\th | \g} \Brac{  \max \{ 0, v_A(\th) - \hat{p}_A(\g), v_B(\th) - \tilde{p}_B(\g) \} }  \\
    &\quad+   (\tilde{p}_B(\g) - p_B^\ast(\g)) \E_{\th | \g} \Brac{  \hat{q}_B(\g, \th)}\\
    &\geq - \hat{s}_A (\hat{p}_A (\g)) + \E_{\th | \g} \Brac{ (v_A(\th) - \hat{p}_A(\g)) \hat{q}_A(\g, \th) + ( v_B(\th) - p_B^\ast (\g)) \hat{q}_B(\g, \th) } \\
    &= I_B (\g ; \hat{s}_A, \hat{p}_A, \hat{q}) - (\bar{G}(\g)/g(\g)) \E_{\th| \g} \Brac{ \hat{q}_A(\g, \th) + \hat{q}_B(\g, \th)} \\
    &= I_B (\g ; \hat{s}_A, \hat{p}_A, \hat{q}) - \bar{G}(\g)/g(\g),
\end{aligned}
\end{equation*}
contrary to \cref{res:profitable_deviation}. 

\paragraph{Part iii} We prove that $\ubar{p}_B^0 \geq p_B^M(\g_0)$. Suppose, for a contradiction, that $\ubar{p}_B^0 < p_B^M (\g_0)$. To reach a contradiction, we show that firm $B$ can profit by inducing higher strike prices for types just right of $\g_0$. Since $p_B^M$ is continuous, there exists $\d > 0$ such that $\ubar{p}_B^0 < p_B^M(\g)$ for all $\g \in (\g_0, \g_0+ \d)$. Define $\tilde{p}_B \colon \G \to [0,\infty]$ by 
\[
\tilde{p}_B (\g) 
=
\begin{cases}
	p_B^M(\g) &\text{if}~ \g \in [\g_0, \g_0 + \d),\\
	\hat{p}_B(\g) &\text{otherwise}.
\end{cases}
\]
Since $\bar{p}_B^0 \geq p_B^M(\g_0)$, the function $\tilde{p}_B$ is weakly decreasing. Define $\tilde{p}_A \colon \G \to [0,\infty]$ by
\[
\tilde{p}_A (\g) 
=
\begin{cases}
	\inf P_A ( p_B^M(\g), \g;  \hat{s}_A) &\text{if}~ \g \in [\g_0, \g_0 + \d),\\
	\hat{p}_A(\g) &\text{otherwise}.
\end{cases}
\]
as follows. By construction, $\tilde{p}_A$ is a selection from the correspondence $\g \mapsto P_A( \tilde{p}_B( \g), \g ; \hat{s}_A)$. Let $\tilde{q}$ be an optimal product-purchase strategy given $\tilde{p}$.  For each $\g \not\in [\g_0, \g_0 + \d)$, we have $\tilde{p} = \hat{p}$, hence $I_B( \g ; \hat{s}_A, \tilde{p}_A, \tilde{q})  = I_B( \g ; \hat{s}_A, \hat{p}_A, \hat{q})$. For each $\g \in [\g_0, \g_0 + \d)$, we have
\begin{equation*}
\begin{aligned}
	&I_B ( \g; \hat{s}_A, \tilde{p}_A, \tilde{q}) - (\bar{G} ( \g)/ g(\g)) \E_{\th | \g} [ \tilde{q}_A( \g, \th)] \\
	&= - \hat{s}_A ( \tilde{p}_A(\g)) + \E_{\th| \g} \Brac{ (v_A(\th) - \tilde{p}_A(\g)) \tilde{q}_A(\g, \th) + ( v_B(\th) - p_B^M(\g)) \tilde{q}_B(\g, \th)} \\
    &= - \hat{s}_A ( \tilde{p}_A(\g)) + \E_{\th| \g} \Brac{ \max\{ 0, v_A(\th) - \tilde{p}_A(\g), v_B(\th) - p_B^M(\g) \} } \\
    &\geq  - \hat{s}_A ( \hat{p}_A(\g)) + \E_{\th| \g} \Brac{ \max\{ 0, v_A(\th) - \hat{p}_A(\g), v_B(\th) - p_B^M(\g) \} }  \\
    &> - \hat{s}_A ( \hat{p}_A(\g)) + \E_{\th| \g} \Brac{ (v_A(\th) - \hat{p}_A(\g)) \hat{q}_A(\g, \th) + ( v_B(\th) - p_B^M(\g)) \hat{q}_B(\g, \th)} \\
    &=I_B ( \g; \hat{s}_A, \hat{p}_A, \hat{q}) - (\bar{G} ( \g)/ g(\g)) \E_{\th | \g} [ \hat{q}_A( \g, \th)],
\end{aligned}
\end{equation*}
where we are using the fact that $\tilde{q}$ is optimal given $\tilde{p}$ and that $\tilde{p}_A(\g)$ is in $P_A (p_B^M ( \g), \g ; \hat{s}_A)$, together with the fact that $\hat{p}_B(\g) <  p_B^M (\g)$. Since $\tilde{p}(\g) \preceq \hat{p}(\g)$, we have $\E_{\th | \g} [ \tilde{q}_A( \g, \th)] \geq \E_{\th | \g} [ \hat{q}_A( \g, \th)]$, so we conclude $I_B ( \g; \hat{s}_A, \tilde{p}_A, \tilde{q}) > I_B ( \g; \hat{s}_A, \hat{p}_A, \hat{q})$, contrary to \cref{res:profitable_deviation}. 

\paragraph{Proof of Claim 2} We compare the limits of each side as $\g \uparrow \g_0$. First, consider the right side of \eqref{eq:ineq_claim2}. By Claim 1, $p_B^\ast (\g_0) > \ubar{p}_B^0$. Let $\ubar{\d}_B = p_B^\ast (\g_0) - \ubar{p}_B^0$. Since $p_B^\ast$ is continuous, it follows that for all $\g$ sufficiently near $\g_0$, we have $p_B^\ast (\g) > \ubar{p}_B^0$, hence $\inf P_A (p_B^\ast (\g), \g ; \hat{s}_A) \leq \bar{p}_A^0$.  Thus, 
\begin{equation} \label{eq:RHS_Delta}
\begin{aligned}
    &\lim_{\g \uparrow \g_0} \frac{ \bar{G}(\g)}{g(\g)} \P_{\th | \g} \bigl( \max \Set{ v_A(\th) - \inf P_A (p_B^\ast(\g), \g ; \hat{s}_A), v_B(\th) - p_B^\ast (\g) } < 0 \bigr) \\
    &\leq \lim_{\g \uparrow \g_0} \frac{ \bar{G}(\g)}{g(\g)} \P_{\th | \g} \Paren{ \max \Set{ v_A(\th) - \bar{p}_A^0, v_B(\th) - p_B^\ast (\g) } < 0 } \\
    &=  \frac{\bar{G}(\g_0)}{g(\g_0)} \int_{( \ubar{p}_B^0 - \bar{p}_A^0)/2}^{ ( \ubar{p}_B^0 - \bar{p}_A^0)/2 + \ubar{\d}_B} f(\th - \g_0) \de \th,
\end{aligned}
\end{equation}
where the equality holds because $\ubar{p}_B^0 + \bar{p}_A^0 = 2v_0$ (by Claim 1).

Next, consider the left side of \eqref{eq:ineq_claim2}. Let $\bar{\d}_B = \bar{p}_B^0 - p_B^\ast (\g_0)$. Since $\bar{p}_B^0 \geq (5/2) \bar{p}_B > (5/2) p_B^\ast(\g_0)$ and $\ubar{p}_B^0 \geq p_B^M(\g_0) = p_B^\ast(\g_0)/2$ (by Claim 1), 
we have $\bar{\d}_B \geq (3/2) p_B^\ast (\g_0) > 2 \ubar{\d}_B$. Therefore,
\begin{equation} \label{eq:LHS_Delta}
\begin{aligned}
    \lim_{\g \uparrow \g_0} \D (\g) 
    &=\int_{(p_B^\ast( \g_0) - \ubar{p}_A^0)/2}^{ (\bar{p}_B^0- \ubar{p}_A^0)/2} \Brac{ v_B(\th) - p_B^\ast (\g_0)  - (v_A(\th) - \ubar{p}_A^0) } f( \th - \g_0) \de \th \\
    &= \int_{(\bar{p}_B^0- \ubar{p}_A^0)/2 - \bar{\d}_B/2}^{ (\bar{p}_B^0 - \ubar{p}_A^0)/2} \Brac{ 2 \th - (p_B^\ast (\g_0) - \ubar{p}_A^0) } f( \th - \g_0) \de \th \\
    &\geq \int_{(\bar{p}_B^0- \ubar{p}_A^0)/2 - \ubar{\d}_B}^{ (\bar{p}_B^0 - \ubar{p}_A^0)/2} \Brac{ 2 \th - (p_B^\ast (\g_0) - \ubar{p}_A^0) } f( \th - \g_0) \de \th.
\end{aligned}
\end{equation}
For all $\th \geq (\bar{p}_B^0 - \ubar{p}_A^0)/2 - \ubar{\d}_B$, we have 
\begin{equation} \label{eq:pointwise_bound_theta}
\begin{aligned}
    2 \th - (p_B^\ast (\g_0) - \ubar{p}_A^0)  
    &\geq (\bar{p}_B^0 - \ubar{p}_A^0) - 2 \ubar{\d}_B - (p_B^\ast (\g_0) - \ubar{p}_A^0) \\
    &\geq \bar{p}_B^0 - 2 p_B^\ast(\g_0) \\
    &> p_B^\ast(\g_0)/2 \\
    &= \bar{G}( \g_0) / g(\g_0),
\end{aligned}
\end{equation}
where the inequality follows from Claim 1 (since $p_B^M(\g_0)= p_B^\ast (\g_0)/2$). Substituting \eqref{eq:pointwise_bound_theta} into  \eqref{eq:LHS_Delta} gives
\[
  \lim_{\g \uparrow \g_0} \D (\g)  > \frac{ \bar{G}(\g_0)}{g(\g_0)} \int_{(\bar{p}_B^0- \ubar{p}_A^0)/2 - \ubar{\d}_B}^{ (\bar{p}_B^0 - \ubar{p}_A^0)/2} f( \th - \g_0) \de \th.
\]
In view of \eqref{eq:RHS_Delta}, it suffices to check that
\begin{equation} \label{eq:f}
    \int_{(\bar{p}_B^0 - \ubar{p}_A^0)/2 - \ubar{\d}_B}^{(\bar{p}_B^0 - \ubar{p}_A^0)/2} f(\th - \g_0) \de \th \geq \int_{( \ubar{p}_B^0 - \bar{p}_A^0)/2}^{( \ubar{p}_B^0 - \bar{p}_A^0)/2 + \ubar{\d}_B} f(\th - \g_0) \de \th.
\end{equation}
Since $f$ is symmetric and single-peaked about $0$, this holds if the average of the endpoints $(\bar{p}_B^0 - \ubar{p}_A^0)/2 - \g_0$ and $( \ubar{p}_B^0 - \bar{p}_A^0)/2  - \g_0$ is weakly negative. Indeed, by assumption, $v_B ( \g_0) - \ubar{p}_B^0 \geq v_A( \g_0) - \ubar{p}_A^0$, so $\g_0 \geq (\ubar{p}_B^0 - \ubar{p}_A^0)/2$. By Claim 1, we have $\bar{p}_B^0 - \ubar{p}_B^0 =  \bar{p}_A^0 - \ubar{p}_A^0$. Therefore, 
\begin{equation*}
\begin{aligned}
    &(\bar{p}_B^0 - \ubar{p}_A^0)/2 - \g_0 + ( \ubar{p}_B^0 - \bar{p}_A^0)/2 - \g_0   \\
    &\leq (\bar{p}_B^0 - \ubar{p}_A^0)/2 + ( \ubar{p}_B^0 - \bar{p}_A^0)/2 - (\ubar{p}_B^0 - \ubar{p}_A^0) \\
    &= (\bar{p}_B^0 - \ubar{p}_B^0)/2 - (\bar{p}_A^0 - \ubar{p}_A^0)/2 \\
    &=0,
\end{aligned}
\end{equation*}
as desired.

\subsection{Proof of Lemma~\ref{res:schedules_equality}}

We first prove that for each firm $i$, 
\begin{equation} \label{eq:s_equality}
    \hat{s}_i (p_i) - \hat{s}_i (\bar{p}_i) = \int_{p_i}^{\bar{p}_i} Q_i^\ast (p_i') \de p_i', \qquad p_i \in [0, \bar{p}_i].
\end{equation}
By symmetry, it suffices to prove \eqref{eq:s_equality} with $i = B$. Recall that the function $Q_B^\ast$ from \cref{res:equilibrium} is only defined almost everywhere. For this proof, it is convenient to choose a particular representative of this equivalence class. Let $(p_B^\ast)^{-1} \colon [0, \bar{p}_B] \to [\ubar{\g}, \bar{\g}]$ be a generalized inverse of $p_B^\ast$. For this proof, define $Q_B^\ast \colon [0, \bar{p}_B] \to [0,1]$ by 
\[
    Q_B^\ast ( p_B) = Q_B \Paren{  p_B, p_A^\ast( (p_B^\ast)^{-1} (p_B)) | (p_B^\ast)^{-1} (p_B)}.
\]

Fix prices $p_B$ and $p_B'$ with $0 < p_B < p_B' < \bar{p}_B$. Let $\g = (p_B^\ast)^{-1} (p_B)$ and $\g' = (p_B^\ast)^{-1}(p_B')$. Since $p_B^\ast$ is continuous, it follows that $p_B^\ast(\g) = p_B$ and $p_B^\ast(\g') = p_B'$. By incentive compatibility for type $\g$, we have
\begin{equation*}
\begin{aligned}
     \hat{s}_B( p_B) - \hat{s}_B(p_B') 
    &\leq   
    \E_{\th | \g} \Brac{ \max \{ 0, v_A( \th) - p_A^\ast(\g), v_B( \th) - p_B \}} \\
    &\quad - \E_{\th | \g} \Brac{ \max \{ 0, v_A( \th) - p_A^\ast(\g), v_B( \th) - p_B' \}} \\
    &= \int_{p_B}^{p_B'} Q_B (  p_B'',p_A^\ast(\g)| \g) \de p_B'' \\
    &\leq Q_B ( p_B, p_A^\ast (\g)| \g) ( p_B' - p_B) \\
    &= Q_B^\ast ( p_B) ( p_B' - p_B),
\end{aligned}
\end{equation*}
where the second inequality holds because the demand $Q_B$ is weakly decreasing in the price of product $B$. A symmetric argument, using incentive compatibility for type $\g'$, gives
\[
 \hat{s}_B( p_B) - \hat{s}_B(p_B') \geq Q_B^\ast (p_B') ( p_B' - p_B). 
\]
So each increment of $\hat{s}_B$ is sandwiched between the upper and lower Riemann rectangles of $Q_B^\ast$. The function $Q_B^\ast$ is weakly decreasing and hence Riemann integrable. By summing these increments over arbitrarily fine partitions, we conclude that for all prices $p_B$ and $p_B'$ satisfying $0 < p_B < p_B' < \bar{p}_B$, we have 
\begin{equation} \label{eq:integral_equality}
    \hat{s}_B (p_B) - \hat{s}_B (p_B') = \int_{p_B}^{p_B'} Q_B^\ast (p_B'') \de p_B''.
\end{equation}

By assumption, $\hat{s}_B$ is lower semicontinuous, so $\hat{s}_B( 0) \leq \lim_{p_B \downarrow 0} \hat{s}_B( p_B)$ and $\hat{s}_B( \bar{p}_B) \leq \lim_{p_B \uparrow \bar{p}_B} \hat{s}_B( p_B)$. By incentive compatibility, we must have $\hat{s}_B( 0) \geq \lim_{p_B \downarrow 0} \hat{s}_B( p_B)$ and $\hat{s}_B( \bar{p}_B) \geq \lim_{p_B \uparrow \bar{p}_B} \hat{s}_B( p_B)$. Therefore, $\hat{s}_B$ is continuous on $[0, \bar{p}_B]$. It follows that \eqref{eq:integral_equality} extends to the endpoints, so \eqref{eq:s_equality} follows. 

Having proven  \eqref{eq:s_equality}, it remains to prove that for each firm $i$, we have $\hat{s}_i( \bar{p}_i) = s_i^\ast (\bar{p}_i)$. Without loss of generality, we prove this result with $i = B$. First, we prove that $\hat{s}_B( \bar{p}_B) \leq s_B^\ast (\bar{p}_B)$. For each type $\g \in (\ubar{\g}, \bar{\g})$,  the vector $(\hat{p}_A(\g), \hat{p}_B(\g)) = (p_A^\ast (\g), p_B^\ast(\g))$ is in $P(\g; \hat{s}_A, \hat{s}_B)$, so
\begin{equation*}
\begin{aligned}
    &- \hat{s}_A (\hat{p}_A(\g)) - \hat{s}_B ( \hat{p}_B(\g)) + \E_{\th| \g} \Brac{ \max\{ 0, v_A(\th) - \hat{p}_A(\g), v_B(\th) - \hat{p}_B(\g) \}}\\
    &\geq - \hat{s}_A(\hat{p}_A(\g)) + \E_{\th | \g} \Brac{ \max\{ 0, v_A(\th) - \hat{p}_A(\g) \}}.
\end{aligned}
\end{equation*}
Pass to the limit as $\g \to \ubar{\g}$. 
Since $\lim_{\g \to \ubar{\g}} \hat{p}_A(\g)  = 0$ and $\lim_{\g \to \ubar{\g}} \hat{p}_B(\g)  = \bar{p}_B$,  we obtain
\begin{equation*}
\begin{aligned}
   \hat{s}_B( \bar{p}_B)  &\leq \E_{\th| \ubar{\g}} \Brac{ \max\{ 0, v_A(\th), v_B(\th) - \bar{p}_B \}} -    \E_{\th | \ubar{\g}} \Brac{ \max\{ 0, v_A(\th) \}} \\
   &=\E_{\th | \ubar{\g}} \Brac{ \Paren{ v_B (\th) - \bar{p}_B -  v_A(\th)_+}_+} \\
   &= s_B^\ast ( \bar{p}_B).
\end{aligned}
\end{equation*}

Next, we prove that $\hat{s}_B( \bar{p}_B) \geq s_B^\ast ( \bar{p}_B)$.
Suppose, for a contradiction, that $\hat{s}_B( \bar{p}_B) < s_B^\ast ( \bar{p}_B)$. We claim that firm $B$ has a profitable deviation to the schedule $\tilde{s}_B = \hat{s}_B + \e$, where $\e = s_B^\ast ( \bar{p}_B) - \hat{s}_B( \bar{p}_B)$.\footnote{This redefinition is over the domain $[0, \infty)$; the null contract remains the same.} By \cref{res:revenue}, firm $B$'s revenue is independent of which best response the consumer plays, so it suffices to show that for each type $\g$ in $(\ubar{\g}, \bar{\g})$, the vector $(\hat{p}_A(\g), \hat{p}_B (\g))$ is in $P ( \g; \hat{s}_A, \tilde{s}_B)$. Fix $\g \in (\ubar{\g}, \bar{\g})$. Since $(\hat{p}_A(\g), \hat{p}_B (\g))$ is in $P ( \g; \hat{s}_A, \hat{s}_B)$ and $\tilde{s}_B$ is a translation of $\hat{s}_B$, it suffices to consider deviations to price pairs $(p_A', \infty)$. The price pair $(\hat{p}_A(\g), \hat{p}_B(\g)) = (p_A^\ast (\g), p_B^\ast(\g))$ is in $P(\g; s_A^\ast, s_B^\ast)$ and, by \eqref{eq:s_equality}, $\tilde{s}_B( \hat{p}_B(\g)) = s_B^\ast ( \hat{p}_B(\g))$, so we have
\begin{equation*}
\begin{aligned}
       u( \hat{p}(\g), \g; s_A^\ast, \tilde{s}_B ) 
       &= u( \hat{p}(\g), \g; s_A^\ast, s_B^\ast )\\
       &\geq \max_{p_A' \in [0, \infty]} \Set { - s_A^\ast ( p_A') + \E_{\th | \g} [ (v_A(\th) - p_A')_+]} \\
       &\geq \max_{p_A' \in [0, \hat{p}_A(\g)]} \Set { - s_A^\ast ( p_A') + \E_{\th | \g} [ (v_A(\th) - p_A')_+]}.
\end{aligned}
\end{equation*}
Subtracting $\hat{s}_A( \bar{p}_A) - s_A^\ast ( \bar{p}_A)$ from each side gives
\begin{equation*}
\begin{aligned}
       u( \hat{p}(\g), \g; \hat{s}_A, \tilde{s}_B )  
       &\geq \max_{p_A' \in [0, \hat{p}_A(\g)]} \Set { - \hat{s}_A ( p_A') + \E_{\th | \g} [ (v_A(\th) - p_A')_+]} \\
       &= \max_{p_A' \in [0, \infty]} \Set { - \hat{s}_A ( p_A') + \E_{\th | \g} [ (v_A(\th) - p_A')_+]} \\
       &= \max_{p_A' \in [0, \infty]} u (p_A', \infty, \g; \hat{s}_A, \tilde{s}_B), 
\end{aligned}
\end{equation*}
where the first equality follows from \cref{res:BR_partial_monotone}.\ref{it:partial_weak_monotonicity} because $(\hat{p}_A(\g), \hat{p}_B(\g))$ is in $P(\g; \hat{s}_A, \hat{s}_B)$. 
   
\subsection{Proof of Proposition~\ref{res:subscription_schedules}}

Without loss of generality, we prove the result for firm $ i = B$. The key observation is that, under our assumption that $v_0 \geq (7/2) \max_{\g' \in \G} (1 / g(\g'))$, duopoly demand for firm $B$ is lower than monopoly demand for firm $B$ in the following strong sense. Recall that $\bar{p}_B^M = 1 /g (\ubar{\g})$ and $\bar{p}_B = 2 / g(\ubar{\g})$. 

\begin{lem}[Demand comparison] \label{res:demand_comparison}
For all types $\g$, the following hold. 
\begin{enumerate}[label = (\roman*), ref = \roman*]
    \item \label{it:Q_high_price} $Q_B ( p_B^\ast (\g), p_A^\ast(\g) | \g) \leq Q_B^M ( \bar{p}_B | \ubar{\g})$ if $p_B^\ast (\g) \geq \bar{p}_B^M$;
    \item \label{it:Q_general} $Q_B ( p_B^\ast (\g), p_A^\ast(\g) | \g) \leq Q_B^M ( \bar{p}_B^M | \ubar{\g})$.
\end{enumerate}
\end{lem}

Using \cref{res:demand_comparison}, we complete the proof. First, we check that $s_B^\ast ( \bar{p}_B^M) < s_B^M ( \bar{p}_B^M)$, that is, 
\begin{equation} \label{eq:s_bound}
   \E_{\th | \ubar{\g}} \Brac{ \Paren{ v_B (\th) - \bar{p}_B -  v_{A}(\th)_+}_+} + \int_{\bar{p}_B^M}^{\bar{p}_B} Q_B^\ast (p_B') \de p_B' < \E_{\th | \ubar{\g}} [ (v_B(\th) - \bar{p}_B^M)_+].
\end{equation}
We have
\begin{equation*}
\begin{aligned}
&\E_{\th | \ubar{\g}} [ (v_B(\th) - \bar{p}_B^M)_+] -  \E_{\th | \ubar{\g}} \Brac{ \Paren{ v_B (\th) - \bar{p}_B -  v_{A}(\th)_+}_+}  \\
&> \E_{\th | \ubar{\g}} [ (v_B(\th) - \bar{p}_B^M)_+] -  \E_{\th | \ubar{\g}} \Brac{ \Paren{ v_B (\th) - \bar{p}_B}_+}  \\
&\geq \E_{\th | \ubar{\g}} \Brac{  (\bar{p}_B - \bar{p}_B^M) [ v_B(\th) \geq \bar{p}_B]} \\
&= 
(\bar{p}_B - \bar{p}_B^M) Q_B^M ( \bar{p}_B | \ubar{\g}).
\end{aligned}
\end{equation*}
Thus, \eqref{eq:s_bound} follows from \cref{res:demand_comparison}.\ref{it:Q_high_price}.  Therefore, for all $p_B$ in $[0, \bar{p}_B^M]$, we have
\[
    s_B^\ast (p_B) = s_B^\ast ( \bar{p}_B^M) + \int_{p_B}^{\bar{p}_B^M} Q_B^\ast (p_B') \de p_B' 
    < s_B^M ( \bar{p}_B^M) + \int_{p_B}^{\bar{p}_B^M} \hat{Q}_B^M (p_B') \de p_B' 
    = s_B^M (p_B),
\]
where the middle inequality holds because $s_B^\ast (\bar{p}_B^M) < s_B^M (\bar{p}_B^M)$, as proven above, and the inequality between the integrals follows from  \cref{res:demand_comparison}.\ref{it:Q_general} since $Q_B^M ( \bar{p}_B^M | \ubar{\g}) \leq Q_B^M ( p_B' | \g)$ for all $p_B' \in [0, \bar{p}_B^M]$ and all types $\g$. 

\subsection{Proof of Lemma~\ref{res:demand_comparison}}

First, we calculate the demands. For any type $\g$, we have
\begin{equation*}
\begin{aligned}
    Q_B ( p_B^\ast (\g), p_A^\ast(\g) | \g)  
    &= \P_{\e}  \Paren{ v_0 + \g + \e - p_B^\ast (\g) \geq v_0 - \g - \e - p_A^\ast (\g) } \\
    &= \P_{\e}  \Paren{ \e \geq (p_B^\ast(\g) - p_A^\ast (\g))/2 - \g }.
\end{aligned}
\end{equation*}
And for any price $p_B$, we have
\begin{equation*}
\begin{aligned}
    Q_B^M ( p_B | \ubar{\g})  
    &= \P_{\e}  \Paren{ v_0 + \ubar{\g} + \e - p_B \geq 0} \\
    &= \P_{\e}  \Paren{ \e \geq p_B - \ubar{\g} - v_0}.
\end{aligned}
\end{equation*}
Thus, $Q_B ( p_B^\ast (\g), p_A^\ast(\g) | \g)  \leq  Q_B^M ( p_B | \ubar{\g})$ if and only if 
\[
    (p_B^\ast(\g) - p_A^\ast (\g))/2 - \g \geq p_B - \ubar{\g} - v_0,
\]
or equivalently, 
\begin{equation} \label{eq:v0_bound}
 \begin{aligned}
      v_0    
      &\geq (p_A^\ast(\g) - p_B^\ast (\g))/2  + p_B + \g -  \ubar{\g} \\
   &= 1/g(\g) - p_B^\ast (\g) + p_B + \g -  \ubar{\g}.
 \end{aligned} 
\end{equation}
 where the equality uses the fact that $p_A^\ast ( \g) + p_B^\ast (\g) = 2 / g(\g)$. Note that
 \[
    (\bar{\g} - \ubar{\g}) \min_{\g' \in \G} g(\g') \leq \int_{\ubar{\g}}^{\bar{\g}} g(\g) \de \g = 1,
\]
so $\bar{\g} - \ubar{\g} \leq \max_{\g' \in \G} (1 / g(\g'))$. We will use this fact below. 

To prove \eqref{it:Q_high_price}, put $p_B = \bar{p}_B$ and note that if $p_B^\ast (\g) \geq \bar{p}_B^M$, we have
\begin{equation*}
\begin{aligned}
1/g(\g) - p_B^\ast (\g) + \bar{p}_B + \g -  \ubar{\g} 
&\leq 
1/g(\g) - \bar{p}_B^M + \bar{p}_B + \g -  \ubar{\g} \\
&= 1/g(\g)  + 1 / g(\ubar{\g}) + (\g -  \ubar{\g}) \\
&\leq 3 \max_{\g' \in \G} (1 / g(\g')) \\
&\leq v_0,
\end{aligned}
\end{equation*}
so $Q_B ( p_B^\ast (\g), p_A^\ast (\g) | \g)  \leq  Q_B^M ( \bar{p}_B | \ubar{\g})$. 
To prove \eqref{it:Q_general}, put $p_B = \bar{p}_B^M$ and note that 
\begin{equation*}
\begin{aligned}
1/g(\g) - p_B^\ast (\g) + \bar{p}_B^M + \g -  \ubar{\g} 
&\leq 
1/g(\g) + \bar{p}_B^M + \g -  \ubar{\g}  \\
&= 1/g(\g) + 1 / g(\ubar{\g}) + (\g -  \ubar{\g})  \\
&\leq 3 \max_{\g' \in \G} (1 / g(\g'))\\
&\leq v_0,
\end{aligned}
\end{equation*}
so $Q_B ( p_B^\ast (\g), p_A^\ast(\g) | \g)  \leq  Q_B^M ( \bar{p}_B^M | \ubar{\g})$.

\subsection{Proof of Proposition~\ref{res:Hotelling}}




Since $f$ and $g$ are both log-concave (by \cref{as:regularity}) and log-concavity is preserved by convolution \citep[Proposition 3.5, p.~60]{SaumardWellner2014}, it follows that $h$ is log-concave. Therefore, $\log H$ and $\log (1 - H)$ are both concave by Pr\'{e}kopa's theorem \citep[Theorem 3.3, p.~58]{SaumardWellner2014}.\footnote{Formally, apply Pr\'{e}kopa's theorem to conclude that the function $q(y) = \int_{-\infty}^{\infty} p(x,y) \de x$ is log-concave with the log-concave functions $p(x,y) = h(x) [ x \leq y]$ and $p(x,y) = h(x) [ x \geq y]$; here, the Iverson bracket $[ \cdot ]$ denotes the indicator function for the predicate it encloses.}
We will use these facts below. 

First, we check that the equation \eqref{eq:spot_pricing_critical} has a unique solution. The difference of the two sides, as a function of $\th$, is
\[
  \D(\th) =  \th - \frac{1 - 2 H(\th)}{h(\th)} = \th - \frac{1 - H(\th)}{h(\th)} + \frac{H(\th)}{h(\th)}.
\]
Since $(1 - H) /h$ is weakly decreasing and $H/h$ is weakly increasing, the function $\D$ is strictly increasing, continuous, and unbounded above and below. Therefore, $\D$ has a unique zero.

In the spot-pricing game, the demand functions are given by 
\begin{equation*}
\begin{aligned}
    Q_A ( p_A, p_B) &= \P ( v_0 - \th - p_A \geq (v_0 + \th - p_B)_+)\\
    &= H \Paren{ \min \{ (p_B - p_A)/2 , v_0 - p_A\}  },
\end{aligned}
\end{equation*}
and
\begin{equation*}
\begin{aligned}
    Q_B ( p_B, p_A) 
    &= \P ( v_0 + \th - p_B \geq (v_0 - \th - p_A)_+)\\
    &= 1  - H \Paren{  \max \{ (p_B -p_A)/2 , p_B - v_0 \} }.
\end{aligned}
\end{equation*}
It can be shown that for each firm $i$ and each fixed price $p_{-i}$, the map $p_i \mapsto \log Q_i (p_i, p_{-i})$ is concave.\footnote{This follows from the rules for the concavity of the composition of functions \citet[p.~84, equation 3.11]{BoydVandenberghe2004}. For $i = A$, the outer function $\log H$ is concave and weakly increasing, and $\min \{ (p_B - p_A)/2 , v_0 - p_A\}$ is concave in $p_A$. For $i = B$, the outer function $\log (1 - H)$ is concave and weakly decreasing, and  $\max \{ (p_B -p_A)/2 , p_B - v_0 \}$ is convex in $p_B$.} Each firm must earn strictly positive revenue in equilibrium, so we can restrict attention to prices in $(0, \infty)$. For each firm $i$, let $\Pi_i (p_i, p_{-i}) = p_i Q_i(p_i, p_{-i})$. The function
\[
    \log \Pi_i (p_i, p_{-i}) = \log p_i + \log Q_i(p_i, p_{-i})
\]
is concave in $p_i$ since it is a sum of concave functions. Therefore, a price vector $p$ is an equilibrium if and only if
\begin{equation} \label{eq:spot_FOC}
  0 \in \pd_i \log \Pi_i (p_i, p_{-i}), \qquad i = A,B
\end{equation}
where $\pd_i$ denotes the $i$-th partial superdifferential.

Assuming $v_0 > 1 /h(\th^\ast)$, we look for solutions of \eqref{eq:spot_FOC} in two different cases. Then we extend our argument to the edge case that $v_0 = 1 / h(\th^\ast)$. 

\paragraph{Case 1} We look for a solution $p$ of \eqref{eq:spot_FOC} satisfying $p_A + p_B < 2 v_0$.  In this case, condition \eqref{eq:spot_FOC} reduces to
\begin{equation*}
\begin{aligned}
     H( (p_B - p_A )/2) &= (1/2) p_A \, h ( (p_B - p_A)/2), \\    
     1 - H ( (p_B  - p_A)/2) &= (1/2) p_B \, h ( (p_B - p_A)/2). 
\end{aligned}
\end{equation*}
Subtracting the first equation from the second, we conclude that $ (p_B - p_A)/2 = \th^\ast$. Therefore, 
\[
    p_A = 2 \frac{ H( \th^\ast)}{h(\th^\ast)}, 
    \qquad
    p_B = 2 \frac{1 -  H( \th^\ast)}{h(\th^\ast)}.
\]
Since $2 / h(\th^\ast) < 2 v_0$, we have $p_A + p_B < 2 v_0$, so it is easily verified that \eqref{eq:spot_FOC} is satisfied. 

\paragraph{Case 2} We show that the condition \eqref{eq:spot_FOC} does not have any solution $(p_A, p_B)$ satisfying $p_A + p_B \geq 2 v_0$.  If $p_A + p_B \geq 2 v_0$, then \eqref{eq:spot_FOC} implies 
\begin{equation*}
\begin{aligned}
    H( v_0 - p_A) &\geq (1/2) p_A h ( v_0 - p_A), \\
    1 - H( p_B - v_0) &\geq (1/2) p_B h (p_B - v_0).
\end{aligned}
\end{equation*}
Let $\th_A = v_0 - p_A$ and $\th_B = p_B - v_0$. Since $p_A +p_B \ge 2v_0$, we have $\th_A \leq \th_B$, so we must have either $\th_A \leq \th^\ast$ or $\th^\ast \leq \th_B$. Without loss, suppose $\th^\ast \leq \th_B$. Thus, 
\[
  2 \frac{1 - H( \th_B)}{ h(\th_B)} \geq v_0 + \th_B,
\]
hence 
\begin{equation} \label{eq:spot_pricing_contradiction}
    v_0 \leq 2 \frac{1 - H(\th_B)}{h(\th_B)} - \th_B \leq 2 \frac{1 - H(\th^\ast)}{h(\th^\ast)} - \th^\ast = \frac{1}{h(\th^\ast)},
\end{equation}
where we have used the fact that $(1 - H)/h$ is weakly decreasing. Thus, we obtain a contradiction with the assumption that $v_0 > 1/ h (\th^\ast)$.

\paragraph{Edge case} Finally, consider the edge case  in which $v_0 = 1/ h (\th^\ast)$. Now, the argument in case 1 shows that \eqref{eq:spot_FOC} has no solution satisfying $p_A + p_B < 2v_0$. The argument in case 2 shows that for any solution $p$ of \eqref{eq:spot_FOC} satisfying $p_A + p_B \geq 2v_0$, we must have $\th_A = \th^\ast = \th_B$. Therefore, 
\[
    p_A = 2 \frac{ H( \th^\ast)}{h(\th^\ast)}, 
    \qquad
    p_B = 2 \frac{1 -  H( \th^\ast)}{h(\th^\ast)}.
\]
Since $p_A + p_B = 2v_0$, \eqref{eq:spot_FOC} is satisfied.

\subsection{Proof of Proposition~\ref{res:exclusive}} \label{sec:exclusive_proof}

Before turning to the main proof, we begin with some preliminaries. 

First, we check that condition \eqref{eq:exclusive_critical_position} has a unique solution. We introduce notation for each side of \eqref{eq:exclusive_critical_position}. Let
\begin{equation} \label{eq:defining_each_side}
\begin{aligned}
   \D_B(\g)  &= \E_{\th | \g} \Brac{( v_B(\th) - p^M_B(\g))_+} - p_{B}^M (\g) Q_{A}^M (  p_A^M (\g) | \g), \\
      \D_A(\g)  &=  \E_{\th | \g} \Brac{ ( v_A(\th) - p^M_A( \g))_+}  - p_{A}^M (\g) Q_{B}^M (  p_B^M (\g) | \g).
\end{aligned}
\end{equation}
Write $\D(\g) = \D_B(\g) - \D_A(\g)$. The function $\D_B$ is continuous and strictly increasing because (i) the conditional distribution of $v_B(\th)$ is strictly increasing in $\g$, with respect to first-order stochastic dominance; (ii) the function $p_B^M$ is weakly decreasing; (iii) the function $Q_A^M$ is strictly decreasing 
in the type and in the price. Symmetrically, $\D_A$ is continuous and strictly decreasing. Thus, $\D$ is continuous and strictly increasing. Our assumption that $\ubar{\g} < 0 < \bar{\g}$ ensures that
\begin{equation} \label{eq:weakened_boundary}
\begin{aligned}
   \lim_{\g \downarrow \ubar{\g}} \D( \g) 
   &= \E_{\th | \ubar{\g}} [ (v_B(\th) - \bar{p}_B^M)_+] - \E_{\th | \ubar{\g}} [ v_A(\th)_+] - \bar{p}_B^M Q_A^M (0 | \ubar{\g}) <0, \\
   \lim_{\g \uparrow \bar{\g}} \D(\g) 
   &= \E_{ \th | \bar{\g}} [ v_B(\th)_+] - \E_{\th | \bar{\g}} [ (v_A(\th) - \bar{p}_A^M)_+] + \bar{p}_A^M Q_B^M (0 | \bar{\g})  >  0.
\end{aligned}
\end{equation}
Therefore, \eqref{eq:exclusive_critical_position} has a unique solution.\footnote{In fact, the assumption $\ubar{\g} < 0 < \bar{\g}$ can be relaxed to the condition that both inequalities in \eqref{eq:weakened_boundary} hold.}

Next, we analyze the structure of the consumer's best response. Given a subscription-schedule pair $(s_A, s_B)$, each type $\g$ chooses strike prices $p_A(\g)$ and $p_B(\g)$ subject to the exclusivity constraint that at most one price is finite. With this exclusivity constraint, a consumer best response can be defined as in the main model. 

\begin{lem}[Structure of best response] \label{res:structure_exclusivity} Let $(s_A, s_B)$ be a pair of exclusive subscription schedules. For each consumer best response $(p, q)$ to $(s_A, s_B)$, there exists a unique pair $(\bar{\g}_A, \ubar{\g}_B)$ satisfying $\ubar{\g} \leq \bar{\g}_A \leq \ubar{\g}_B \leq \bar{\g}$ such that the following hold. 
\begin{enumerate}
    \item For each type $\g \in [\ubar{\g}, \bar{\g}_A)$, we have $p_A(\g) < \infty$ and $p_B(\g) = \infty$. 
    \item For each type $\g \in (\bar{\g}_A, \ubar{\g}_B)$, we have $p_A(\g) = \infty$ and $p_B(\g) = \infty$. 
    \item For each type $\g \in (\ubar{\g}_B, \bar{\g}]$,  we have $p_B(\g) < \infty$ and $p_A(\g) = \infty$. 
\end{enumerate}
\end{lem}

With these preliminaries established, we turn to the main proof. Denote the strategy profile in \cref{res:exclusive} by $(s_A^{\mathrm{E}}, s_B^{\mathrm{E}}, p^{\mathrm{E}}, q^{\mathrm{E}})$.\footnote{Technically, with our null-contract convention,
\[
p^{\mathrm{E}}( \g) = 
\begin{cases} 
    ( p_A^M(\g), \infty) &\text{if}~ \g < \g^{\dagger}, \\
    (\infty, p_B^M(\g)) &\text{if}~\g \geq \g^{\dagger}. 
\end{cases}
\]} The proof has three parts.  
To show that $(s_A^{\mathrm{E}}, s_B^{\mathrm{E}}, p^{\mathrm{E}}, q^{\mathrm{E}})$ is an equilibrium, we check that the consumer is playing a best response, and then we check that each firm is playing a best response. Finally, we show that $(s_A^{\mathrm{E}}, s_B^{\mathrm{E}}, p^{\mathrm{E}}, q^{\mathrm{E}})$ is the essentially unique Pareto-dominant equilibrium. 



\paragraph{Equilibrium verification: Consumer best response} We check that the consumer strategy $(p^{\mathrm{E}}, q^{\mathrm{E}})$ is a best response to the exclusive schedule pair $(s_A^{\mathrm{E}}, s_B^{\mathrm{E}})$. 

It is clear that the product-purchase strategy $q^{\mathrm{E}}$ is optimal, given the contract-selection strategy $p^{\mathrm{E}}$. We check that $p^{\mathrm{E}}$ is optimal in the first period. 

First, we confirm that every type finds it optimal to enter some contract. It suffices to check that the interim utility of type $\g^{\dagger}$ is nonnegative, for then types to the left of $\g^{\dagger}$ get weakly higher utility at firm $A$, and types to the right of $\g^{\dagger}$ get weakly higher utility at firm $B$. The interim utility of type $\g^{\dagger}$ equals the common value of the two sides of \eqref{eq:exclusive_critical_position}. Writing $p^{\dagger}_A = p_A^M(\g^{\dagger})$ and $p^{\dagger}_B = p_B^M ( \g^{\dagger})$, we have
\begin{equation} \label{eq:hatp_sum}
    p^{\dagger}_A + p^{\dagger}_B = \frac{G(\g^{\dagger})}{g(\g^{\dagger})} + \frac{1 - G(\g^{\dagger})}{g(\g^{\dagger})} = \frac{1}{g(\g^{\dagger})}.
\end{equation}
We use the notation $[\cdot]$ to denote the indicator function for the predicate it encloses. Summing the two sides of \eqref{eq:exclusive_critical_position} and applying \eqref{eq:hatp_sum}, we claim that
\begin{equation*}
\begin{aligned}
   &\E_{\th| \g^{\dagger}} \bigl[ v_A( \th) [ v_A(\th) \geq p^{\dagger}_A] \bigr]  - (1/ g(\g^{\dagger})) Q_A^M (p^{\dagger}_A | \g^{\dagger}) \\
   &\quad + \E_{\th| \g^{\dagger}} \bigl[ v_B( \th) [ v_B(\th) \geq p^{\dagger}_B] \bigr] -  (1/ g(\g^{\dagger})) Q_B^M (p^{\dagger}_B | \g^{\dagger})  \\
   &= \Paren{ \E_{\th| \g^{\dagger}} \bigl[ v_A( \th)| v_A(\th) \geq p^{\dagger}_A \bigr] - 1/ g(\g^{\dagger}) } Q_A^M (p^{\dagger}_A | \g^{\dagger}) \\
   &\quad + \Paren{ \E_{\th| \g^{\dagger}} \bigl[ v_B( \th)| v_B(\th) \geq p^{\dagger}_B \bigr] - 1 /g(\g^{\dagger}) } Q_B^M (p^{\dagger}_B | \g^{\dagger}) \\
   &\geq \Paren{ \E_{\th| \g^{\dagger}} \bigl[ v_A( \th)\bigr] - 1/ g(\g^{\dagger}) } Q_A^M (p^{\dagger}_A | \g^{\dagger}) \\
   &\quad + \Paren{ \E_{\th| \g^{\dagger}} \bigl[ v_B( \th)\bigr] - 1 /g(\g^{\dagger}) } Q_B^M (p^{\dagger}_B | \g^{\dagger}) \\
   & \geq \Paren{ v_0 - |\g^{\dagger}| - 1/ g(\g^{\dagger}) } Q_A^M (p^{\dagger}_A | \g^{\dagger}) \\
   &\quad + \Paren{ v_0 - |\g^{\dagger}| - 1 /g(\g^{\dagger}) } Q_B^M (p^{\dagger}_B | \g^{\dagger}) \\
   &\geq 0,
\end{aligned}
\end{equation*}
where the last inequality uses our assumption that $v_0 \geq 1 / g(\g^{\dagger}) + |\g^{\dagger}|$.

By following the argument for global incentive compatibility in the proof of  Proposition~\ref{res:monopoly_benchmark} (\cref{sec:proof_monopoly_benchmark}), we conclude that (a) for each type $\g \leq \g^{\dagger}$, the strike price $p_A^M (\g)$ is the optimal choice for type $\g$ from $s_A^{\mathrm{E}}$, and (b) for each type $\g \geq \g^{\dagger}$, the strike price $p_B^M(\g)$ is the optimal choice for type $\g$ from $s_B^{\mathrm{E}}$. By \eqref{eq:exclusive_critical_position}, type $\g^{\dagger}$ is indifferent between having each firm as his exclusive supplier, so it follows from \cref{res:structure_exclusivity} that $p^{\mathrm{E}}$ is optimal for the consumer.

\paragraph{Equilibrium verification: Firm best response} We prove a stronger result, namely that neither firm can profit by deviating to a direct stochastic mechanism and then selecting the consumer's best response. By symmetry, it suffices to check that firm $B$ does not have such a profitable deviation. 

For each type $\g$, let 
\[
  U_A^{\mathrm{E}} (\g) = \max_{p_A' \in [0, \infty]}  \Set{ - s_A^E (p_A') + \E_{\th|\g} \Brac{ (v_A(\th) - p_A')_+}}.
\]
By \cref{res:structure_exclusivity}, firm $B$'s problem reduces to choosing a cutoff type $\ubar{\g}_B$ together with an allocation rule $q_B \colon [\ubar{\g}_B, \bar{\g}] \times \Th \to [0,1]$ and a transfer rule $t_B \colon [\ubar{\g}_B, \bar{\g}] \times \Th \to \R$ to solve
\begin{equation} \label{eq:firm_B_exclusive_monopoly}
\begin{aligned}
    &\maz && \int_{\ubar{\g}_B}^{\bar{\g}} \E_{\th| \g} \Brac { t_B (\g, \th)} g(\g) \de \g, \\
    &\text{subject to} && \ubar{\g}_B \in \G,\\
    &&&\th \in \argmax_{\th' \in \Th}  \Set{ v_B (\th) q_B ( \g, \th') - t_B(\g, \th')}, \quad (\g, \th) \in [\ubar{\g}_B, \bar{\g}] \times \Th \\
    &&& \g \in \argmax_{\g'}\, \E_{\th | \g} \Brac{ \max_{\th' \in \Th}  \Set {v_B (\th) q_B ( \g', \th') - t_B(\g', \th') }}, \quad \g \in [\ubar{\g}_B, \bar{\g}] \\
    &&& \E_{\th| \ubar{\g}_B} \Brac{ v_B (\th) q_B ( \ubar{\g}_B, \th) - t_B(\ubar{\g}_B, \th)} = \max\{ U_A^{\mathrm{E}}( \ubar{\g}_B), 0\}.
\end{aligned}
\end{equation}
The last constraint ensures that type $\ubar{\g}_B$ is indifferent between choosing firm $B$  and his outside option of choosing firm $A$ or neither firm. We impose equality to ensure that the set of types choosing firm $B$ is $[\ubar{\g}_B, \bar{\g}]$.\footnote{Technically, if $\ubar{\g}_B = \ubar{\g}$, then it is feasible for type $\ubar{\g}_B$ to strictly prefer firm $B$, but in this case, it is still optimal to induce equality in this participation constraint.} 

We solve firm $B$'s problem as a nested maximization problem, where $\ubar{\g}_B$ is chosen in the outer problem and the function $(q_B, t_B) \colon [\ubar{\g}_B, \bar{\g}] \times \Th \to [0,1] \times \R$ is chosen in the inner problem. The inner problem takes essentially the same form as the monopoly problem in \eqref{eq:firm_B_monopoly}. Following the steps in the proof of \cref{res:monopoly_benchmark},  we see that the relaxed inner problem reduces to choosing $q_B$ to maximize the objective 
\[
 \int_{\ubar{\g}_B}^{\bar{\g}} \E_{\th | \g} \Brac{ \Paren{ v_B(\th) - \frac{1 - G(\g)}{g(\g)}} q_B( \g, \th) } g(\g) \de \g -  \max\{ U_A^{\mathrm{E}}(\ubar{\g}_B), 0\} (1 - G (\ubar{\g}_B)).
\]
The expression inside the integral is pointwise maximized by 
\[
    q_B(\g, \th) = [ v_B( \th) \geq (1 - G(\g))/g(\g)].
\]
Therefore, the value of the inner problem, $V_B(\ubar{\g}_B)$, is given by
\[
    V_B(\ubar{\g}_B) = \int_{\ubar{\g}_B}^{\bar{\g}}  \E_{\th | \g} \Brac{ (v_B(\th) - p_B^M(\g) )_+} g(\g) \de \g  - \max \{ U_A^{\mathrm{E}}(\ubar{\g}_B), 0\} (1 - G (\ubar{\g}_B)).
\]

From the analysis of the monopoly problem in \eqref{eq:firm_B_monopoly}, it can be verified that the maximum of $V_B$ is achieved over the interval where $U_A^{\mathrm{E}}$ is nonnegative.\footnote{\label{ft:coverage}Otherwise, strictly reduce $\ubar{\g}_B$, and note that the previous solution is feasible in the new problem.} Therefore, we can analyze $V_B$ over this interval. Whenever $U_A^{\mathrm{E}}$ is nonnegative, its derivative is  $-\bar{Q}_A$, where the function $\bar{Q}_A \colon \G \to [0,1]$ is given by
\[
    \bar{Q}_A (\g) = 
    \begin{cases}
        Q_A^M ( p_A^M(\g) | \g) &\text{if}~\g \leq \g^{\dagger}, \\
        Q_A^M ( p_A^M (\g^{\dagger})| \g) &\text{if}~\g \geq \g^{\dagger}.
    \end{cases}
\]
In this case, differentiating $V_B$ gives
\[
    V_B' (\ubar{\g}_B) = -   \E_{\th | \ubar{\g}_B} \Brac{ (v_B(\th) - p_B^M(\ubar{\g}_B) )_+}g(\ubar{\g}_B)  + \bar{Q}_A (\ubar{\g}_B) (1 - G( \ubar{\g}_B)) +  U_A^{\mathrm{E}}(\ubar{\g}_B) g (\ubar{\g}_B).
\]
We check that the derivative $V_B'$ is single-crossing from above. To see this, we show that the ratio $V_B' (\ubar{\g}_B)/ g(\ubar{\g}_B)$ is strictly decreasing in $\ubar{\g}_B$. We have
\begin{equation} \label{eq:derivative_ratio}
    \frac{ V_B' (\ubar{\g}_B)}{g (\ubar{\g}_B)} = -  \E_{\th | \ubar{\g}_B} \Brac{ (v_B(\th) - p_B^M(\ubar{\g}_B) )_+} + \bar{Q}_A (\ubar{\g}_B) p_B^M(\ubar{\g}_B) +  U_A^{\mathrm{E}}(\ubar{\g}_B).
\end{equation}
This expression is differentiable except possibly at $\g = \g^{\dagger}$. To simplify notation, let $\tilde{Q}_B (\g) = Q_B^M( p_B^M(\g) |\g)$. Wherever the derivative of \eqref{eq:derivative_ratio} exists, it is given by
\begin{equation*}
\begin{aligned}
    &- \tilde{Q}_B (\ubar{\g}_B) + (p_B^M)'(\ubar{\g}_B) \tilde{Q}_B (\ubar{\g}_B) + \bar{Q}_A'(\ubar{\g}_B) p_B^M (\ubar{\g}_B) + \bar{Q}_A( \ubar{\g}_B) (p_B^M)' (\ubar{\g}_B) - \bar{Q}_A (\ubar{\g}_B) \\
    &= -( \bar{Q}_A ( \ubar{\g}_B) + \tilde{Q}_B ( \ubar{\g}_B)) \bigl[ 1 - (p_B^M)'(\ubar{\g}_B) \bigr] + \bar{Q}_A'(\ubar{\g}_B) p_B^M (\ubar{\g}_B) \\
    &< 0.
\end{aligned}
\end{equation*}
where the inequality holds because $Q_A^M + Q_B^M$ is strictly positive and the derivatives $p_B^M$ and $\bar{Q}_A'$ are weakly negative. 

Thus, $V_B$ is maximized where the derivative vanishes. Using \eqref{eq:exclusive_critical_position}, it is easily verified that the derivative vanishes at $\ubar{\g}_B = \g^{\dagger}$.


\paragraph{Pareto domination} Consider an arbitrary equilibrium $(\hat{s}_A, \hat{s}_B, \hat{p}, \hat{q})$ of the exclusive contracting game. We prove that $(s_A^{\mathrm{E}}, s_B^{\mathrm{E}}, p^{\mathrm{E}}, q^{\mathrm{E}})$ gives each firm weakly higher revenue than $(\hat{s}_A, \hat{s}_B, \hat{p}, \hat{q})$; moreover, if the revenue is not strictly higher for some firm, then $(\hat{s}_A, \hat{s}_B, \hat{p}, \hat{q})$ must be equivalent to $(s_A^{\mathrm{E}}, s_B^{\mathrm{E}}, p^{\mathrm{E}}, q^{\mathrm{E}})$. 

Using the argument in \cref{ft:coverage}, we conclude that the associated thresholds $\bar{\g}_A$ and $\ubar{\g}_B$ defined in \cref{res:structure_exclusivity} must be equal. Denote this common threshold by $\g_0$. It can be further verified that $\g_0$ is in $(\ubar{\g}, \bar{\g})$. Following the argument above in the equilibrium verification, it follows that each type $\g < \g_0$ chooses $p_A^M(\g)$ from firm $A$ and each type $\g > \g_0$ chooses $p_B^M(\g)$ from firm $B$. For each type $\g$ and firm $i$, let 
\[
  \hat{U}_i (\g) = \max_{p_i' \in [0, \infty]}  \Set{ - \hat{s}_i (p_i') + \E_{\th|\g} \Brac{ (v_i(\th) - p_i')_+}}.
\]
Each function $\hat{U}_i$ is convex, being a maximum over convex functions. From the envelope theorem, the subgradients of these functions satisfy
\[
    \pd \hat{U}_A( \g_0) \subseteq [- Q_A^M ( p_A^M(\g_0) | \g_0), 0] 
    \quad
    \text{and}
    \quad
    \pd \hat{U}_B( \g_0) \subseteq [0, Q_B^M ( p_B^M (\g_0) | \g_0) ].
\]
For each firm $i$, let $\hat{V}_i$ denote the value function in the inner problem, given the utility function $\hat{U}_{-i}$ at the other firm. By the argument in the equilibrium verification, we have
\begin{equation*}
\begin{aligned}
    \hat{V}_B( \ubar{\g}_B) &= \int_{ \ubar{\g}_B}^{\bar{\g}} \E_{\th| \g} [( v_B(\th) - p_B^M (\g))_+] g(\g) \de \g - \max\{ \hat{U}_A(\ubar{\g}_B), 0\} (1 - G(\ubar{\g}_B)), \\
    \hat{V}_A( \bar{\g}_A) &= \int_{ \ubar{\g}}^{\bar{\g}_A} \E_{\th| \g} [ (v_A(\th) - p_A^M (\g))_+] g(\g) \de \g - \max\{ \hat{U}_B(\bar{\g}_A),0\} G( \bar{\g}_A). 
\end{aligned}
\end{equation*}
For each $i$, let $\pd \hat{V}_i$ denote the supergradient of $\hat{V}_i$.\footnote{We are not claiming that $\hat{V}_i$ is globally concave. The supergradient is defined for all functions, though in general, it may be empty.} Since $(\hat{s}_A, \hat{s}_B, \hat{p}, \hat{q})$ is an equilibrium, we must have $\hat{U}_A( \g_0) = \hat{U}_B(\g_0)$ and $0 \in \pd \hat{V}_i (\g_0)$ for each $i = A, B$. Therefore, the triple $(\g_0, \hat{U}_A(\g_0), \hat{U}_B(\g_0)) \in \G \times \R^2$ satisfies
the system 
\begin{equation} \label{eq:bad_equilibria}
\begin{aligned}
\hat{U}_A( \g_0) &= \hat{U}_B (\g_0), \\
        0 \leq \E_{\th | \g_0} \Brac{ (v_B(\th) - p_B^M(\g_0) )_+} -\hat{U}_A(\g_0) &\leq Q_A^M ( p_A^M(\g_0) | \g_0) p_B^M (\g_0), \\
    0 \leq \E_{\th | \g_0} \Brac{ (v_A(\th) - p_A^M(\g_0) )_+} - \hat{U}_B(\g_0) &\leq Q_B^M ( p_B^M(\g_0) | \g_0) p_A^M (\g_0).
\end{aligned}
\end{equation}
If $\g_0 = \g^{\dagger}$ and $\hat{U}_A (\g_0) = \hat{U}_B (\g_0) = \D_B(\g^\dagger)$, then the equilibrium $(\hat{s}_A, \hat{s}_B, \hat{p}, \hat{q})$ is equivalent to $(s_A^{\mathrm{E}}, s_B^{\mathrm{E}}, p^{\mathrm{E}}, q^{\mathrm{E}})$. Otherwise, we claim that $(\hat{s}_A, \hat{s}_B, \hat{p}, \hat{q})$ is strictly Pareto dominated by $(s_A^{\mathrm{E}}, s_B^{\mathrm{E}}, p^{\mathrm{E}}, q^{\mathrm{E}})$. If $\g_0 = \g^{\dagger}$, and $\hat{U}_A (\g_0) = \hat{U}_B (\g_0) > \D_B(\g^\dagger)$, then this is immediate. So suppose $\g_0 \neq \g^{\dagger}$. Without loss, we may assume that $\g_0 > \g^{\dagger}$; the case $\g_0 < \g^{\dagger}$ is symmetric. Thus, $\D_B( \g_0) > \D_B (\g^{\dagger}) > 0$. Note that $\D_B'(\g) > Q_B^M(p_B^M(\g) | \g)$. We integrate this bound after applying \eqref{eq:bad_equilibria} to get
\begin{equation*}
\begin{aligned}
    \hat{U}_B(\g_0) 
    &= \hat{U}_A(\g_0) \\
    &\geq \D_B(\g_0) \\
    &> \D_B(\g^{\dagger}) + \int_{\g^{\dagger}}^{\g_0} Q_B^M ( p_B^M(\g') | \g') \de \g' \\
    &=  U_B^{\mathrm{E}} (\g^{\dagger}) + \int_{\g^{\dagger}}^{\g_0} Q_B^M ( p_B^M(\g') | \g') \de \g' \\ 
    &= U_B^{\mathrm{E}} (\g_0) \\
    &> U_A^{\mathrm{E}} (\g_0),
\end{aligned}
\end{equation*}
where the last inequality follows from  \cref{res:structure_exclusivity} since $\g_0 > \g^{\dagger}$. We conclude that each firm's revenue is strictly lower under  $(\hat{s}_A, \hat{s}_B, \hat{p}, \hat{q})$ than under $(s_A^{\mathrm{E}}, s_B^{\mathrm{E}}, p^{\mathrm{E}}, q^{\mathrm{E}})$. Indeed, in $(s_A^{\mathrm{E}}, s_B^{\mathrm{E}}, p^{\mathrm{E}}, q^{\mathrm{E}})$, each firm $i$'s revenue is weakly higher than its maximal revenue (in the inner problem) from choosing $\g_0$ as the critical type, and giving type $\g_0$ utility $\max \{ U_{-i}^{\mathrm{E}} (\g_0) , 0 \}$.

\subsection{Proof of Lemma~\ref{res:structure_exclusivity}}

For any subscription schedule $s_A$ for firm $A$ that is not identically $\infty$, the consumer's interim utility function $U_A$ is weakly decreasing, strictly so whenever the interim demand is strictly positive. Similarly, for any subscription schedule for firm $B$ that is not identically $\infty$, the consumer's interim utility function $U_B$ is weakly increasing, strictly so whenever the interim demand is strictly positive. Therefore, the difference $U_B( \g) - U_A(\g)$ is weakly increasing, strictly so if the interim demand at either firm is strictly positive. Let
\begin{equation*}
\begin{aligned}
    \bar{\g}_A &= \sup \{ \g  \in [\ubar{\g}, \bar{\g}] : U_A(\g) > U_B(\g) \}, \\    \ubar{\g}_B &= \inf \{ \g  \in [\ubar{\g}, \bar{\g}] : U_B(\g) > U_A(\g) \}, \\
\end{aligned}
\end{equation*}
where $\sup \varnothing = \ubar{\g}$ and $\inf \varnothing = \bar{\g}$. 

\subsection{Proof of Proposition~\ref{res:efficiency}}

By \cref{res:equilibrium},
\[
    q^{\mathrm{NE}} (\g, \th)
    =
    \begin{cases}
        (1,0) &\text{if}~ v_B(\th) - v_A(\th) < 2 ( p_B^M(\g) - p_A^M(\g)), \\
        (0,1) &\text{if}~ v_B(\th) - v_A(\th) \geq 2 ( p_B^M(\g) - p_A^M(\g)).
    \end{cases}
\]
By \cref{res:exclusive},
\[
    q^{\mathrm{E}} (\g, \th) 
    =
    \begin{cases}
        (0,0) &\text{if}~ v_B(\th) < p_B^M(\g), \\
        (0,1) &\text{if}~ v_B(\th) \geq p_B^M(\g).
    \end{cases}
\]
Fix $\g > 0$; the case $\g < 0$ is symmetric. We separate into two cases according to the position $\th$. 
\begin{itemize}
    \item Suppose $v_B(\th) - v_A(\th) \leq 2 ( p_B^M(\g) - p_A^M(\g))$. Since $p_B^M(\g) < p_A^M(\g)$, it follows that $v_B(\th) < v_A(\th)$. Therefore, the allocation $q^{\mathrm{NE}} (\g, \th) = (1,0)$ yields strictly higher surplus than the allocation $q^{\mathrm{E}} (\g, \th)$, which is either $(0,0)$ or $(0,1)$. 
    \item Suppose $v_B(\th) - v_A(\th) > 2 ( p_B^M(\g) - p_A^M(\g))$. We know $v_A(\th)  + v_B(\th)  = 2 v_0$. Adding these inequalities gives
    \[
        2 v_B(\th) > 2 p_B^M(\g) + 2 ( v_0 - p_A^M(\g)) > 2 p_B^M(\g),
    \]
    where the second inequality holds because $p_A^M(\g) \leq 1/g(\g) \leq v_0$. Therefore, the two allocations agree: $q^{\mathrm{NE}} (\g, \th) = q^{\mathrm{E}} (\g, \th) = (0,1)$. 
\end{itemize}

\subsection{Proof of Proposition~\ref{res:multi-product_monopoly}}

For the proof, we use a second-order version of the envelope theorem. To state this result, we introduce the setting of the envelope theorem in \cite{milgrom2002envelope}. Consider a set $X$ and a bounded function $u \colon X \times [0,1] \to \R$. Define the value function $V \colon [0,1] \to \R$ by $V(t) = \sup_{x \in X} u(x, t)$. For each $t \in [0,1]$, let $X^\ast (t) = \argmax_{x \in X} u(x, t)$. 

\begin{lem}[Second-order envelope theorem] \label{res:second_order_envelope} Suppose that for each $x$, the function $u(x, \cdot)$ is convex and twice-differentiable. The right-derivative $V_+'$ of $V$ exists at each point in $[0,1)$. Let $v = V_+'$. Fix $t \in (0,1)$. If $v$ is differentiable at $t$, then for all $x^\ast$ in $X^\ast(t)$, we have
\[
    v'(t) \geq u_{22} (x^\ast, t). 
\]
\end{lem}

With this preparation, we turn to the proof. By assumption, $G$ is symmetric, so $\ubar{\g} = - \bar{\g}$. For each $\k \in (0, F(2 \ubar{\g}))$, let 
\[
    C(\k) = \Brac{ \frac{ 2 f(0)}{\k} \vee \sup \{  |f'(\e)|/ f(\e): F^{-1}( F(2 \ubar{\g}) -\k) \leq \e \leq F^{-1} ( F( 2 \bar{\g}) + \k)  \} }.
\]
Let $C = \inf_{0 < \k < F( 2 \ubar{\g})} C(\k)$. Finally, let
\[
    \bar{v} (F,G) = C \max_{\g'} \frac{1}{g^2 (\g')}.
\]

Consider an incentive-compatible subscription mechanism $(p,s) \colon \G \to [0, \infty]^2 \times \R$. For any types $\g, \g' \in \G$, let
\[
    u(\g' | \g) = \E_{\th | \g} \Brac{ \max\{ v_A(\th) - p_A( \g'), v_B(\th) - p_B(\g'), 0 \}} - s(\g'). 
\]
The consumer's interim utility function is given by $U(\g) = \sup_{\g' \in \G} u(\g' |\g)$. The function $U$ is the pointwise maximum of $1$-Lipschitz functions, so $U$ is continuous. Let $\g_0$ be a minimizer of $U$ over $\G$. For each type $\g$, let  
\[
    \D(\g) = \P_{\th | \g} \Paren{ \max\{ v_A(\th) - p_A(\g), v_B(\th) - p_B( \g) \} \leq 0 }.
\]
In words, $\D(\g)$ is the probability that type $\g$ does not purchase either good under the mechanism $(p,s)$. 

Define the function $U_0 \colon \R \to \R_+$ by 
\[
    U_0(\g) =  \int_{0}^{\g} (1 - 2F(-y)) \de y.
\]

The next result implies that among all fully covered mechanisms, the symmetric fully covered mechanism minimizes expected information rent. 

\begin{lem}[Centering utility] \label{res:centering_utility} We have
\[
     \min_{\g' \in \G} \E_{\g} [ U_0 (\g - \g')] = \E_{\g} [ U_0 (\g)].
\]
\end{lem}

The heart of the proof is the following result, which controls how a reduction in coverage can decrease expected information rent. 


\begin{lem}[Utility difference] \label{res:util_difference_bound} We have
\[
  \E_{\g} [ U_0 (\g - \g_0)] - \E_{\g} [U(\g)] \leq \frac{C}{g^2(\g_0)} \E_{\g} [\D(\g)].
\]
\end{lem}

To prove \cref{res:util_difference_bound}, we use the second-order envelope theorem (\cref{res:second_order_envelope}) to obtain a lower bound on the second derivative of $U$. 

Now we combine these results to complete the proof. Suppose that $v_0 \geq \bar{v}(F,G)$. In particular, $v_0 \geq C /g^2 (\g_0)$. Let $\pi (p,s)$ denote the monopolist's expected revenue from the mechanism $(p,s)$. Let $\pi^\ast$ denote the monopolist's expected revenue from the mechanism in \cref{res:multi-product_monopoly}. Decomposing this revenue as surplus net of information rent, we get
\begin{equation*}
\begin{aligned}
   \pi(p,s) &\leq \E_{\g} \Brac{ \E_{\th | \g}  \max\{ v_A(\th), v_B(\th) \} } - v_0 \E_{\g} [ \D(\g)] - \E_{\g} [U(\g)] \\
    &\leq
    \E_{\g} \Brac{ \E_{\th | \g} \max\{ v_A(\th), v_B(\th) \} } - \frac{C}{g(\g_0)^2} \E_{\g} [ \D(\g)] - \E_{\g} [U(\g)] \\
    &\leq 
    \E_{\g} \Brac{ \E_{\th | \g} \max\{ v_A(\th), v_B(\th) \} } - \E_{\g} [U_0 (\g - \g_0)] \\
    &\leq 
    \E_{\g} \Brac{ \E_{\th | \g} \max\{ v_A(\th), v_B(\th) \} } - \E_{\g} [U_0(\g)] \\
    &= \pi^\ast,
\end{aligned}
\end{equation*}
where the third inequality follows from \cref{res:util_difference_bound}, and the final inequality follows from \cref{res:centering_utility}.

\subsection{Proof of Lemma~\ref{res:second_order_envelope}}

Fix $\e > 0$. Since $v$ is differentiable at $t$, there exists $\d > 0$ such that for all $h \in (-\d, \d)$, we have
\[
    |v(t + h) - v(t) - v'(t) h| \leq \e |h|.
\]
For each $k \in (-\d, \d)$, we have
\[
    V(t + k) - V(t) -  v(t) k - (1/2) v'(t) k^2 
    = \int_{0}^{k} (v(t + h)  -v(t) - v'(t) h) \de h,
\]
so 
\[
    |V(t + k) - V(t) -  v(t) k - (1/2) v'(t) k^2| \leq (1/2) \e k^2.
\]
Divide by $k^2$ and pass to the limit in $k$. Then pass to the limit in $\e$ to  conclude that
\begin{equation} \label{eq:V_limit}
    (1/2) v'(t)  =  \lim_{k \to 0} \frac{V(t + k) - V(t) -  v(t) k}{k^2}.
\end{equation}
Fix $x^\ast \in X^\ast (t)$. Applying a similar argument to $u(x^\ast,  \cdot)$ in place of $V$, we conclude that
\begin{equation} \label{eq:u_limit}
  (1/2) u_{22}(x^\ast, t)  =  \lim_{k \to 0} \frac{u (x^\ast, t + k) - u (x^\ast, t) -  u_{2} (x^\ast, t) k}{k^2}.
\end{equation}
Since $V(t) = u(x^\ast, t)$ and $v(t) = u_{2} (x^\ast, t)$ (by the Milgrom--Segal envelope theorem), subtracting \eqref{eq:u_limit} from \eqref{eq:V_limit} gives
\[
(1/2) [ v'(t) - u_{22} (x^\ast, t)]
= \lim_{k \to 0} \frac{ V(t + k) - u (x^\ast, t + k)}{k^2} \geq 0.
\]

\subsection{Proof of Lemma~\ref{res:centering_utility}}

We provide a direct proof, but the argument is essentially the proof of the antitone Hardy-Littlewood inequality. Since the density $f$ is symmetric (by \cref{as:regularity}), the function $U_0$ is convex and symmetric about $0$. Therefore, using the layer-cake representation, we have
\begin{equation*}
\begin{aligned}
    \E_{\g} [U_0(\g - \g_0)]
    &= \int_{0}^{\infty} \P ( U_0(\g - \g_0) \geq z) \de z  \\
    &= \int_{0}^{\infty} [1 - \P  (- U_0^{-1} (z) \leq \g - \g_0 \leq U_0^{-1}(z) )] \de z \\ 
    &= \int_{0}^{\infty} [1 - \P  (\g_0 - U_0^{-1} (z) \leq \g \leq \g_0 +  U_0^{-1}(z))] \de z.
\end{aligned}
\end{equation*}
To complete the proof, we show that for each fixed $\b > 0$, the function 
\[
    p( \g_0) =  \P  (\g_0 - \b \leq \g \leq \g_0 + \b)
\] 
is maximized at $\g_0 = 0$. To see this, note that 
\[
    p'(\g_0) = g ( \g_0 + \b) - g (\g_0 - \b). 
\]
The function $g$ is symmetric (by assumption) and single-peaked about $0$ (by \cref{as:regularity}). If $\g_0 < 0$, then $|\g_0 + \b| < | \g_0 - \b|$, so $p'(\g_0) > 0$. If $\g_0 > 0$, then $| \g_0 - \b|> |\g_0 + \b|$, so $p'(\g_0) < 0$. 

\subsection{Proof of Lemma~\ref{res:util_difference_bound}}

 We will first analyze types to the right of $\g_0$, and then we will apply a symmetric argument to types to the left of $\g_0$. Let $U_+'$ denote the right-derivative of the function $U$. Define the function $J \colon [\g_0, \bar{\g}) \to \R$ by 
\[
    J(\g) = \min \{ U_+'(\g), 1 - 2 F(\g_0 -\g) \}. 
\]
By construction, we have $0 \leq J(\g) \leq 1$ for all $\g \in [\g_0, \bar{\g})$. 

We claim that for every point $\g$ at which $J$ and $U_+'$ are both differentiable, we have 
\begin{equation} \label{eq:differential_claim}
    J'(\g) \geq 2 f(\g_0 - \g) - C \D( \g).
\end{equation}
To prove this claim, fix $\g \in [\g_0, \bar{\g})$ at which $J$ and $U_+'$ are both differentiable. In particular, $U_+'$ is continuous at $\g$, so $U$ is differentiable at $\g$. Fix $\k \in (0, F(2 \ubar{\g}))$. We prove that 
\begin{equation} \label{eq:differential_inequality}
     J'(\g) \geq 2 f(\g_0 - \g) - C(\k) \D( \g) .
\end{equation} 
 If $J(\g) = 1 - 2 F (\g_0 - \g)$, then $J'(\g) = 2 f( \g_0 - \g)$, which implies \eqref{eq:differential_inequality}. Therefore, we may assume that $J(\g) < 1 - 2 F( \g_0 - \g)$. In fact, we know that $0 \leq J(\g) < 1 - 2 F( \g_0 - \g)$. By the Milgrom--Segal envelope theorem, 
\[
     J(\g) = U_+'(\g) = U'(\g) = Q_B (p(\g) |\g) - Q_A( p(\g) | \g). 
\]
We separate into two cases. 

First, suppose $p_A(\g) + p_B(\g) \leq 2 v_0$. Let $\e^\ast = (p_B(\g) - p_A(\g))/2 - \g$. We have 
\[
   Q_B (p(\g) |\g) - Q_A( p(\g) | \g) =  1 - 2 F(\e^\ast), 
\]
so 
\[   0 \leq 1 - 2 F(\e^\ast) <  1 - 2F( \g_0  - \g).
\]
Thus, $\g_0 - \g  < \e^\ast \leq 0$. By the second-order envelope theorem (\cref{res:second_order_envelope}),
\[
    J'(\g) \geq u_{22} (\g | \g) = 2 f( \e^\ast ) > 2 f(\g_0 - \g),
\]
where the last inequality holds because $f$ is single-peaked about $0$. Hence, \eqref{eq:differential_inequality} is satisfied. 

Second, suppose $p_A(\g) + p_B(\g) > 2 v_0$. Let $\e_A = v_0 - p_A(\g) - \g$ and $\e_B = p_B(\g) - v_0 - \g$. Thus, $\e_A < \e_B$. We have
\[
    Q_B (p(\g) |\g) - Q_A( p(\g) | \g) =  1 - F(\e_B)  - F(\e_A), 
\]
so 
\[
    0 \leq 1 - F(\e_B)  - F(\e_A) < 1 - 2F( \g_0  - \g).
\]
Thus, $\e_A  + \e_B \leq 0$ and $\g_0 - \g < \e_B$. Since $\e_A < \e_B$,  we have $\e_A < 0$. By the second-order envelope theorem (\cref{res:second_order_envelope}), 
\begin{equation} \label{eq:J_deriv_uncovered}
    J'(\g) \geq u_{22} (\g|\g) = f(\e_A) + f(\e_B). 
\end{equation}
 If $\g_0 - \g \leq \e_A$, then $f(\g_0 - \g) \leq f(\e_A) < f(\e_B)$ because $f$ is single-peaked around $0$. In this case, \eqref{eq:differential_inequality} is satisfied. Therefore, we may assume that $\e_A < \g_0 - \g$. Hence, $\e_A < \g_0 - \g < \e_B$. If $\D(\g) \geq \k$, then $C(\k) \D( \g) \geq 2 f(0) \geq 2 f(\g_0 - \g)$, so \eqref{eq:differential_inequality} is immediate. Suppose that $\D(\g) < \k$. We have
\begin{equation} \label{eq:f_fprime}
\begin{aligned}
    & 2 f(\g_0 - \g) - (f (\e_B) + f ( \e_A))  \\
    &\leq  \int_{\e_A}^{\g_0 - \g} |f' (z)| \de z 
    + \int_{\g_0 - \g}^{\e_B} |f'(z)| \de z \\
    &= \int_{\e_A}^{\g_0 - \g} \frac{|f' (z)|}{f(z)} f(z) \de z 
    +  \int_{\g_0 - \g}^{\e_B} \frac{|f'(z)|}{f(z)} f(z) \de z \\
    &\leq  \sup \{ |f'(\e)|/f(\e) : \e_A \leq \e \leq \e_B \}  \int_{\e_A}^{\e_B} f(z) \de z \\
    &\leq C(\k) \D(\g),
\end{aligned}
\end{equation}
where the last inequality follows from the definition of $C(\k)$ since $2 \ubar{\g} \leq \g_0 - \g \leq 2 \bar{\g}$ and $F(\e_B) - F(\e_A) = \D(\g) < \k$. Combining \eqref{eq:f_fprime} with \eqref{eq:J_deriv_uncovered} gives \eqref{eq:differential_inequality}.

Having proven the claimed inequality \eqref{eq:differential_claim}, we can now complete the proof by repeated integration. Since $U$ is convex, the function $U_+'$ is weakly increasing. Thus, $J$ is weakly increasing as well. Therefore, $U_+'$ and $J$ are each differentiable almost everywhere. For every $\g \in (\g_0, \bar{\g}]$, integrating the (almost-everywhere) derivative of $J$ from $\g_0$ to $\g$ gives
\begin{equation} \label{eq:Lebesgue}
\begin{aligned}
    U_+'(\g) &\geq
    J(\g)  \\
    &\geq J(\g) - J(\g_0) \\
    &\geq \int_{\g_0}^{\g} J'(z) \de z \\
    &\geq \int_{\g_0}^{\g} [2f (\g_0 - z)
    -  C \D(z)] \de z \\
    &= 1 - 2 F( \g_0 - \g) - C \int_{\g_0}^{\g} \D(z) \de z,
\end{aligned}
\end{equation}
where the third inequality follows from a version of Lebesgue's fundamental theorem of calculus for weakly increasing functions,\footnote{We have equality if $J$ is absolutely continuous, but $J$ may not even be continuous.} and the fourth inequality follows from  the (almost everywhere) inequality \eqref{eq:differential_claim}.

From \eqref{eq:Lebesgue}, integrating $U_+'(y)$ from $y = \g_0$ to $y  = \g$ gives
\begin{equation} \label{eq:util_double_integral}
    U(\g) \geq U(\g) - U(\g_0) \geq U_0(\g - \g_0) - C  \int_{\g_0}^{\g} \Paren{ \int_{\g_0}^{y} \D(z) \de z } \de y.
\end{equation}
Next, we integrate \eqref{eq:util_double_integral} over $[\g_0, \bar{\g}]$ to get
\begin{equation} \label{eq:right_bound}
\begin{aligned}
    &\int_{\g_0}^{\bar{\g}} [U_0(\g - \g_0) - U(\g)] g(\g) \de \g   \\
    &\leq   C \int_{\g_0}^{\bar{\g}} \int_{\g_0}^{\g} \int_{\g_0}^{y} \D(z) g(\g) \de z \de y \de \g \\
    &=   C \int_{\g_0}^{\bar{\g}} \int_{z}^{\bar{\g}} \D(z) [1 - G(y)] \de y \de z  \\
    &= C \int_{\g_0}^{\bar{\g}}  \D(z) \Paren{ \int_{z}^{\bar{\g}} \frac{1 - G(y)}{g(z)} \de y} g(z) \de z   \\
    &\leq C \Paren{ \frac{ 1 - G(\g_0)}{g(\g_0)}}^2  \int_{\g_0}^{\bar{\g}}  \D(z) g(z) \de z,
\end{aligned}
\end{equation}
where we have used the fact that $(1 - G)/g$ is weakly decreasing (since $g$ is log-concave).

To obtain \eqref{eq:right_bound}, we established a pointwise lower bound on $U_+'$ over the interval $[\g_0, \bar{\g}]$ and then we integrated. Symmetrically, we can get a pointwise upper bound on $U_+'$ over the interval $[\ubar{\g}, \g_0]$ and then integrate to obtain
\begin{equation} \label{eq:left_bound}
\int_{\ubar{\g}}^{\g_0} [U_0(\g - \g_0) - U(\g)] g(\g) \de \g \leq C \Paren{ \frac{G(\g_0)}{g(\g_0)}}^2  \int_{\ubar{\g}}^{\g_0}  \D(z) g(z) \de z.
\end{equation}
Combining \eqref{eq:right_bound} and \eqref{eq:left_bound}, we conclude that
\begin{equation*}
\begin{aligned}
    \E_{\g} [ U_0 (\g - \g_0)] - \E_{\g} [U(\g)] 
    &\leq \frac{C}{g(\g_0)^2} \max\{ (1 - G(\g_0))^2, G^2(\g_0) \}  \E_{\g} [\D(\g)] \\
    &\leq \frac{C}{g(\g_0)^2} \E_{\g} [\D(\g)],
\end{aligned}
\end{equation*}
as desired.

\subsection{Proof of Proposition~\ref{res:MM_welfare}}

For each setting $j \in \{ \mathrm{NE}, \mathrm{SP}, \mathrm{E}, \mathrm{MM} \}$, let $U^j$ denote the consumer's interim utility function in setting $j$. By assumption, $G$ is symmetric, so the spot-pricing equilibrium price vector is given by $p^{\mathrm{SP}} = (1/h(0), 1/h(0))$, and in the exclusive equilibrium, $\g^{\dagger} = 0$. Therefore, from the envelope theorem, each function $U^j$ is minimized at $0$, and we have
\begin{align*}
    U^{\mathrm{E}} (\g) - U^{\mathrm{E}}(0) 
    &= \begin{cases} 
        \int_{\g}^{0} Q_A^M( p_A^M(\g') | \g') \de \g' &\text{if}~\g < 0, \\
            \int_{0}^{\g} Q_B^M( p_B^M(\g') | \g') \de \g'  &\text{if}~\g > 0, 
        \end{cases} \\
    U^{\mathrm{NE}}(\g) - U^{\mathrm{NE}}(0) 
    &= \int_{0}^{\g} [Q_B (2 p^M(\g')| \g') - Q_A(2 p^M(\g')| \g')] \de \g', \\
     U^{\mathrm{SP}} (\g)  - U^{\mathrm{SP}} (0) &=  \int_{0}^{\g} [ Q_B ( p^{\mathrm{SP}} | \g') - Q_A (p^{\mathrm{SP}} | \g') ] \de \g', \\
     U^{\mathrm{MM}}(\g)  &= \int_{0}^{\g} [ Q_B ( 0,0 | \g') - Q_A (0,0 | \g') ] \de \g'.
\end{align*}

To establish the consumer surplus comparison, we check that for each $j \in \{ \mathrm{NE}, \mathrm{SP}, \mathrm{E} \}$, and each type $\g \neq 0$, we have
\begin{equation} \label{eq:info_rent_lower_bound}
    U^j(\g) > U^{\mathrm{MM}} (\g).
\end{equation}
By symmetry, it suffices to prove \eqref{eq:info_rent_lower_bound} for $\g > 0$. For  each $\g > 0$, we have
\begin{equation*}
\begin{aligned}
    Q_B^M( p_B^M(\g) | \g) 
    &> Q_B (2 p^M(\g)| \g) - Q_A(2 p^M(\g)| \g) \\
    &> Q_B ( p^{\mathrm{SP}} | \g) - Q_A (p^{\mathrm{SP}}| \g)\\
    &= Q_B ( 0,0 | \g) - Q_A ( 0,0 | \g),
\end{aligned}
\end{equation*}
where the equality follows from full coverage. Therefore, 
\begin{equation*}
\begin{aligned}
    U^{\mathrm{E}}(\g) -  U^{\mathrm{E}}(0) 
    &> U^{\mathrm{NE}} (\g) - U^{\mathrm{NE}}(0) \\
    &>U^{\mathrm{SP}} (\g) - U^{\mathrm{SP}}(0) \\
    &= U^{\mathrm{MM}} ( \g) - U^{\mathrm{MM}}(0) \\
    &= U^{\mathrm{MM}}(\g).
\end{aligned}
\end{equation*}
Then \eqref{eq:info_rent_lower_bound} follows upon noting that $U^{\mathrm{E}}(0)$ and $U^{\mathrm{NE}}(0)$ are nonnegative (by the participation constraints) and $U^{\mathrm{SP}}(0) > 0$.

Finally, we use the consumer surplus comparison to derive the producer surplus comparison. For each setting $j$, let $\mathrm{TS}^j = \mathrm{CS}^j + \mathrm{PS}^j$. The allocation under the multi-product monopolist is efficient, so 
\[
    \mathrm{TS}^{\mathrm{MM}} \geq \max \{ \mathrm{TS}^{\mathrm{SP}}, \mathrm{TS}^{\mathrm{NE}}, \mathrm{TS}^{\mathrm{E}} \}.
\]
We showed above that 
\[
\mathrm{CS}^{\mathrm{MM}} < \min\{ \mathrm{CS}^{\mathrm{NE}}, \mathrm{CS}^{\mathrm{SP}} , \mathrm{CS}^{\mathrm{E}} \}.
\]
Subtracting gives
\begin{equation*}
\begin{aligned}
 \mathrm{PS}^{\mathrm{MM}}  
 &= \mathrm{TS}^{\mathrm{MM}}  - \mathrm{CS}^{\mathrm{MM}} \\
 &> \max \{ \mathrm{TS}^{\mathrm{SP}}, \mathrm{TS}^{\mathrm{NE}}, \mathrm{TS}^{\mathrm{E}} \} - \min\{ \mathrm{CS}^{\mathrm{NE}}, \mathrm{CS}^{\mathrm{SP}} , \mathrm{CS}^{\mathrm{E}} \} \\
 &\geq \max\{ \mathrm{PS}^{\mathrm{NE}}, \mathrm{PS}^{\mathrm{SP}}, \mathrm{PS}^{\mathrm{E}} \}.
\end{aligned}
\end{equation*}

\subsection{Proof of Proposition~\ref{res:surplus}}

For each $\s > 0$, we introduce notation for the $\s$-scaled environment.  For each $j \in \{ \mathrm{NE}, \mathrm{E}, \mathrm{SP} \}$, let $\mathrm{CS}^j (\s)$ and $\mathrm{PS}^j (\s)$ denote the (expected) consumer and producer surplus under setting $j$ in the $\s$-scaled environment. Let $\mathrm{TS}^j ( \s) = \mathrm{CS}^j (\s) + \mathrm{PS}^j (\s)$. We will compute the limits of these expressions as $\s$ tends to $0$. 

For each $\s >0$, let $G_\s$ and $g_{\s}$ denote the distribution and density of the random variable $\s \g$. Observe that $\min_{x \in [\s \ubar{\g} , \s \bar{\g}]} g_\s (x) = \min_{\g' \in \G} g(\g')/\s$. Therefore, $\max_{x \in [\s \ubar{\g} , \s \bar{\g}]} (1/ g_\s (x))$ tends to $0$ as $\s$ tends to $0$, so the inequality $v_0 \geq (7/2) \max_{x \in [\s \ubar{\g} , \s \bar{\g}]} 1/ g_\s (x) $ holds for all $\s$ sufficiently small. 

For each $\s >0$, let $H_\s$ and $h_\s$ denote the distribution and density of the random variable $\s \g + \e$. Define $\th^\ast (\s)$ by
\[
    \th^\ast (\s) = \frac{ 1 - 2 H_\s (\th^\ast (\s))}{h_\s (\th^\ast (\s))}.
\]
The distribution $F$ is symmetric, so as $\s$ tends to $0$, the critical position $\th^\ast (\s)$ tends to $0$, hence  $h_\s (\th^\ast (\s))$ tends to $f(0)$. By assumption, $v_0 >1 / f(0)$, so it follows that $v_0 > 1 / h_\s (\th^\ast (\s))$ for all $\s$ sufficiently small. 

With these preliminaries established, we calculate the limiting contracts as follows. 
\begin{itemize}
    \item Under non-exclusive contracting, consider the set  $S_i^{\mathrm{NE}}(\s)$ of all contracts selected from firm $i$ in equilibrium. As $\s$ tends to $0$, the set  $S_i^{\mathrm{NE}}(\s)$ converges in the Hausdorff metric to the singleton containing the contract $ (0, \E_{\th \sim F} [ (v_i(\th) - v_{-i}(\th)_+)_+] )$.
    \item Under spot pricing, the equilibrium prices $p_A^\ast (\s)$ and $p_B^\ast (\s)$ both converge to $1/f(0)$ as $\s$ tends to $0$.
    \item Under exclusive contracting, for each firm $i$, consider the set  $S_i^{\mathrm{E}} (\s)$ of all  non-null contracts selected from firm $i$ in equilibrium. As $\s$ tends to $0$, the set $S_i^{\mathrm{E}} (\s)$ converges in the Hausdorff metric to the singleton containing  $(0,0)$. 
\end{itemize}

We first consider the ranking of the limiting total surplus. We have 
\[
     \lim_{\s \to 0} \mathrm{TS}^{\mathrm{NE}}( \s) =  \lim_{\s \to 0} \mathrm{TS}^{\mathrm{SP}}( \s) =  \E_{
    \th \sim F} [ \max \{ v_A(\th)_+, v_B(\th)_+ \}]
\]
and
\[
   \lim_{\s \to 0} \mathrm{TS}^{\mathrm{E}}( \s) = \E_{\th \sim F} \Brac{ v_A(\th)_+} =  \E_{\th \sim F} \Brac{ v_B(\th)_+}.
\]
Thus, 
\[
\lim_{\s \to 0} \mathrm{TS}^{\mathrm{E}}( \s) \leq \lim_{\s \to 0} \mathrm{TS}^{\mathrm{NE}}( \s) \leq \lim_{\s \to 0} \mathrm{TS}^{\mathrm{SP}}( \s),
\]
so the producer surplus ranking follows from the consumer surplus ranking. 

For the consumer surplus ranking, we have
 \begin{equation*}
\begin{aligned}
     \lim_{\s \to 0} \mathrm{CS}^{\mathrm{NE}}( \s) &=  \E_{
    \th \sim F} [ \min \{ v_A(\th)_+, v_B(\th)_+ \}], \\
   \lim_{\s \to 0} \mathrm{CS}^{\mathrm{SP}}( \s) &= \E_{\th \sim F} \Brac{ \max \{ v_A(\th)_+ -1/f(0), v_B (\th)_+ - 1/f(0) \} }, \\
   \lim_{\s \to 0} \mathrm{CS}^{\mathrm{E}}( \s) &= \E_{\th \sim F} \Brac{ v_A(\th)_+} =  \E_{\th \sim F} \Brac{ v_B(\th)_+}.
\end{aligned}
 \end{equation*}

Next, we compare these limits. First, we compare the exclusive and non-exclusive limits. If $\th$ is distributed according to $F$, then the strict inequalities $v_A( \th)_+ > v_B( \th)_+$ and $v_B( \th)_+ > v_A(\th)_+$ both hold with positive probability. Thus,
\begin{equation*}
\begin{aligned}
     \lim_{\s \to 0} \mathrm{CS}^{\mathrm{E}} (\s)  
     &=  \E_{\th \sim F} [ v_B(\th)_+] \\
     &> \E_{\th \sim F} \Brac{ \min\{ v_A(\th)_+, v_B (\th)_+ \}}\\
     &= \lim_{\s \to 0} \mathrm{CS}^{\mathrm{NE}}( \s).
\end{aligned}
\end{equation*}

Second, we compare non-exclusive and spot pricing. We have
\begin{equation} \label{eq:NE_SP}
\begin{aligned}
 &\lim_{\s \to 0} \mathrm{CS}^{\mathrm{NE}}( \s) - \lim_{\s \to 0} \mathrm{CS}^{\mathrm{SP}} (\s)  \\
 & = 1 / f(0) - \E_{\th \sim F} \Brac{ ( v_A(\th) - v_B(\th)_+)_+} - \E_{\th \sim F} \Brac{  (v_B(\th) - v_A(\th)_+)_+} \\
 &= 1 / f(0) - 2  \E_{\th \sim F} \Brac{  (v_B(\th) - v_A(\th)_+)_+},
\end{aligned}
\end{equation}
where the last equality holds because the distribution $F$ is symmetric. Since $v_A(\th) < 0$ with positive probability under $F$, we have 
\begin{equation} \label{eq:ineq_F}
\begin{aligned}
    \E_{\th \sim F} \Brac{  (v_B(\th) - v_A(\th)_+)_+} 
    &< \E_{\th \sim F} [ (v_B(\th) -  v_A(\th) )_+] \\
     &= \int_{0}^{\infty} 2 \th f(\th) \de \th \\
     &= 2 \int_{0}^{\infty} (1 - F(\th)) \de \th \\
     &=    2 \int_0^{\infty} \frac{ 1 - F(\th)}{f(\th)} f(\th) \de \th  \\
     &\leq   2 \int_0^{\infty} \frac{ 1 - F(0)}{f(0)} f(\th) \de \th \\
     &= \frac{1}{2 f(0)},
\end{aligned}
\end{equation}
where the last inequality holds because $(1 - F) /f$ is weakly decreasing, and the last equality holds because $F(0) = 1/2$. Substituting \eqref{eq:ineq_F} into \eqref{eq:NE_SP}, we conclude that $\lim_{\s \to 0} \mathrm{CS}^{\mathrm{NE}}( \s) > \lim_{\s \to 0} \mathrm{CS}^{\mathrm{SP}} (\s)$. 


\subsection{Proof of Proposition~\ref{res:welfare_distribution}}

Since $G$ is symmetric, the spot-pricing equilibrium price vector is given by $p^{\mathrm{SP}} = (1/h(0), 1/h(0))$, and in the exclusive equilibrium, $\g^{\dagger} = 0$. Therefore, from the envelope theorem, each function $U^j$ is minimized at $0$, and we have 
\begin{align*}
    U^{\mathrm{E}} (\g) - U^{\mathrm{E}}(0) 
    &= \begin{cases} 
        \int_{\g}^{0} Q_A^M( p_A^M(\g') | \g') \de \g' &\text{if}~\g < 0, \\
            \int_{0}^{\g} Q_B^M( p_B^M(\g') | \g') \de \g'  &\text{if}~\g > 0, 
        \end{cases} \\
    U^{\mathrm{NE}}(\g) - U^{\mathrm{NE}}(0) 
    &= \int_{0}^{\g} [Q_B (2 p^M(\g')| \g') - Q_A(2 p^M(\g')| \g')] \de \g', \\
     U^{\mathrm{SP}} (\g)  - U^{\mathrm{SP}} (0) &=  \int_{0}^{\g} [ Q_B ( p^{\mathrm{SP}} | \g') - Q_A (p^{\mathrm{SP}} | \g') ] \de \g'.
\end{align*}
By the symmetry of $G$ and $F$, we conclude that each utility function is convex and symmetric about $0$. To complete the proof, it suffices to observe that for each $\g > 0$, we have
\[
    Q_B^M( p_B^M(\g) | \g) > Q_B (2 p^M(\g)| \g) - Q_A(2 p^M(\g)| \g) > Q_B ( p^{\mathrm{SP}} | \g) - Q_A (p^{\mathrm{SP}}| \g), 
\]
and for each $\g < 0$, we have
\[
    Q_A^M( p_A^M(\g) | \g) > Q_A(2 p^M(\g)| \g)  -  Q_B (2 p^M(\g)| \g) > Q_A (p^{\mathrm{SP}} | \g) - Q_B ( p^{\mathrm{SP}}| \g). 
\]
These inequalities are immediate by full coverage and the monotonicity properties of the demand functions.

\end{document}